\documentclass[twocolumn]{aastex631} 

\newcommand{\logg} {\log \textsl{\textrm{g}}}
\usepackage{xcolor,tikz}
\usepackage{gensymb}

\begin{document}

\title{Massive White Dwarfs in the 100 pc Sample: Magnetism, Rotation, Pulsations, and the Merger Fraction}

\email{gjewett@ou.edu}

\author[0009-0009-9105-7865]{Gracyn Jewett} 
\affiliation{Homer L. Dodge Department of Physics and Astronomy, University of Oklahoma, 440 W. Brooks St., Norman OK, 73019, USA}

\author[0000-0001-6098-2235]{Mukremin Kilic} 
\affiliation{Homer L. Dodge Department of Physics and Astronomy, University of Oklahoma, 440 W. Brooks St., Norman OK, 73019, USA}

\author[0000-0003-2368-345X]{Pierre Bergeron} 
\affiliation{Département de Physique, Université de Montréal, C.P. 6128, Succ. Centre-Ville, Montréal, QC H3C 3J7, Canada}

\author[0000-0001-7143-0890]{Adam Moss} 
\affiliation{Homer L. Dodge Department of Physics and Astronomy, University of Oklahoma, 440 W. Brooks St., Norman OK, 73019, USA}

\author[0000-0002-9632-1436]{Simon Blouin}
\affiliation{Department of Physics and Astronomy, University of Victoria, Victoria BC V8W 2Y2, Canada}

\author[0000-0002-4462-2341]{Warren R.\ Brown}
\affiliation{Center for Astrophysics, Harvard \& Smithsonian, 60 Garden Street, Cambridge, MA 02138, USA}

\author[0000-0002-9878-1647]{Alekzander Kosakowski}
\affiliation{Department of Physics and Astronomy, Texas Tech University, Lubbock, TX 79409, USA}

\author[0000-0002-2998-7940]{Silvia Toonen}
\affiliation{Anton Pannekoek Institute for Astronomy, University of Amsterdam, 1090 GE Amsterdam, The Netherlands}

\author[0000-0001-7077-3664]{Marcel A. Ag\"{u}eros}
\affiliation{Department of Astronomy, Columbia University, 550 West 120th Street, New York, NY 10027, USA}

\begin{abstract}
We present a detailed model atmosphere analysis of massive white dwarfs with $M > 0.9~M_\odot$ and $T_{\rm eff}\geq11,000$ K in the Montreal White Dwarf Database 100 pc sample and the Pan-STARRS footprint. We obtained follow-up optical spectroscopy of 109 objects with no previous spectral classification in the literature. Our spectroscopic follow-up is now complete for all 204 objects in the sample. We find 118 normal DA white dwarfs, including 45 massive DAs near the ZZ Ceti instability strip. There are no normal massive DBs: the six DBs in the sample are strongly magnetic and/or rapidly rotating. There are  20 massive DQ white dwarfs in our sample, and all are found in the crystallization sequence. In addition, 66 targets are magnetic (32\% of the sample). We use magnetic white dwarf atmosphere models to constrain the field strength and geometry using offset dipole models. We also use magnetism, kinematics, and rotation measurements to constrain the fraction of merger remnant candidates among this population. The merger fraction of this sample increases from 25\% for 0.9--$1~M_{\odot}$ white dwarfs to 49\% for 1.2--$1.3~M_{\odot}$. However, this fraction is as high as $78_{-7}^{+4}$\% for 1.1--$1.2~M_{\odot}$ white dwarfs. Previous works have demonstrated that 5--9\% of high-mass white dwarfs stop cooling for $\sim8$ Gyr due to the $^{22}$Ne distillation process, which leads to an overdensity of Q-branch stars in the solar neighborhood. We demonstrate that the over-abundance of the merger remnant candidates in our sample is likely due to the same process. 

\end{abstract}

\keywords{Magnetic fields (994) --- Stellar evolution (1599) --- White dwarf stars (1799)}
 
\section{Introduction} \label{sec:intro}

White dwarfs are the end state for the vast majority of stars \citep{fontaine01}, and studying populations of white dwarfs can reveal their origins. While we expect the majority of single white dwarfs to have evolved in isolation, population synthesis models predict that 10--30\% of single white dwarfs have a binary origin \citep{Temmink}. This is consistent with the observed discrepancies between the binary fractions of the local white dwarf sample \citep[$\sim$25\%,][]{holberg,toonen17} and their typical A-star progenitors \citep[$\sim$45\%,][]{derosa14}.  
This discrepancy implies a significant fraction of binaries are lost to mergers during their post-main-sequence evolution.

Observations allow us to constrain 
the evolutionary pathways that can produce a single white dwarf from a binary system. The most common pathway is the merger of a post-main-sequence and a main-sequence star, but for white dwarfs with a mass larger than $0.9~M_{\odot}$ (referred to as massive white dwarfs in this article), the dominant channel is the merger of double white dwarfs. \citet{Temmink} estimate that 30--50\% of single massive white dwarfs form through a binary merger, a fraction that is significantly higher than that for the more common $0.6~M_{\odot}$ white dwarfs.

To better understand the overall population of massive white dwarfs in the solar neighborhood, and specifically the merger rate among single massive white dwarfs, we have undertaken a spectroscopic survey of $M\geq0.9~M_{\odot}$ white dwarfs in the Montreal White Dwarf Database \citep[MWDD,][]{dufour} 100 pc sample and the Pan-STARRS footprint.

The identification of binary merger remnants in the local white dwarf population is possible, but not trivial. Perhaps the most well known
examples of merger remnants are the hot carbon-dominated atmosphere DQ white dwarfs with $T_{\rm eff} \approx$ 18,000-24,000 K \citep{dufour08}. Hot DQs are massive ($M\geq0.8~M_{\odot}$), and they have unusual atmospheric composition, high incidence of magnetism, rapid rotation, and relatively large
tangential velocities that are more consistent with a kinematically old population. These properties indicate a merger origin
\citep{dunlap15,coutu19,kawka23}. Warm DQs with $T_{\rm eff} \sim$ 10,000 - 18,000 K display many similarities with the hot DQ population
\citep{coutu19,koester19}. A recent analysis of the DAQ and warm DQ white dwarfs by \citet{kilic24} show that hot and warm
DQs are related and that both populations are likely white dwarf merger remnants.

Besides hot and warm DQs, other merger products among the local white dwarf sample can be identified based on their magnetism, kinematics, or rapid rotation. \citet{Garcia} demonstrate that the differentially rotating convective outer layers of a double white dwarf merger remnant can indeed produce a strong magnetic field \citep[see also][]{Briggs}. However, there are other explanations for the emergence of magnetic fields in white dwarfs, including fossil fields and crystallization induced dynamos \citep{Ferrario15,isern17,schreiber21,ginzburg22}. On the other hand, the presence of a strong field in a relatively hot (young) and massive white dwarf is a strong indication of a merger origin \citep{bagnulo22}. 

\citet{cheng20} show that binary merger remnants are expected to have a higher velocity dispersion because they are older, and use a sample of high mass white dwarfs identified in Gaia DR2 to estimate a double white dwarf merger fraction of $\sim$20\% among 0.8--1.3 $M_{\odot}$ white dwarfs. Merger products are also expected to rotate rapidly. \citet{Schwab} predicts that single white dwarfs formed from double white dwarf mergers have rotational periods of $\sim$10 min. Rapid rotation is common among hot DQ white dwarfs \citep{williams16}, and we have recently seen a surge in the number of rapidly rotating white dwarfs found with strong magnetic fields and/or with unusual atmospheric compositions
\citep[e.g.,][]{pshirkov20,caiazzo21,kilic21,moss23}. 

Empirical constraints on the fraction of mergers among the local white dwarf population are valuable for understanding the common-envelope evolution and the contribution of double white dwarfs to the type Ia supernovae rate \citep[e.g.,][]{cheng20}. A recent analysis
of the 25 most massive white dwarfs in the 100 pc sample found that as much as $56^{+9}_{-10}$\% of $M=1.3~M_{\odot}$ white dwarfs may form through mergers \citep{kilic23a}. This value is higher than the values predicted by the default binary population synthesis models, and favor efficient orbital shrinkage during the common envelope evolution. 

\begin{figure}
\includegraphics[width=3.5in, clip, trim={0.2in 0.1in 0.4in 0.4in}]{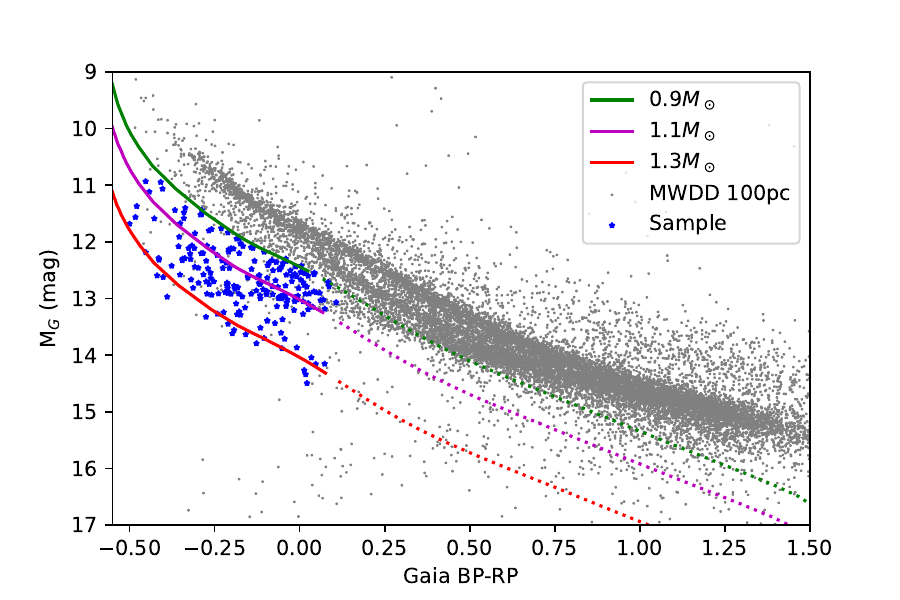}
\caption{Color magnitude diagram of the 100 pc Montreal White Dwarf Database sample in the Pan-STARRS footprint (gray points)
along with the $M>0.9~M_\odot$ white dwarfs with $T_{\rm eff}\geq 11,000$ K (blue stars) selected for spectroscopic follow-up. The green, magenta, and red lines show the evolutionary sequences for 0.9, 1.1, and $1.3~M_{\odot}$ white dwarfs for reference. The solid portions of the sequences represent $T_{\rm eff}\geq11,000$ K, as in our target selection.} 
\label{fig:100pc}
\end{figure}

\begin{deluxetable*}{crrrrrrccc}
\tabletypesize{\footnotesize}
\tablecolumns{10} \tablewidth{0pt}
\tablecaption{Observational properties of our massive white dwarf sample based on Gaia Data Release 3. The full table is available in the online version of this article.} \label{tab:mini gaia}
\tablehead{\colhead{Object name} & \colhead{Gaia ID} & \colhead{RA} & \colhead{DEC} & \colhead{Parallax} & \colhead{$\mu$$_{\rm RA}$ } & \colhead{$\mu$$_{\rm DEC}$ } & \colhead{G} & \colhead{$G_{\rm BP}$} & \colhead{$G_{\rm RP}$}\\
& & ($^{\circ}$) & ($^{\circ}$) & (mas) & (mas yr$^{-1}$) & (mas yr$^{-1}$) & (mag) & (mag) & (mag)}
\startdata
J0006$+$3104 & 2861452348130844160 & 1.65808 & 31.07098 & 10.19 $\pm{0.06}$ & 21.7 & $-$25.8 & 16.80 & 16.69 & 17.00 \\
J0012$-$0606 & 2443419990050464128 & 3.08603 & $-$6.10606 & 14.62 $\pm{0.07}$ & 123.5 & 14.2 & 16.35 & 16.31 & 16.42 \\
J0029$+$3648 & 366784816895496064 & 7.49632 & 36.80948 & 17.01 $\pm{0.06}$ & 90.9 & $-$45.9 & 16.42 & 16.31 & 16.66 \\
J0039$-$0357 & 2527618112309283456 & 9.78326 & $-$3.95607 & 11.30 $\pm{0.22}$ & 54.5 & $-$41.8 & 18.61 & 18.62 & 18.67 \\
J0043$-$1000 & 2377863773908424448 & 10.94092 & $-$10.00754 & 32.12 $\pm{0.04}$ & $-$145.6 & $-$134.9 & 14.53 & 14.43 & 14.70 \\
J0045$-$2336 & 2348747743931814656 & 11.36596 & $-$23.60878 & 21.17 $\pm{0.07}$ & 283.6 & $-$145.4 & 16.64 & 16.65 & 16.64 \\
J0049$-$2525 & 2345323551189913600 & 12.32153 & $-$25.43257 & 10.03 $\pm{0.25}$ & 22.5 & $-$28.3 & 19.03 & 19.08 & 19.04 \\
J0050$+$3138 & 360858960322547968 & 12.57920 & 31.64609 & 13.27 $\pm{0.13}$ & $-$90.1 & $-$38.7 & 18.08 & 18.06 & 18.17 \\
J0050$-$0326 & 2529337507976700928 & 12.69082 & $-$3.44882 & 12.58 $\pm{0.08}$ & $-$23.6 & $-$18.3 & 16.79 & 16.67 & 17.01 \\
J0050$-$2826 & 2342438501397962112 & 12.71700 & $-$28.43495 & 11.07 $\pm{0.11}$ & 69.3 & 16.1 & 17.81 & 17.86 & 17.80 \\
\enddata
\end{deluxetable*}

Here, we present the results of our spectroscopic survey of $M\geq0.9~M_{\odot}$ white dwarfs in the MWDD 100 pc sample and the Pan-STARRS footprint. 
We limit our follow-up to objects with estimated temperatures higher than 11,000 K so that helium lines can be detected, if present in the atmosphere. Figure \ref{fig:100pc} shows the Gaia color magnitude diagram of this sample. We identify 212 massive white dwarf candidates, 109 of which lacked spectral classification in the literature. 

We discuss the selection criteria for the sample and the follow-up observations in Section 2, and we present the model atmosphere analysis in Section 3. In Section 4, we present white dwarfs with photometric variability. We discuss the magnetic objects, kinematics, and the merger fraction in Section 5, and state our conclusions in Section 6. 

\section{Sample Selection and Observations}

\subsection{Sample Selection}

We selected $M > 0.9~M_\odot$ white dwarf candidates from the MWDD 100 pc sample. We limited our selection to the Pan-STARRS footprint so that we can take advantage of the Pan-STARRS $grizy$ photometry in our model fits. The Montreal database selection is based on the Gaia Data Release 2, and includes white dwarf candidates within 100 pc of the Sun (but allowing for the error on the parallax measurement) with $>10\sigma$ significant parallax and photometry \citep{dufour}. 

We limited our sample to objects with $T_{\rm eff}\geq11,000$ K so that helium lines would be visible, if present in the atmosphere. This selection results in 212 candidates, including 13 objects in common with the ultramassive white dwarf sample presented in \citet{kilic23a}. We are missing 12 objects from that sample, either because they have effective temperatures below our cutoff, or they are outside of the Pan-STARRS footprint.

We further remove eight objects from our sample. Two objects, Gaia DR2 161053615673941248 (WDJ044831.34+320652.18) and Gaia DR2 3966679722679277824 (LSPM J1121+1417) are IR-faint white dwarfs. These objects were originally included in the sample because of their unusual blue colors, but they are clearly much cooler than 11,000 K \citep{bergeron22}. One object, Gaia DR2 166587938734739328 (WDJ041642.45+321120.76), is 1.9 arcsec away from a late type star and is missing photometry in the redder Pan-STARRS bands.Three additional DA white dwarfs, Gaia DR2 692134843040270080 (WDJ090734.27+273903.44), Gaia DR2 4281190419601308672 (WDJ185450.45+041125.90), and Gaia DR2 2024985481361040384 (WDJ193618.58+263255.78), fall below our mass cut after a detailed model atmosphere analysis (see below). Finally, Gaia DR2 129352114170007680 (WDJ025431.45+301935.38) and Gaia DR2 1845487489350432128 (WDJ205351.74+270555.07) are DC white dwarfs where the surface temperature is either very close to or below our temperature cutoff depending on the assumed composition \citep{kilic21}. Hence, our final list includes 204 targets. Table \ref{tab:mini gaia} presents the observational properties of our massive white dwarf sample, including the Gaia source ID, astrometry, and photometry for each star.  

\subsection{Spectroscopy}
\label{obs}

We used five telescopes to obtain follow-up optical spectroscopy of 109 targets with missing spectral types in the literature.
We used the Apache Point Observatory (APO) 3.5 m telescope equipped with the Kitt Peak Ohio State Spectrograph \citep[KOSMOS,][]{martini14} to obtain spectroscopy of the majority of our targets. We used the blue grism with a $2\arcsec$ slit in the high or center slit positions,
which cover the wavelength ranges 4150--7050 and 3800--6600 \AA, respectively. We binned the CCD by 2$\times$2. We started our follow-up program using the high slit position, but switched to the center slit soon after to increase the blue coverage of the spectra. This setup provides spectra with a resolution of 1.4 \AA\ per pixel. 

At the Fred Lawrence Whipple Observatory (FLWO) 1.5 m telescope, we used the FAst Spectrograph for the Tillinghast Telescope \citep[FAST,][]{fabricant98} with  the 300 l mm$^{-1}$ grating and the $1.5\arcsec$ slit to obtain a spectral resolution of 3.6 \AA\ over the wavelength range 3500–7400 \AA.

At the MDM Observatory 2.4 m Hiltner telescope, we used the Ohio State Multi-Object Spectrograph (OSMOS) \citep{martini11} with the Blue VPH grism and the $1.2\arcsec$ inner slit to obtain a spectral resolution of 3.3 \AA\ over the wavelength range 3975--6865 \AA. These observations were done as part of the MDM OSMOS queue. 

At the 6.5 m Multi-Mirror Telescope (MMT), we acquired a spectrum of J0655+2939 using the Blue Channel spectrograph \citep{schmidt89} with the 500 l~mm$^{-1}$ grating and the 1.25 arcsec slit.  This set-up provided 4.5~\AA\ spectral resolution over $3850 < \lambda < 7000$~\AA. 

We obtained follow-up optical spectroscopy of 11 targets using the Gemini North and South 8 m telescopes equipped with the Gemini Multi-Object Spectrograph (GMOS) as part of the queue programs GN-2023A-Q-329 and GS-2023A-Q-328. We observed 10 of these targets at Gemini South with the B600 grating and a $1\arcsec$ slit, providing wavelength coverage from 3670 to 6800 \AA\ and a resolving power of $R=844$. We observed one target at Gemini North using the B480 grating, which provides a resolving power of $R = 761$. We obtained spectra for several additional targets at the 6.5m Magellan telescope with the MagE spectrograph. We used the $0.85\arcsec$ slit, providing wavelength coverage from about 3400 to 9400 \AA\ with a resolving power of $R = 4800$.

\section{Model Atmosphere Analysis}
\label{sec:model}

\subsection{Fitting Method}

We use the photometric technique as detailed in \citet{bergeron19}, \citet{Genest}, and \citet{blouin19}.
We use the SDSS $u$ (if available) and Pan-STARRS $grizy$ photometry along with the Gaia DR3 parallaxes to constrain
the effective temperature and the solid angle, $\pi (R/D)^2$, where $R$ is the radius of the star and $D$ is its distance. We include Galex photometry, when available, to use as an indicator of the atmospheric composition, but not in the fits themselves. 
Since the distance is precisely known from Gaia parallaxes, we can constrain the radius of the star directly, and therefore the
mass based on the evolutionary models for white dwarfs. Since our sample is restricted to 100 pc, we ignore reddening.

We convert the observed magnitudes into average fluxes using the appropriate zero points, and compare with the average
synthetic fluxes calculated from model atmospheres with the appropriate chemical composition. A $\chi^2$ value is defined
in terms of the difference between the observed and model fluxes over all bandpasses, properly weighted by the photometric uncertainties, which is then minimized using the nonlinear least-squares method of Levenberg-Marquardt \citep{press86} to obtain the best fitting parameters. Here we supplement our model grid with the warm DQ/DAQ white dwarf models from \citet{blouin19} and \citet{kilic24}. We also rely on the evolutionary models described in \citet{bedard20} with CO cores, $q({\rm He})\equiv \log M_{\rm   He}/M_{\star}=10^{-2}$, $q({\rm H})=10^{-4}$ and $10^{-10}$, which are representative of H- and  He-atmosphere white dwarfs, respectively.

\subsection{Summary of the Spectroscopic Survey}

\begin{figure}
\includegraphics[width=3.2in, clip, trim={0.7in 0.3in 0.8in 1.7in}]{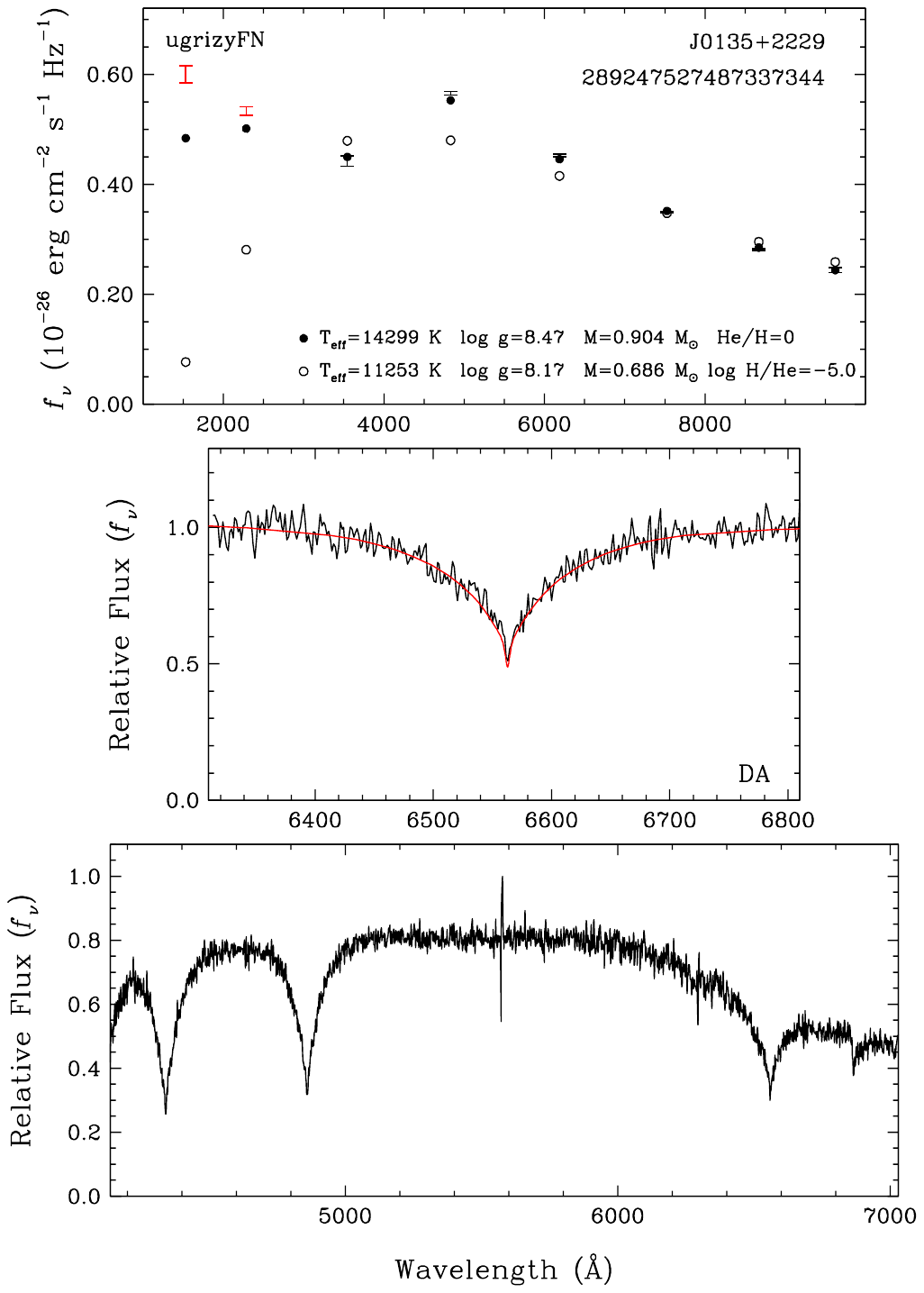}
\caption{Model fit to the DA white dwarf J0135+2229. The top panel shows the best-fitting pure hydrogen (filled dots) and helium-dominated (open
circles) atmosphere white dwarf models to the photometry (error bars). Gaia Source ID, object name, and the photometry used in the
fitting are included in this panel. The middle panel shows the predicted spectrum (red line) in the H$\alpha$ region based on the pure
hydrogen atmosphere solution. The bottom panel shows the entire spectrum. All fits are available in the online version of this article. }
\label{fig:da}
\end{figure}

\begin{deluxetable*}{cccccrc}
\tabletypesize{\footnotesize}
\tablecolumns{7} \tablewidth{0pt}
\tablecaption{The physical parameters for massive white dwarfs with $M>0.9~M_{\odot}$ and $T_{\rm eff}>11,000$ K in the 100 pc MWDD sample and the Pan-STARRS footprint. The full table is available in the online version of this article.} \label{tab:mini phys}
\tablehead{\colhead{Object name} & \colhead{Composition} & \colhead{Spectral Type} & \colhead{$T_{\rm eff}$} & \colhead{Mass} & \colhead{Cooling Age} & \colhead{Merger} \\
& & & (K) & ($M_\odot$) & (Gyr) & Evidence }
\startdata
J0006$+$3104 & H                 & DC     & 25442 $\pm$522 & 1.138 $\pm$0.010 & 0.20 $\pm$0.01 &  MR\\
J0012$-$0606 & H                 & DA     & 13730 $\pm$119 & 0.902 $\pm$0.006 & 0.55 $\pm$0.01  & \\
J0029$+$3648 & H                 & DA     & 25858 $\pm$313 & 1.284 $\pm$0.004 & 0.42 $\pm$0.02  & \\
J0039$-$0357 & H                 & DA     & 11871 $\pm$214 & 1.271 $\pm$0.009 & 2.09 $\pm$0.06  & \\
J0043$-$1000 & log(H/He)=$-$2.00 &  DBAH    & 18381 $\pm$371 & 1.077 $\pm$0.014 & 0.43 $\pm$0.03 & MR\\
J0045$-$2336 & log(H/He)=$-$4.00 & DQ     & 11540 $\pm$43 & 1.126 $\pm$0.004 & 1.80 $\pm$0.02 &  AV\\
J0049$-$2525 & H                 & DA     & 13018 $\pm$460 & 1.312 $\pm$0.010 & 1.72 $\pm$0.10  & \\
J0050$+$3138 & log(H/He)=$-$3.00 & He-DA & 12519 $\pm$221 & 1.215 $\pm$0.009 & 1.67 $\pm$0.05 &  \\
J0050$-$0326 & H                 & DC     & 23916 $\pm$355 & 1.213 $\pm$0.006 & 0.33 $\pm$0.02 &  MR\\
J0050$-$2826 & H                 & DA     & 11320 $\pm$155 & 1.061 $\pm$0.011 & 1.72 $\pm$0.08   & \\
\enddata
\tablecomments{Merger Evidence: A = Atmospheric composition, M = Magnetism, R = Rapid rotation, V = (large tangential) Velocity.}
\end{deluxetable*}

Our follow-up spectroscopy observations show that there are  170 DA white dwarfs in our sample, including two He-DAs, and 50 objects that are either confirmed or suspected to be magnetic. The He-DA spectral type represents the two DAs that show relatively weak Balmer lines for their effective temperatures that are best explained by helium-dominated atmospheres (see also \citealt{rolland}). There are only  6 massive DB white dwarfs, but none are normal: one is a rapidly rotating DBA \citep{pshirkov20} and  five are magnetic. There are 8 DC white dwarfs; given their effective temperatures above 11,000 K, those must be strongly magnetic so that their absorption features are shifted and distorted to the point where the spectrum becomes essentially a featureless continuum. The remaining 20 targets in our sample have C-rich atmospheres. There are 14 warm DQ/DQA, 5 DAQ white dwarfs,
and 1 hot DQ. Table \ref{tab:mini phys} presents the spectral types and the best-fitting model parameters, including the
effective temperature, mass, and the cooling age (assuming CO cores) for each target, which we now discuss in turn. 

\subsection{DA White Dwarfs}
The majority of the massive white dwarfs in the solar neighborhood are DA white dwarfs. Figure \ref{fig:da} shows the model fits for one of these targets. The top panel shows the predicted fluxes from the best-fitting pure hydrogen (filled dots) and helium-dominated (open circles) atmosphere models. The latter include trace amounts of hydrogen, since the 
split in the Gaia white dwarf sequence requires the presence of hydrogen or other electron donors in helium dominated atmosphere white dwarfs \citep{bergeron19,blouin23a,blouin23b,Camisassa}. 
The black error bars show the SDSS $u$ and Pan-STARRS $grizy$ photometry, and the red show Galex FUV and NUV photometry. We label the Gaia DR3 source ID, object name, and the photometry used in the fitting, also in this panel. The middle panel highlights the H$\alpha$ region of the spectrum with the predicted spectrum based on the pure hydrogen solution. The bottom panel shows the entire spectrum available for this source.

J0135+2229 is a relatively warm DA with several Balmer lines visible in its APO spectrum, and a significant Balmer jump also visible between its UV and optical photometry. J0135+2229 has GALEX FUV and NUV (red error bars) photometry available. Note that the UV photometry is not used in the model fits, but it is extremely valuable for discerning the atmospheric composition. The $ugrizy$ photometry and Gaia DR3 parallax indicate a pure H atmosphere with $T_{\rm eff} = 14,299 \pm 208$ K and $M = 0.904 \pm 0.011~M_{\odot}$. The predicted H$\alpha$ line profile for these parameters provide an excellent match to the observed spectrum, indicating that this is a pure hydrogen atmosphere white dwarf. 

\begin{figure*}
\centering
\includegraphics[width=3.1in, clip, trim={0.7in 0.3in 0.8in 0.6in}]{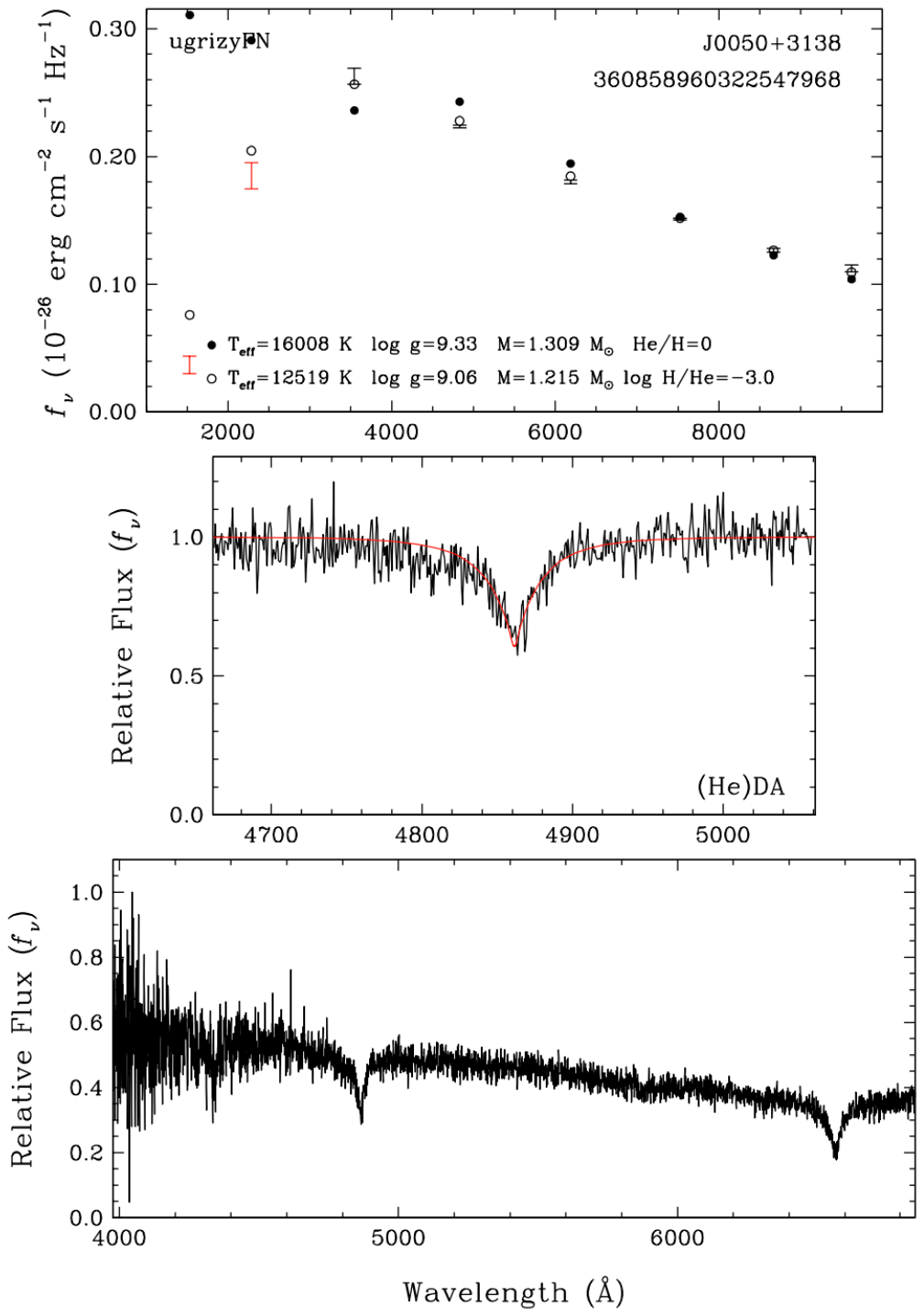}
\includegraphics[width=3.2in, clip, trim={0.7in 0.3in 0.8in 0.6in}]{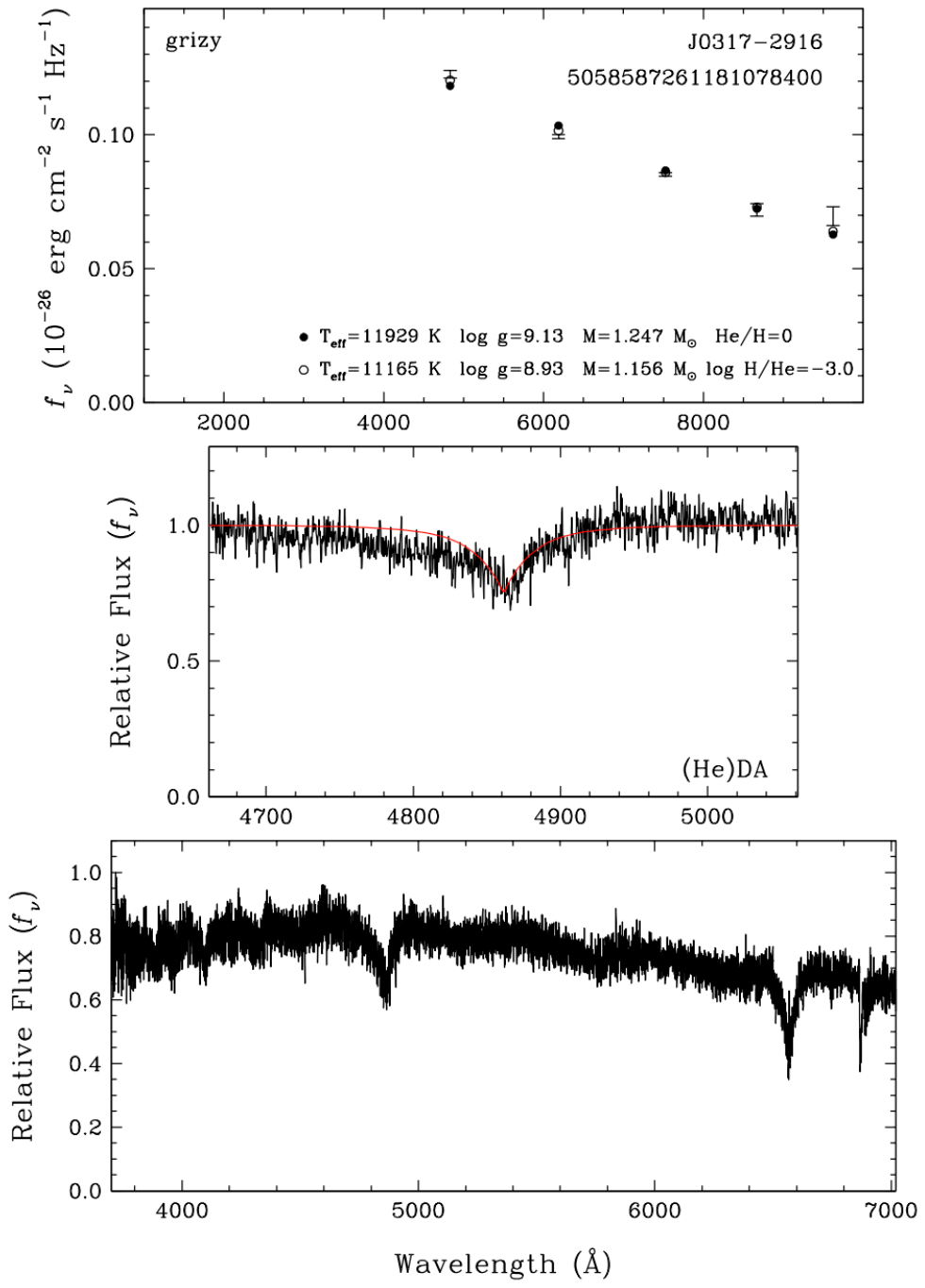}
\caption{Model fits to the two He-rich DA white dwarfs in our sample. The top panels show the best-fitting pure hydrogen (filled dots)
and mixed helium/hydrogen (open circles) atmosphere white dwarf models to the photometry (error bars). The middle panels compare the
observed H$\beta$ line profiles with those predicted from the mixed atmosphere solutions (red line). The bottom panels show the entire spectral
range of our observations.}
\label{fig:heda}
\end{figure*}

The fits shown in Figure \ref{fig:da} are representative of the entire DA white dwarf population in our sample. Because our sample is restricted
to $T_{\rm eff}\geq11,000$ K, all of the DA white dwarfs show relatively strong Balmer lines and the Balmer jump (if the SDSS
$u$-band or GALEX UV photometry is available).  The UV photometry is generally consistent with the pure hydrogen atmosphere model predictions for the DA white dwarfs in
our sample \citep[see for example Figure 2 in][]{wall23}. However, the FUV photometry is brighter than expected for J0135+2229. The source of this discrepancy
is unclear, though there are two FUV measurements available in the GALEX database, and they also differ from each other at the 2$\sigma$ level. 

\subsection{{He-DA} White Dwarfs}

There are two DA white dwarfs in our sample where the Balmer lines are significantly weaker than expected for pure H atmosphere
white dwarfs. Figure \ref{fig:heda} shows the model fits for these two stars. 
For J0050+3138 (left panels), the pure H atmosphere solution requires a surface temperature around 16,000 K. However,
the observed spectral energy distribution, especially in the Galex FUV and NUV bands, is incompatible with that solution, and instead
favors a helium-dominated atmosphere. Both the optical and UV photometry and the observed Balmer line profiles for J0050+3138
can be explained by a mixed atmosphere model with $\log {\rm H/He}=-3$, $T_{\rm eff} = 12,519 \pm 221$ K, and $M= 1.215 \pm 0.009~M_{\odot}$.  
The middle panel in  Figure \ref{fig:heda} shows that this model provides an excellent match to the H$\beta$ line profile, which is also
asymmetric. Such asymmetric hydrogen absorption features are seen in other He-DA white dwarfs as well \citep{caron}. Hydrogen lines in these stars are
heavily broadened through van der Waals interactions in helium dominated atmospheres. 

UV photometry for J0317$-$2916 (right panels) is unavailable. Even though we cannot clearly distinguish between the pure H and mixed H/He
atmosphere solutions based on the available photometry, the observed spectrum is very similar to J0050+3138 (left panels). The relatively
weak H$\alpha$ and H$\beta$ lines and the asymmetry of the H$\beta$ line (middle panel) all point to a mixed composition. A model with
$\log {\rm H/He}=-3$, $T_{\rm eff} = 11,165 \pm 346$ K and $M = 1.156 \pm 0.020~M_{\odot}$ provides an excellent match to the observed spectral
energy distribution of J0317$-$2916. 

\subsection{Magnetic White Dwarfs} 

\begin{figure*}
\hspace{-0.3in}
\includegraphics[width=2.5in, clip, trim={0.3in 0.3in 0.5in 0.6in}]{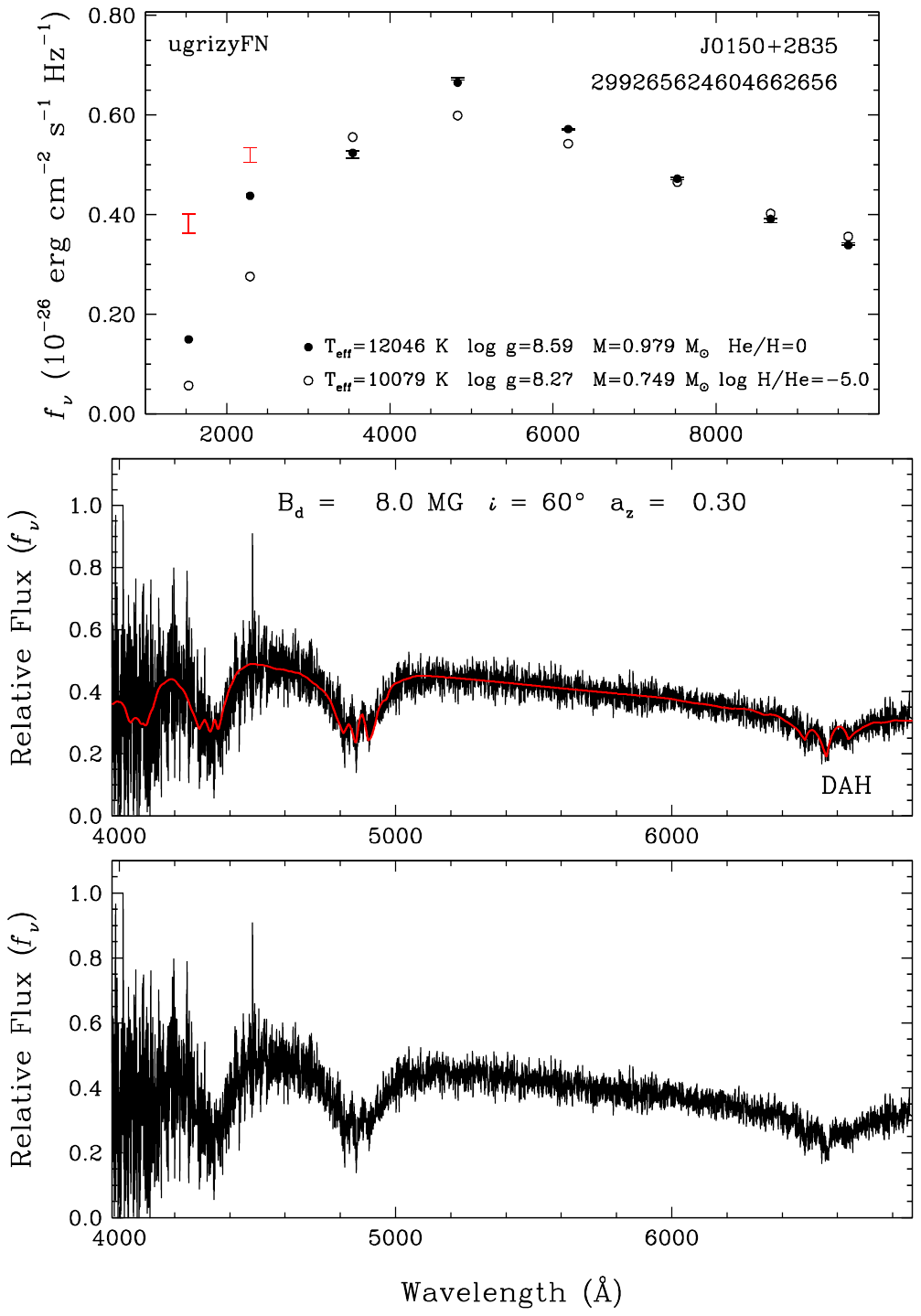}
\includegraphics[width=2.5in, clip, trim={0.3in 0.3in 0.5in 0.6in}]{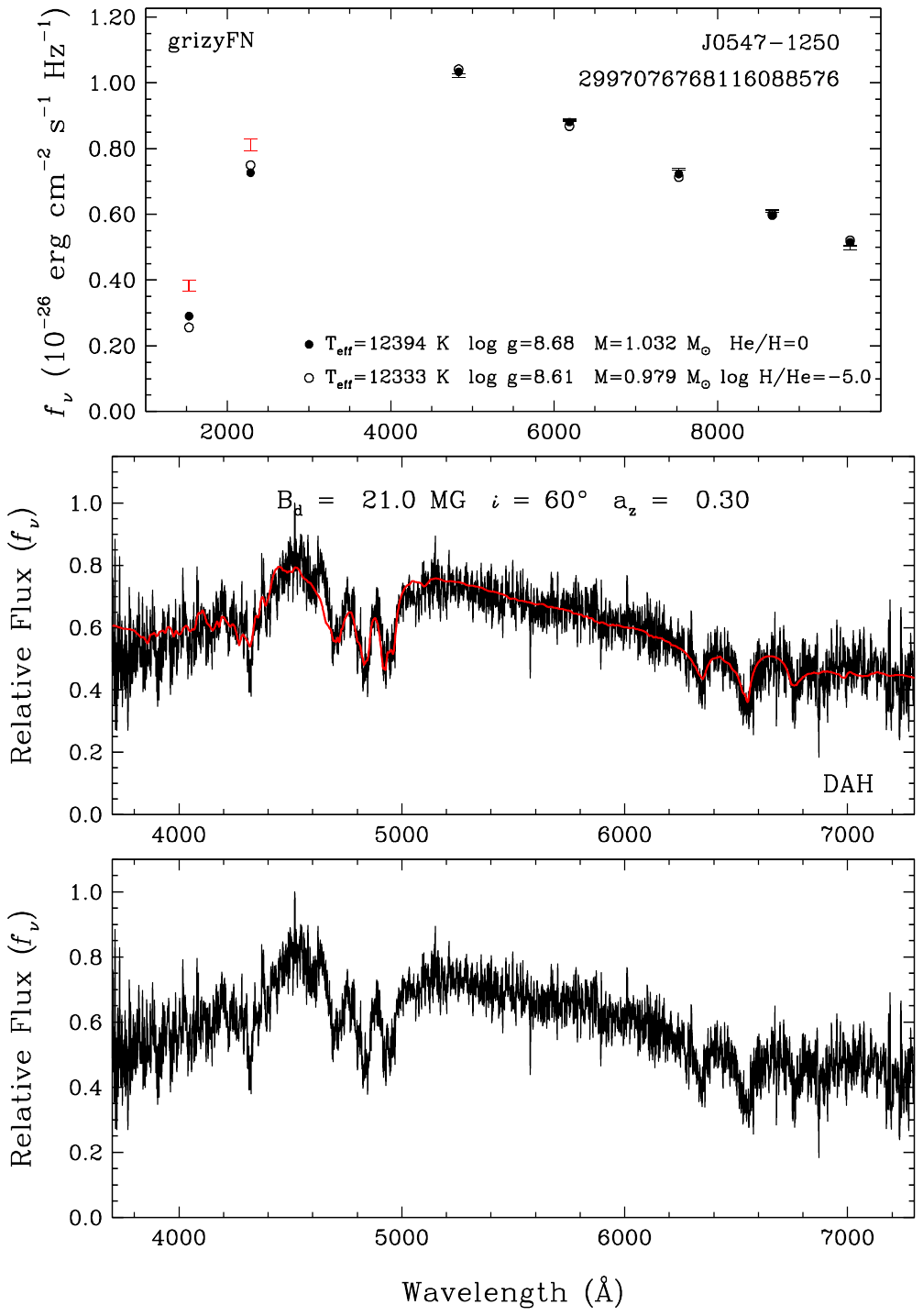}
\includegraphics[width=2.5in, clip, trim={0.3in 0.3in 0.5in 0.6in}]{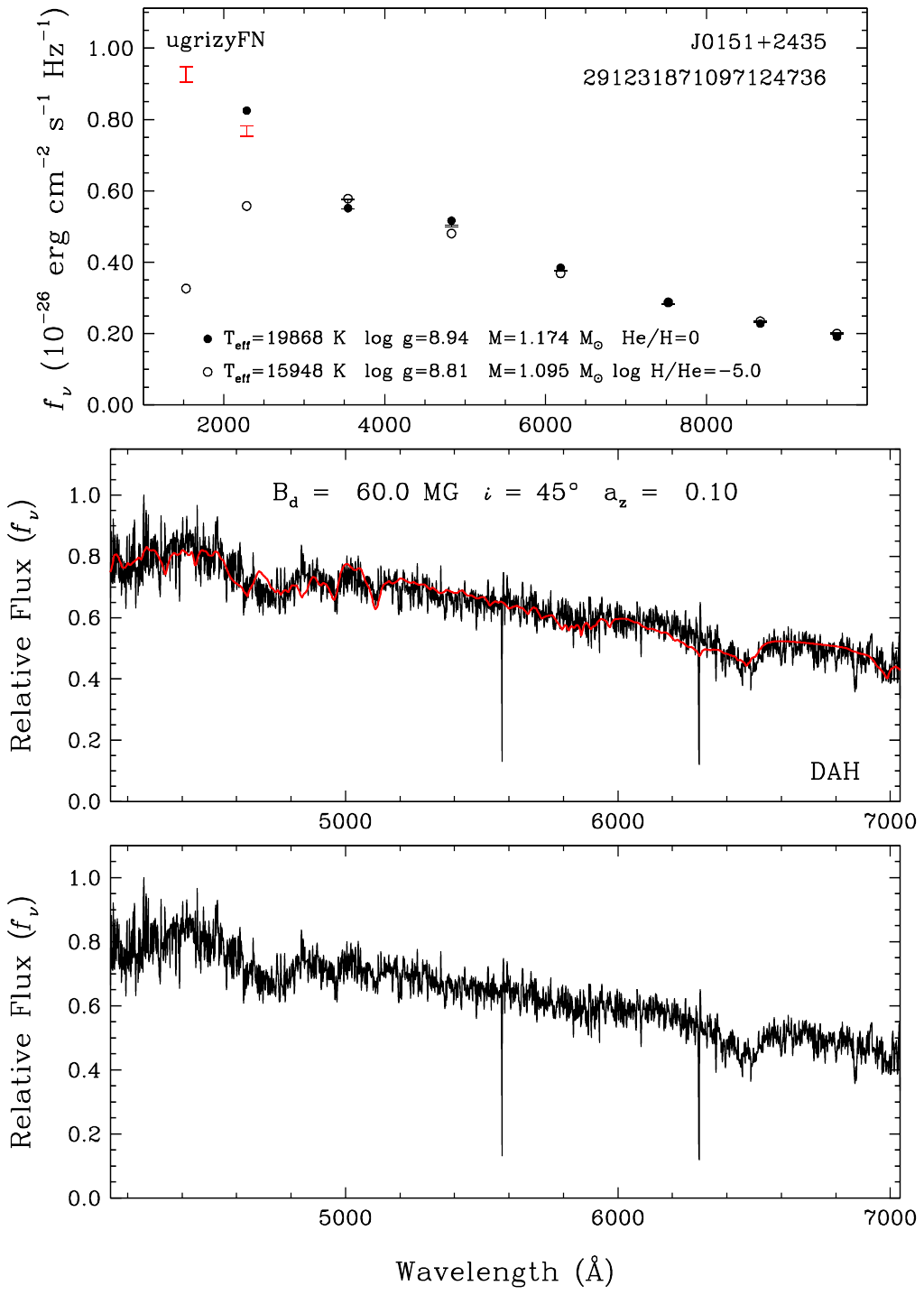}
\caption{Example model fits to three of the magnetic DA white dwarfs in our sample. Top row: results from our photometric fits, which we use to constrain the effective temperature and surface gravity of each object. Middle row: results from the spectroscopic fits, and the parameters for the best-fitting offset dipole models. Bottom row: The full, available spectrum. Model fits for all magnetic white dwarfs are available in the online version of this article.} 
\label{fig:dah}
\end{figure*}

There are  50 DA and  5 DB white dwarfs in our sample that are either confirmed or suspected to be magnetic,  plus 3 magnetic DQs, and 8 magnetic DCs. We model the magnetic DAH and DBH white dwarfs using a theoretical approach similar to that described in \citet{moss24} where the total line opacity is calculated as the sum of the individual Stark-broadened Zeeman components. We use the line displacements and oscillator strengths of the Zeeman components of H$\alpha$ through H$\delta$ and the neutral He lines kindly provided to us by S.~Jordan \citep[see also][]{hardy23}. For both H and He lines, the total line opacity is normalized to that resulting from the zero-field solution. The specific intensities at the surface, $I(\nu,\mu,\tau_{\nu}=0)$, are obtained by solving the radiative transfer equation for various field strengths and values of $\mu$ ($\mu= \cos \theta$, where $\theta$ is the angle between the angle of propagation of light and the normal to the surface of the star). The details of these magnetic models are further discussed in \citet{moss24}.

We created a model grid of magnetic spectra for each object using offset dipole models and the effective temperature and surface gravity obtained from the photometric method. These grids sample three parameters: the magnetic dipole field strength ($B_d$), the inclination between the line of sight and the dipole axis ($i$), and the dipole offset ($a_z$, in stellar radii). We used separate grids for magnetic DA and DB white dwarfs. Using this method, we are able to find excellent solutions for DA white dwarfs with $B<100$ MG. Above this limit the spectra become mostly featureless, or the observed absorption features are relatively broad, making it difficult to find a unique field geometry. Without spectropolarimetry, we are therefore limited in our ability to constrain the strengths of the largest fields observed in white dwarfs. 

Figure \ref{fig:dah} shows our best photometric and spectroscopic fits to three of the DAH white dwarfs in our sample, all with $B<100$ MG. The Zeeman split H$\alpha$ and H$\beta$ lines are clearly visible for J0150+2835 and J0547$-$1250 (left and middle panels). Our magnetic DA model fits produce excellent solutions for both objects, constraining the field strength to 8 and 21 MG, respectively. For J0151+2435 (right panels), the absorption features are more complex. However, thanks to the UV photometry from Galex,
the photometric fit clearly favors a pure hydrogen solution, and our magnetic model fits under the assumption of a pure hydrogen
atmosphere provide an excellent match to the observed spectrum for a dipole field strength of 60 MG.

\begin{deluxetable*}{ccccccc}
\tabletypesize{\scriptsize}
\tablecolumns{7} \tablewidth{0pt}
\tablecaption{Results from the model fits to the magnetic white dwarfs in our sample. 
We present the results from both our offset dipole fits ($B_d, i, a_z$) and the literature ($B_{d, Lit}$), if available.} \label{tab:fields}
\tablehead{\colhead{Object name} & \colhead{Spectral Type} & \colhead{$B_d$} & \colhead{$B_{d, Lit}$} & \colhead{i} & \colhead{$a_z$} & \colhead{\textbf{Reference}}\\
& & (MG) & (MG) & ($^{\circ}$) & ($R_{\star}$) }
\startdata 
J0043$-$1000 &  DBAH  & 34 & 39 & 60 & 0.19 & \cite{hardy23}\\
J0118$-$0156 & DAH  & \nodata & \nodata & \nodata & \nodata & \\
J0150$+$2835 & DAH  & 8 & \nodata & 60 & 0.30 & \\
J0151$+$2435 & DAH  & 60 &\nodata & 45 & 0.10 & \\
J0211$+$2115 & DAH  & 207 & 166 & 45 & 0.40 & \cite{Kulebi}\\
J0216$+$3541 & DBAH  & 235 & \nodata & 30 & $-$0.30 & \\
J0230$+$3842 & DAH  & 65 & \nodata & 30 & 0.30 & \\
J0249$-$1831 & DAH  & 30 & \nodata & 15 & 0.30 & \\
J0256$-$1515 & DAH  & 45 & \nodata & 45 & 0.30 & \\
J0257$+$0308 & DAH  & 65 & \nodata & 60 & 0.30 & \\
J0319$+$4628 & DAH  & 189 & \nodata & 60 & 0.36 & \\
J0326$+$1331 & DAH  & 4 & \nodata & 15 & 0.30 & \\
J0507$+$2645 & DAH  & 5 & \nodata & 60 & 0.30 & \\
J0547$-$1250 & DAH  & 21 & \nodata & 60 & 0.30 & \\
J0547$+$1501 & DAH  & 8 & \nodata & 60 & 0.30 & \\
J0601$+$3726 & DAH  & 3.4 & 2.3 & 60 & 0.30 & \cite{hardy23}\\
J0602$+$4652 &  DQH?  & \nodata & \nodata & \nodata & \nodata & \\
J0607$+$3415 & DAH  & 3.2 & \nodata & 45 & 0.30 & \\
J0625$+$1902 & DAH & 168 & \nodata & 30 & 0.34 & \\
J0705$-$2046 & DAH  & 30 & \nodata & 45 & 0.30 & \\
J0803$+$1229 & DAH  & 35 & 39 & 15 & 0.30 & \cite{hardy23}\\
J0842$-$0222 & DAH? & \nodata & \nodata & \nodata & \nodata & \\
J0851$+$1201 & DAH  & 1.8 & 2 & 60 & 0.30 & \cite{hardy23}\\
J0951$-$1517 & DAH  & 382 & \nodata & 45 & 0.30 & \\
J1014$-$0417 & DAH  & 2 & \nodata & 60 & 0.30 & \\
J1034$+$0327 & DAH  & 13 & 13.3 & 75 & 0.30 & \cite{amorim}\\
J1046$-$0518 & DBH  & \nodata & 820 & \nodata & \nodata & \cite{schmidtmag}\\
J1054$+$5523 & DAH: & 0.2 & \nodata & 30 & 0.30 & \\
J1105$+$5225 & DAH? & \nodata & \nodata & \nodata & \nodata & \\
J1107$+$8122 & DAH  & 30 & \nodata & 45 & 0.30 & \\
J1214$-$1724 & DBH  & 62 & \nodata & 15 & 0.35 & \\
J1217$+$0828 & DAH  & 3.2 & 3.5 & 60 & 0.30 & \cite{hardy23}\\
J1333$+$6406 & DAH  & 13 & 13 & 75 & 0.30 & \cite{hardy23}\\
J1339$-$0713 & DAH  & 30 & & 45 & 0.30 & \\
J1440$-$1951 & DAH? & \nodata & \nodata & \nodata & \nodata & \\
J1459$-$0411 & DAH & 52 & \nodata & 30 & 0.30 & \\
J1543$+$3021 & DAH  & 160 & \nodata  & 15 & 0.30 & \\
J1548$+$2451 & DAH  & 6.4 & 7 & 60 & 0.20 & \cite{hardy23}\\
J1621$+$0432 & DAH  & \nodata & \nodata  & \nodata & \nodata & \\
J1630$+$2724 & DAH  & 35 & 35.6 & 45 & 0.30 & \cite{amorim}\\
J1659$+$4401 & DAH  & 3.8 & 4 & 60 & 0.30 & \cite{hardy23}\\
J1707$+$3532 & DAH  & 2.8 & 2 & 60 & 0.30 & \cite{hardy23}\\
J1719$-$1446 & DAH? & \nodata & \nodata  & \nodata & \nodata & \\
J1723$+$0836 & DAH  & 44 & \nodata & 15 & 0.15 & \\
J1849$+$6458 &  DQH? & \nodata & \nodata & \nodata & \nodata & \\
J1900$+$7039 & DAP  & 164 & 320 & 15 & 0.20 & \cite{angelmag}\\
J1901$+$1458 & DAH  & \nodata & 600-900 & \nodata & \nodata & \cite{caiazzo21}\\
J1924$-$2913 & DAH  & 30 & \nodata & 30 & 0.00 & \\
J2012$+$3113 & DBP  & \nodata & 520 & \nodata & \nodata & \cite{berdmag}\\
J2035$-$1835 & DAH  & 104 & \nodata & 15 & $-$0.10 & \\
J2100$+$5142 & DAH  & 60 & \nodata & 15 & 0.10 & \\
J2111$+$1102 & DAH  & 3.6 & 2.8 &  60 & 0.30 & \cite{amorim}\\
J2148$-$1629 & DAH  & 3.2 & \nodata & 75 & 0.30 & \\
J2204$+$2543 & DAH  & \nodata & \nodata & \nodata & \nodata & \\
J2221$+$4406 & DAH  & 11 & \nodata & 45 & 0.30 & \\
J2255$+$0710 & DAH & 230 & \nodata & 0 & 0.21 & \\
J2257$+$0755 & DAH  & 8.8 & 16 & 0 & 0.10 & \cite{Kulebi}\\
\enddata
\end{deluxetable*}  

A significant fraction of magnetic white dwarfs are variable on a short time scale \citep{brinkworth,moss23}, and therefore our spectra most likely represent the average of the magnetic field distribution across the entire surface. It is beyond the scope of this paper to obtain phase resolved spectroscopy, and our field strength estimates are likely a proxy for the overall magnetic field strength of each star. It is also possible that we are missing additional magnetic white dwarfs with lower field strengths, as low resolution spectroscopy is insensitive to fields smaller than about 100 kG in massive DA white dwarfs \citep[see for example][]{kilic2015}. 

Table \ref{tab:fields} shows the results from our analysis of the magnetic DAH and DBH white dwarfs in the sample. In addition to the dipole field strength ($B_d$), the inclination ($i$), and the offset ($a_z$) from our best-fitting models, we also provide the dipole field strength as reported in the literature for the same stars. We do not report model fits to stars where the spectra are almost featureless, since we do not trust the constraints based on the relatively broad and shallow features observed in these stars. 

There are several magnetic white dwarfs that require further work to get a better fit to their spectra. The first one, J0043$-$1000, is the well-known patchy atmosphere
DBAH white dwarf Feige 7 \citep{liebert77,achilleos92} where variable surface abundances of hydrogen and helium are observed as the star rotates. J0216+3541 is similar, as it also
shows both hydrogen and helium features. While we assign a helium composition to J1046$-$0518 and J2012+3133, we suspect that they also have patchy atmospheres. For J0602+4652 and J1849+6458 it is impossible to match the observed features using the pure hydrogen or the helium solution, so we suggest that these objects are potentially magnetic DQ white dwarfs. We plan on investigating these objects in future work.

Eighteen targets have a field measurement in the literature \citep{hardy23,Kulebi,caiazzo21,schmidtmag,angelmag,berdmag,amorim}. Our results are remarkably similar to the overall literature values. The largest difference is for J2257+0755, where our model fit indicates an 8.8 MG field, whereas \cite{Kulebi} obtained 17.39 MG for the same star. However, they obtained an inclination angle of 74.9\degree and an offset of 0.15. It is likely that due to the large degeneracies in spectral fitting of magnetic white dwarfs, we can achieve excellent spectral fits even with different parameters.

Interestingly, none of our magnetic model fits require a negative dipole offset ($a_z<0$). In fact, the majority of our targets are best-explained with offset dipoles with $a_z=+0.3$. This is similar to the results from \cite{hardy23}, who found a positive offset for 131 of the 140 magnetic white dwarfs with good solutions. In their sample, 97 of the 140 objects have $a_z\geq+0.25$. It is not clear why this offset is generally positive. One possibility is that positive $a_z$ makes the central $\pi$ component appear deeper/stronger than the split $\sigma$ components \citep[see Figure 4 in][]{bergeron92}, which may make it easier to identify the shifted features in magnetic white dwarfs in the presence of noise. However, a more likely explanation is that these are not simple offset dipoles, and that the models we use only serve as a proxy to the true field geometry. More work is required for understanding  the source of this bias in $a_z$. 

\subsection{DC White Dwarfs}

There are 8 DC white dwarfs in our sample with featureless optical spectra. These objects are J0006+3104, J0050$-$0326, J0327+2227, J0707+5611, J0718+3731, J1010$-$2427, J1537+8419, and J2026+1848. They have best-fitting
$T_{\rm eff}>14,000$ K, regardless of their atmospheric composition (pure hydrogen or helium-dominated atmospheres). Hence,
the only way for these stars to have a featureless spectrum is if they are strongly magnetic: their absorption features
are shifted and spread out beyond recognition. 

More importantly, five of these objects also show photometric variability due to rapid rotation in the Transiting Exoplanet Survey Satellite (TESS) or the Zwicky Transient Facility (ZTF) data. J0006+3104, J0050$-$0326, J0327+2227, J0707+5611, and J0718+3731 show significant photometric variations with periods of 23.15, 40.31, 58.6, 63, and 11.27 min, respectively (see below). Such rotation periods are common among magnetic white dwarfs \citep{kawka20}, and the variability is likely due to changes in the magnetic field structure and/or an inhomogeneous atmosphere \citep{moss23,moss24}.  

\subsection{Warm DQs}

The most common DQ white dwarfs in the solar neighborhood are cool DQs with temperatures below 10,000 K. The trace amounts
of carbon seen in these stars is well explained by the convective dredge-up of carbon from the deep interior in helium dominated atmospheres \citep{pelletier86,bedard22}. \citet{blouin23a,blouin23b} show that carbon dredge-up is ubiquitous in hydrogen-deficient white dwarfs, and that dredge-up of optically undetectable traces of carbon is crucial for explaining the bifurcation seen in the Gaia H-R diagram. 

Hot and warm DQs are remarkably different than the cool DQs \citep{dufour08}. Hot DQs with $T_{\rm eff}\approx18,000$--24,000 K and warm DQs with $T_{\rm eff}\approx11,000$--18,000 K are massive, they have unusually C-rich or C-dominated atmospheres, have large tangential velocities, and some show evidence of rapid rotation, or magnetism \citep{koester19,coutu19,kawka23}. These properties all point to a common origin in white dwarf mergers \citep{dunlap15}. 

We identify 20 DQ white dwarfs (including the two magnetic DQH candidates discussed above) in our survey, 12 of which are new discoveries. There is only 1 hot DQ known in this sample, J1819$-$1208 \citep{kilic23a}. Hence, our survey has more than doubled the sample of warm DQs within 100 pc.  In addition, there was only 1 DAQ white dwarf known prior to this work, J0551+4135 \citep{hollands}. We identified four additional DAQ white dwarfs within 100 pc, three of which are presented in a separate publication by \citet{kilic24}.

The last DAQ to be identified is J0655+2939, which provides an interesting story. J0655+2939 was classified as a DA white dwarf by \citet{kilic20} based on a relatively noisy spectrum that clearly showed Balmer lines. \citet{wall23} compared the best-fitting parameters for DA white dwarfs derived from optical only and optical + UV data, and identified five outliers in the 100 pc sample, including J0655+2939 (see their Figure 2). The other four outliers are either magnetic or have mixed atmospheres. J0655+2939 is more than a factor of two fainter than expected in the FUV band assuming a pure H atmosphere composition. \citet{wall23} further investigated this object by obtaining a new spectrum at the APO 3.5m telescope, and confirmed the DA spectral type. They concluded that the source of the FUV flux discrepancy is unclear. Looking at this relatively noisy spectrum shown in their Figure 3, we noticed that there may be weak C features near H$\beta$. This prompted us to observe J0655+2939 once more, but at a bigger telescope. We were able to obtain a better quality spectrum at the 6.5m MMT in 2024 May using the same setup as described in Section \ref{obs}.

\begin{figure}
\includegraphics[width=3.2in, clip, trim={0.5in, 3in, 0.5in, 1.5in}]{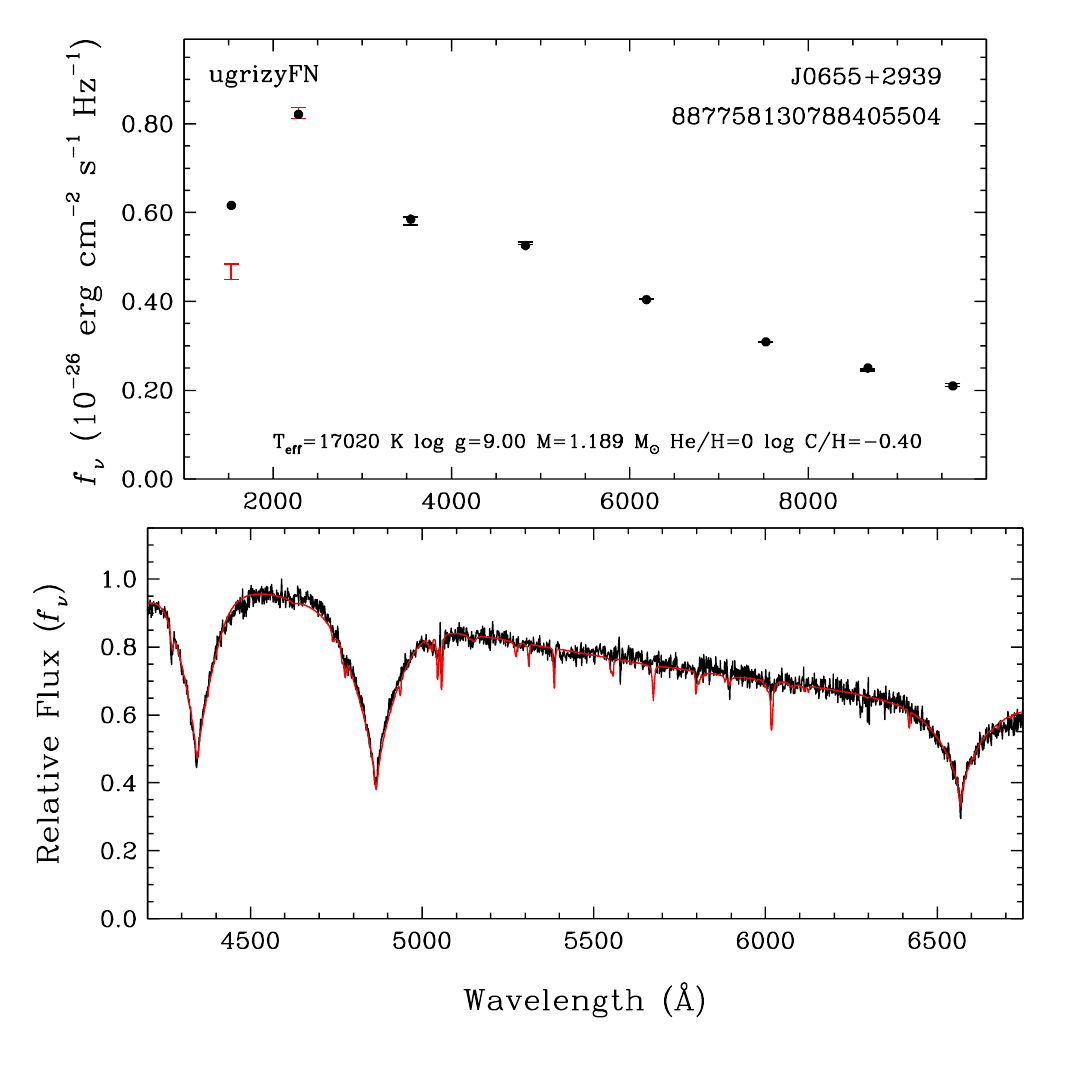}
\caption{Model atmosphere fits to the MMT spectum of the newly discovered DAQ J0655+2939. The top panel shows the photometric fit, and the bottom panel shows the spectroscopic fit.}
\label{figdaq}
\end{figure}

\begin{figure*}
\includegraphics[width=2.5in, clip, trim={0.3in 0.3in 0.5in 0.6in}]{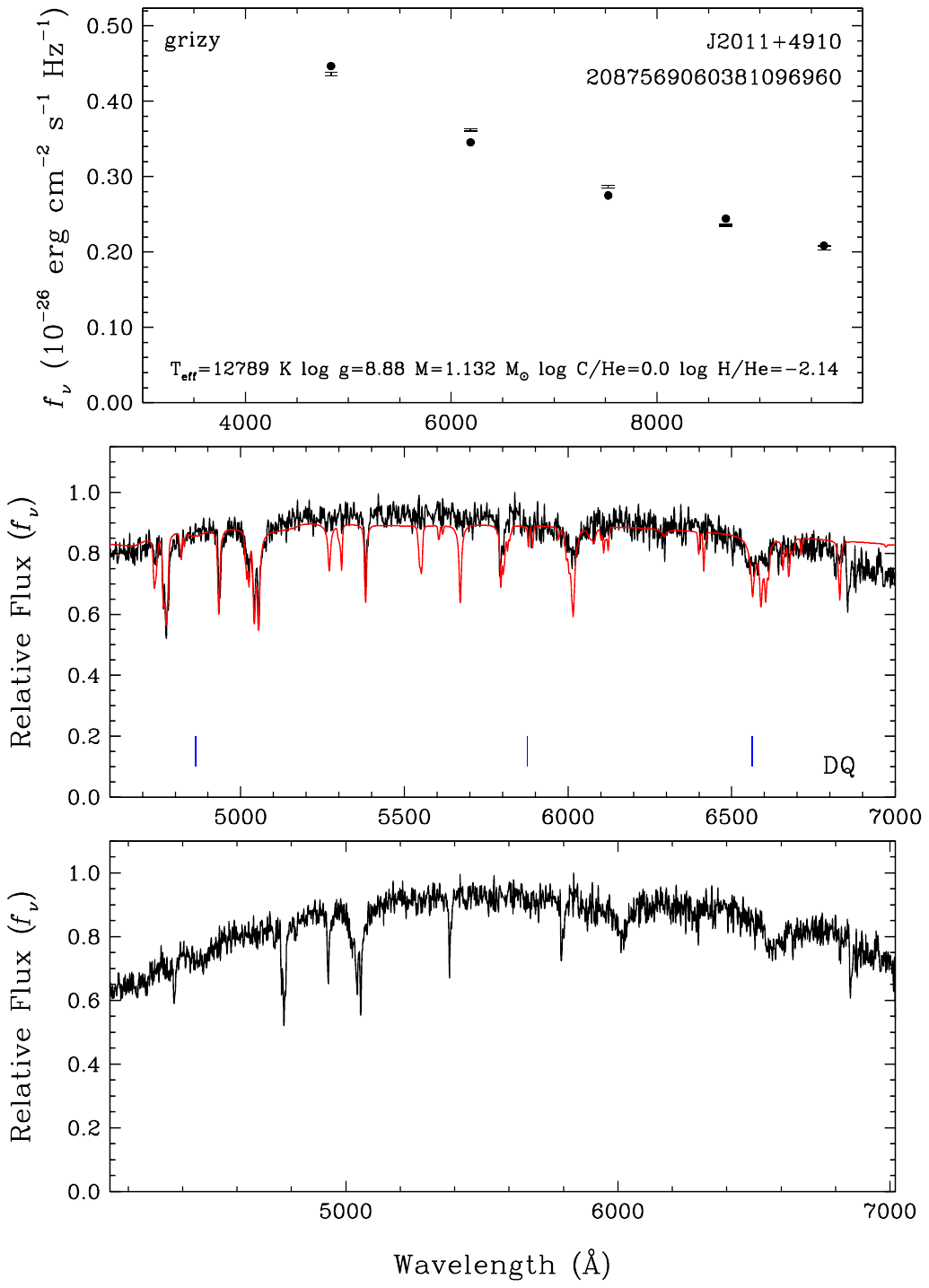}
\includegraphics[width=2.5in, clip, trim={0.3in 0.3in 0.5in 0.6in}]{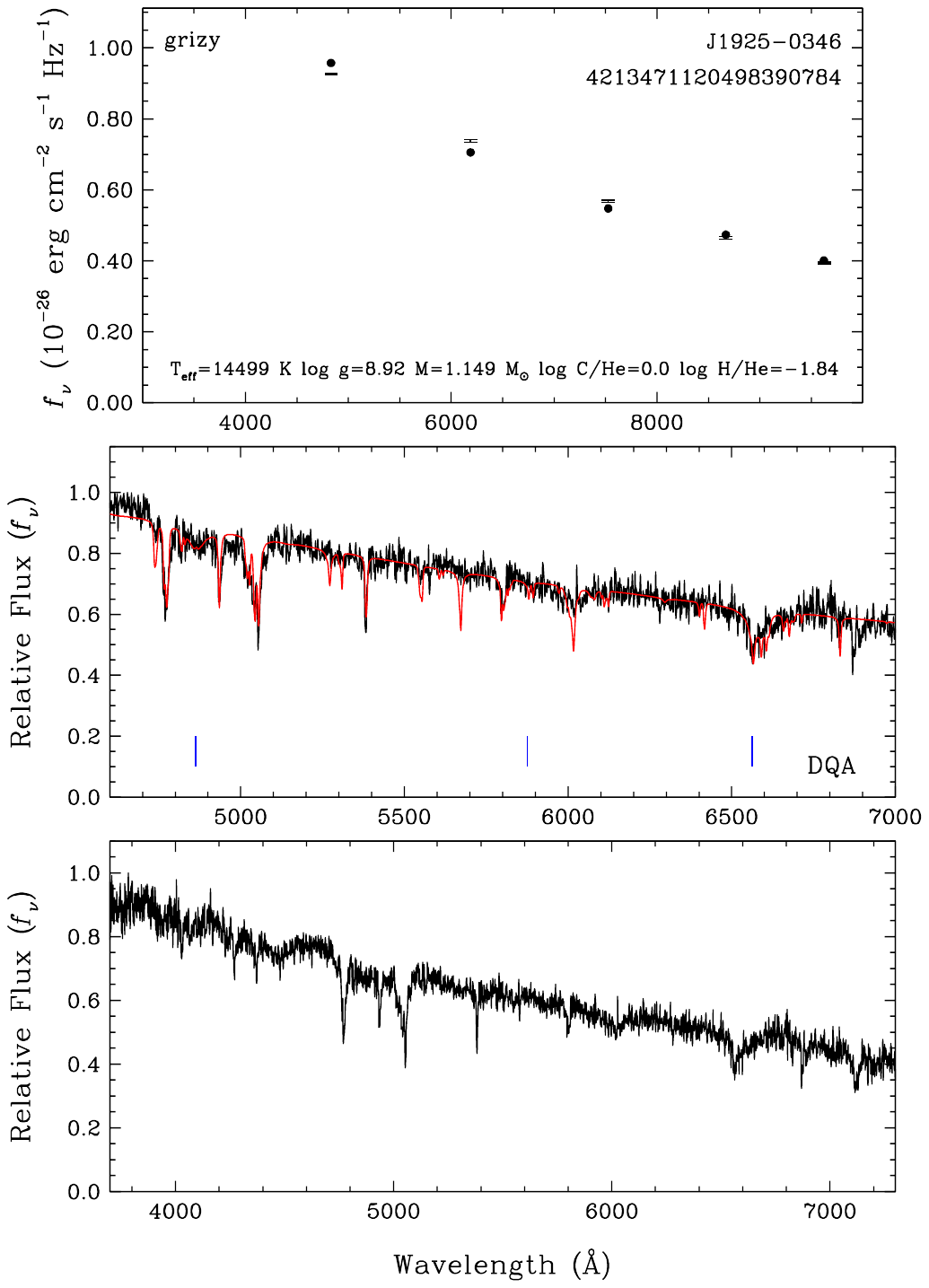}
\includegraphics[width=2.5in, clip, trim={0.3in 0.3in 0.5in 0.6in}]{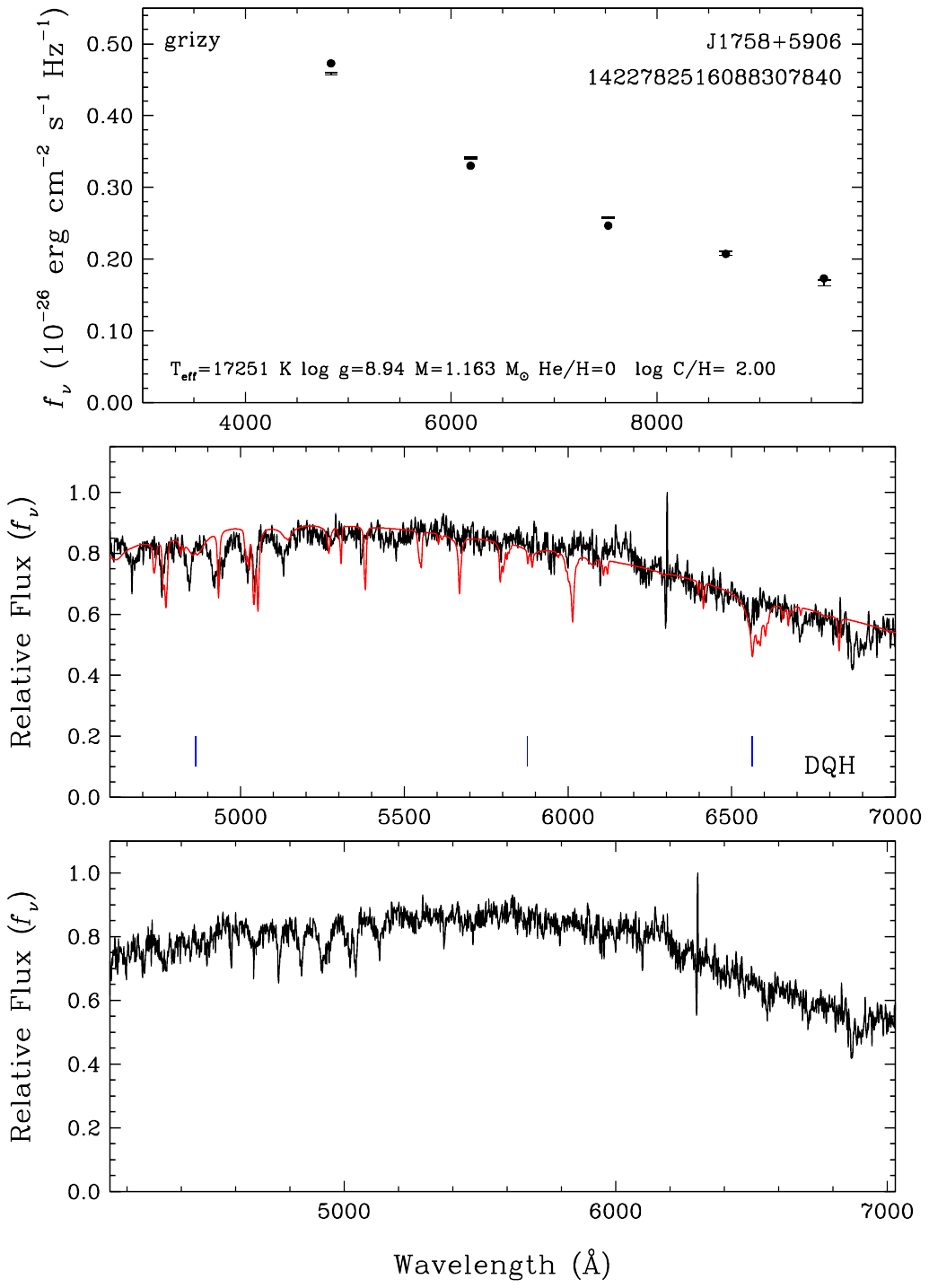}
\caption{Example model atmosphere fits for a warm DQ (left), a DQA (middle), and a magnetic DQH (right panels). 
The top and middle panels show the photometric and the spectroscopic fits, respectively. The blue lines in the middle
panels mark the locations of the H$\alpha$, H$\beta$, and the \ion{He}{1} $\lambda$5876 \AA. Model fits for all warm DQs are available in the online version of this article.}
\label{figdq}
\end{figure*}

Figure \ref{figdaq} presents the MMT spectrum of J0655+2939 along with our model atmosphere fits. There is a simple explanation for the discrepant GALEX FUV photometry; J0655+2939 is a DAQ white dwarf with a hydrogen + carbon atmosphere. The best-fitting DAQ model has $\log$ C/H = $-$0.40, $T_{\rm eff} = 17020 \pm 412$ K, and $M=1.189 \pm 0.009~M_{\odot}$. This model provides a much better fit to the entire spectral energy distribution, including the FUV photometry.

\citet{kilic24} demonstrated that there is a range of hydrogen abundances among the warm DQ population, and the distinction between DAQ and warm DQ white dwarfs is artificial. The DAQs simply represent the most hydrogen rich stars among the warm DQ population, but otherwise they belong to the same population. In fact, most warm DQs do show hydrogen lines in their spectra \citep{koester19,coutu19}. 

For the model atmosphere analysis of warm DQs, we rely on a distinct model atmosphere grid based on the calculations of \citet{blouin19}. The model grid covers the range $T_{\rm eff} = 10,000\ {\rm K}\ (500\ {\rm K})\ 16,000$ K, $\logg = 7.0\ (0.5)\ 9.0$, $\log {\rm He/H} = 1.0\ (1.0)\ 4.0$, and $\log {\rm C/He} = -5.0\ (0.5)\ 1.0$. We also rely on the evolutionary models described in \citet{bedard20} with CO cores, $q({\rm He})=10^{-2}$ and $q({\rm H})=10^{-10}$, which are representative of He-atmosphere (or thin H-atmosphere) white dwarfs. 

\citet{koester19}, \citet{hollands}, and \citet{kilic24} discuss in detail the issues with the atomic data for carbon. We exclude the carbon lines with quality flags D and E in the NIST database, and the two absorption features in the models near 5268 and 5668 \AA\ from our fits, as the latter are not observed. In addition, \citet{kilic24} show that the helium abundance is unconstrained in warm DQs, and that atmosphere models with no helium result in model fits that are just as good as the atmosphere models including helium. Because we do not observe a He line at 5876 \AA\ in warm DQs, we can only set an upper limit on the He abundance.
For this analysis, we adopt a grid with a fixed value of $\log$ C/He = 0, and then fit for H/He (or H/C if a helium-poor composition is assumed) to match H$\alpha$. This method enables us to constrain the hydrogen abundance using spectroscopic observations. 

We show representative fits to three warm DQs in Figure \ref{figdq}, including a DQ (left), a DQA (middle), and a magnetic DQH (right panels). For the DQ and the DQA, we assume equal amounts of C and He in the atmosphere. The blue lines mark the locations of the H$\alpha$, H$\beta$, and the \ion{He}{1} $\lambda$5876 \AA\ features. Our model fits indicate that J1925$-$0346 and J2011+4910 are warm DQs with $M=1.13$--1.15 $M_{\odot}$ and $\log{\rm H/He}\sim-2$. The former is slightly more hydrogen rich, which depicts itself as a stronger H$\alpha$ feature, hence the DQA classification. We classify J2011+4910 as a DQ rather than a DQA because of its very weak H$\alpha$ line, whereas the DQAs clearly show H$\alpha$ in their spectra.

The parameters of the magnetic DQ J1758+5906 shown in the right panel are outside of our model grid for DQA white dwarfs. Hence, we used the DAQ model grid to fit this object. Since these models are not magnetic, the model fit to the J1758+5906 is not perfect, but
it provides a decent match to the overall C features, indicating that this is a C-dominated atmosphere white dwarf with a temperature above 17,000 K, hence very close to the artificial separation between the hot and warm DQ white dwarfs. 

We do not have an optical spectrum of the DQ white dwarf J0045$-$2336 \citep[G268-40,][]{koester82} available. We assume the same composition as the other warm DQs in our sample, $\log$ C/He = 0, and adopt a grid with the smallest trace of hydrogen, which result in the best-fitting parameters of $T_{\rm eff} = 11,540 \pm 43$ K and $M=1.126 \pm 0.004~M_{\odot}$ for J0045$-$2336. These parameters should be used with caution.  We provide the model fits for all other DQ and DQA white dwarfs in the online version of this article, but we adopt the parameters
of the hot DQ and the 4 of the DAQs in our sample from \cite{kilic23a} and \cite{kilic24}, respectively as they use better optical spectra for these objects.

\section{Photometric Variability}

\subsection{Rotation Periods} \label{sec:rotate}

To search for photometric variability among our massive white dwarf sample, we checked both the Transiting Exoplanet Survey Satellite (TESS) 20 s and 2 min cadence data, and the Zwicky Transient Facility (ZTF) data for each object. Table \ref{tab:rotation} presents the photometric periods for our targets from TESS, ZTF, and the literature. We exclude previously known pulsating DAV white dwarfs from this list, as they are discussed below. 

Eight of our targets have rotation periods reported in the literature with periods ranging from 5.9 to 131.6 min. These include 2 DAH, 2 DAQ, 1 DBA, 1 DBAH, 1 DC (which has to be magnetic, given its effective temperature), and 1 DA. Interestingly, all but one of these objects with previous rotation measurements are either magnetic or have unusual atmospheric composition. Note that even though DBA white dwarfs are common among the DB white dwarf population, DB white dwarfs themselves are unusually rare among massive white dwarfs. The DA white dwarf J1529+2928 is the only `normal' white dwarf in this sample that shows photometric variability with a period of 38 min. Even though J1529+2928
is near the ZZ Ceti instability strip, the observed period is too long to be due to pulsations, and it is clearly due to spots in this otherwise normal white dwarf \citep{kilic2015}.

\begin{deluxetable}{cccc}
\tabletypesize{\scriptsize}
\tablecolumns{6} \tablewidth{0pt}
\tablecaption{Photometric rotation periods for our targets from TESS, ZTF, and the literature.} \label{tab:rotation}
\tablehead{\colhead{Object} & \colhead{Spectral} & \colhead{Period} & \colhead{Source}\\
& Type & (min) & }
\startdata
J0006$+$3104 & DC & 23.15 & TESS  \\
J0043$-$1000 &  DBAH & 131.6 & \cite{liebert77}  \\
J0050$-$0326 & DC & 40.31 & TESS  \\
J0118$-$0156 & DAH & 54.69 & TESS \\
J0256$-$1515 & DAH & 29.16 & TESS  \\
J0327$+$2227 & DC &  58.6 & ZTF  \\
J0707$+$5611 & DC & 63 & \cite{kilic23a}\\
J0718$+$3731 & DC & 11.27 & TESS  \\
J0831$-$2231 & DAQ & 10.7  & \cite{kilic24}\\
J1154$+$3650 & DA & 35.6 & ZTF  \\
J1214$-$1724 & DBH & 107.36 & TESS  \\
J1529$+$2928 & DA &  38  & \cite{kilic2015}\\
J1543$+$3021 & DAH & 78.33 & TESS \\
J1659$+$4401 & DAH & 42.24 & TESS  \\
J1707$+$3532 & DAH & 34.72 & ZTF  \\
J1719$-$1446 & DAH? & 5.5 & ZTF  \\
J1832$+$0856 & DBA &  5.9  & \cite{pshirkov20}\\
J1901$+$1458 & DAH & 6.9  & \cite{caiazzo21}\\
J2100$+$5142 & DAH & 149.1 & ZTF \\
J2204$+$2543 & DAH & 415.22 & TESS  \\
J2257$+$0755 & DAH &  22.8   & \cite{williams}\\
J2340$-$1819 & DAQ &  12.1 & \cite{kilic24}\\
\enddata
\tablecomments{All five DCs in this table are warmer than 11000 K regardless of their atmospheric composition. Hence, they must be strongly magnetic.}
\end{deluxetable} 

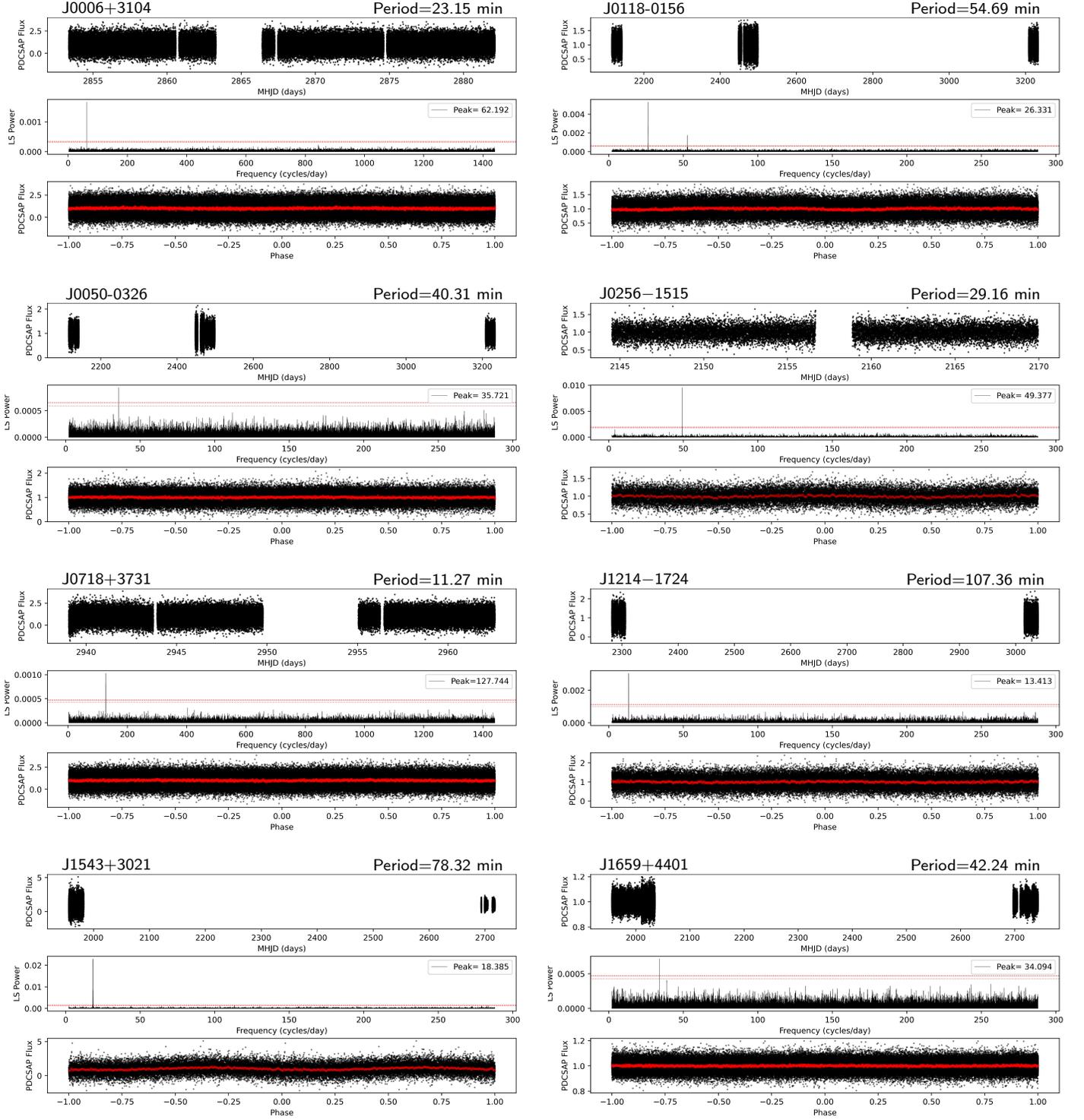
\begin{figure*}
\begin{tikzpicture}
\node at (0,0) {\includegraphics[width = 3.6in, clip, trim={0in 0in 5.8in .6in}]{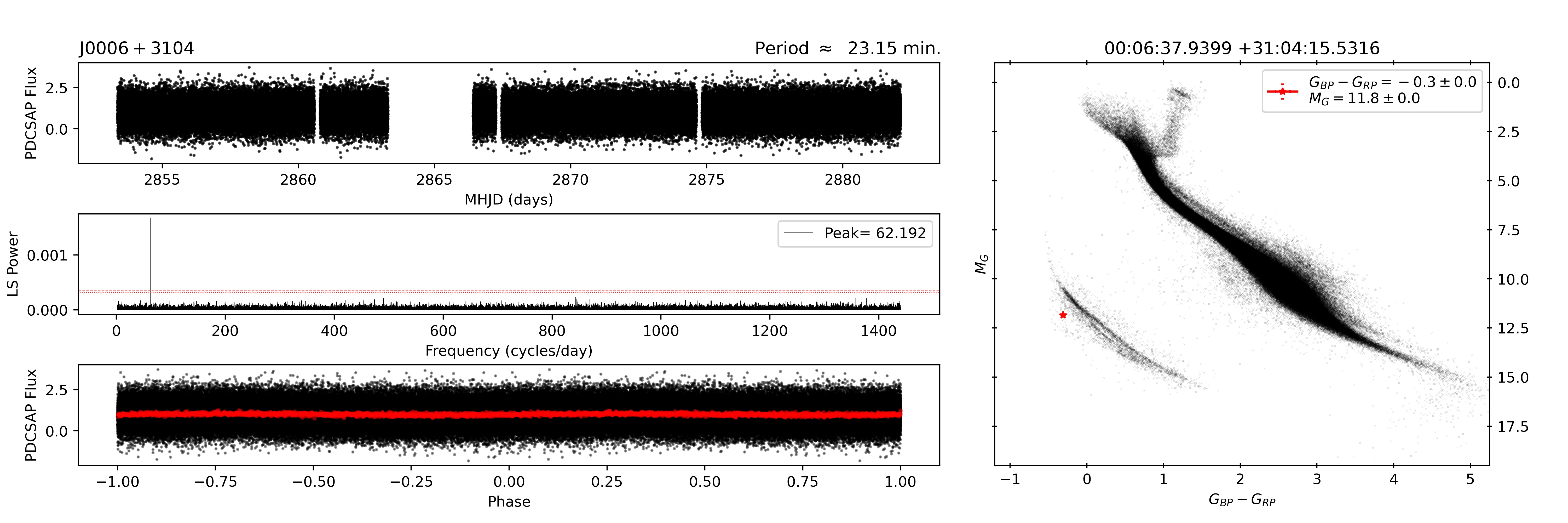}};
\node at (-2.8,2.35) {\textsf{\footnotesize J0006+3104}};
\node at (3, 2.35) {\textsf{\footnotesize Period=23.15 min}};
\node at (9.5,0) {\includegraphics[width = 3.6in, clip, trim={0in 0in 5.8in .6in}]{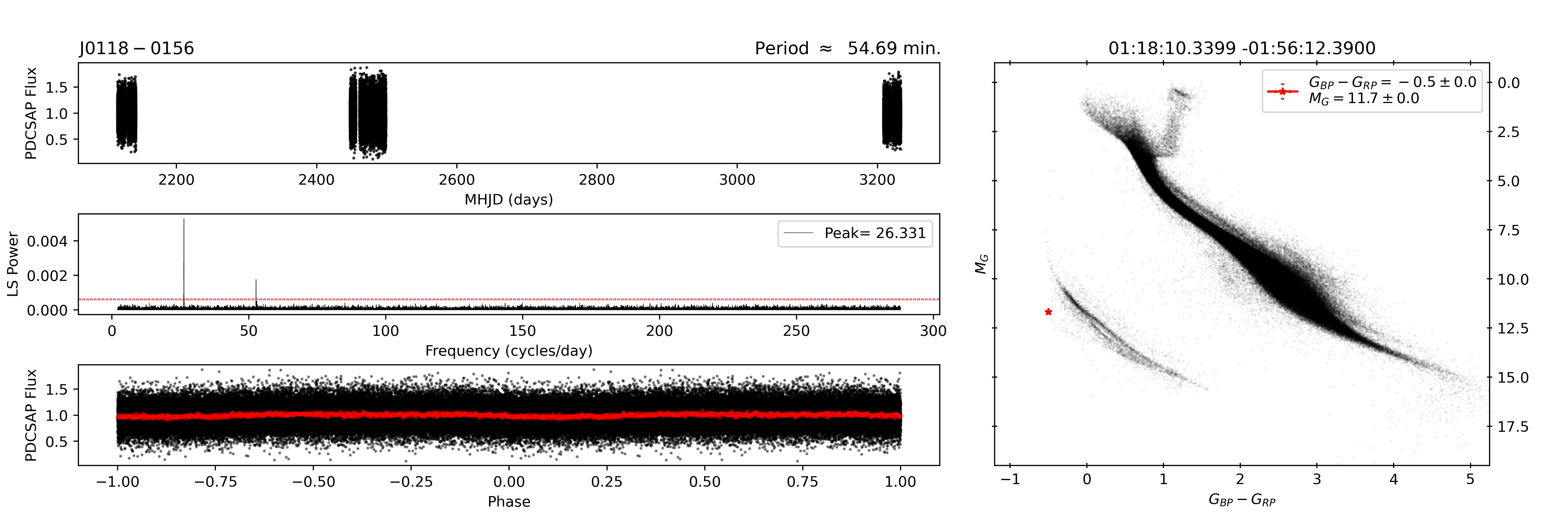}};
\node at (6.6,2.35) {\textsf{\footnotesize J0118-0156}};
\node at (12.4,2.35) {\textsf{\footnotesize Period=54.69 min}};
\node at (0,-5) {\includegraphics[width = 3.6in, clip, trim={0in 0in 5.8in .6in}]{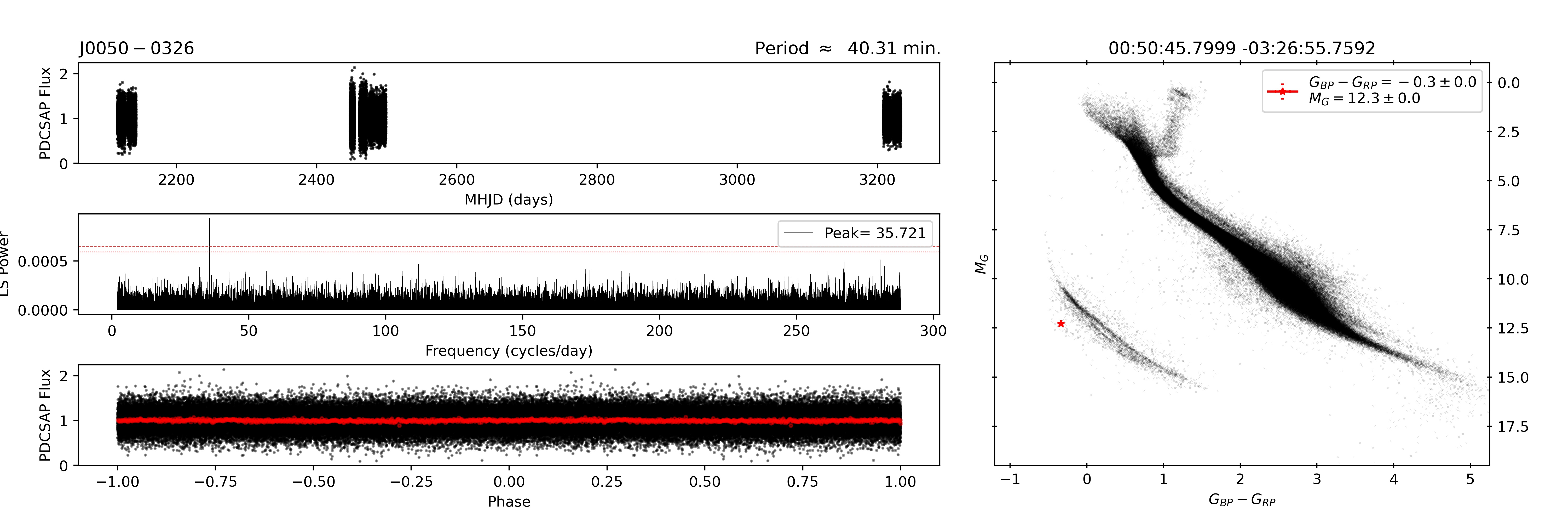}};
\node at (-2.8,-2.65) {\textsf{\footnotesize J0050-0326}};
\node at (3,-2.65) {\textsf{\footnotesize Period=40.31 min}};
\node at (9.5,-5) {\includegraphics[width = 3.6in, clip, trim={0in 0in 5.8in .6in}]{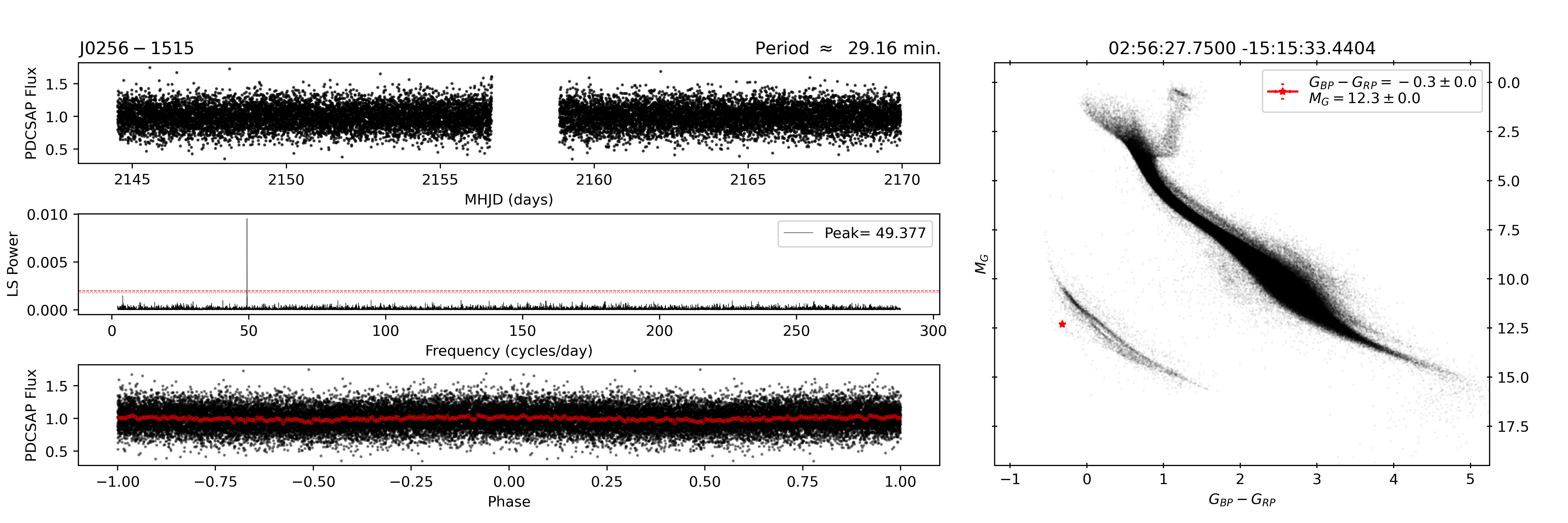}};
\node at (6.6,-2.65) {\textsf{\footnotesize J0256$-$1515}};
\node at (12.4, -2.65) {\textsf{\footnotesize Period=29.16 min}};
\node at (0,-10) {\includegraphics[width = 3.6in, clip, trim={0in 0in 5.8in .6in}]{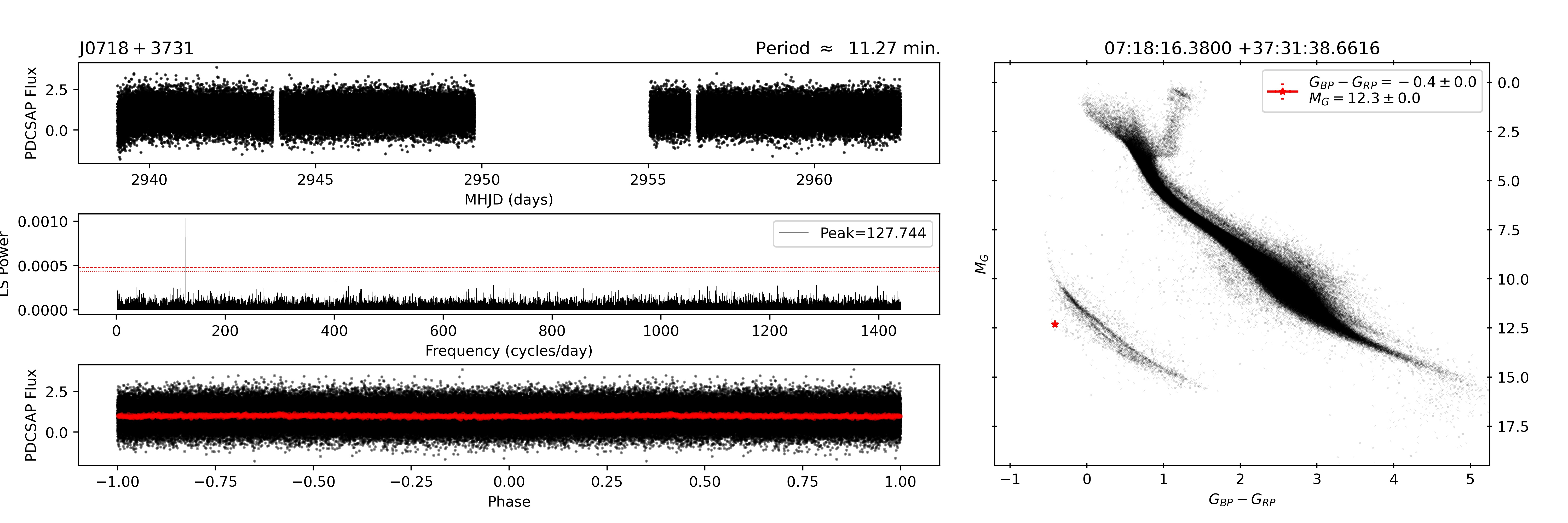}};
\node at (-2.8,-7.65) {\textsf{\footnotesize J0718+3731}};
\node at (3,-7.65) {\textsf{\footnotesize Period=11.27 min}};
\node at (9.5,-10) {\includegraphics[width = 3.6in, clip, trim={0in 0in 5.8in .6in}]{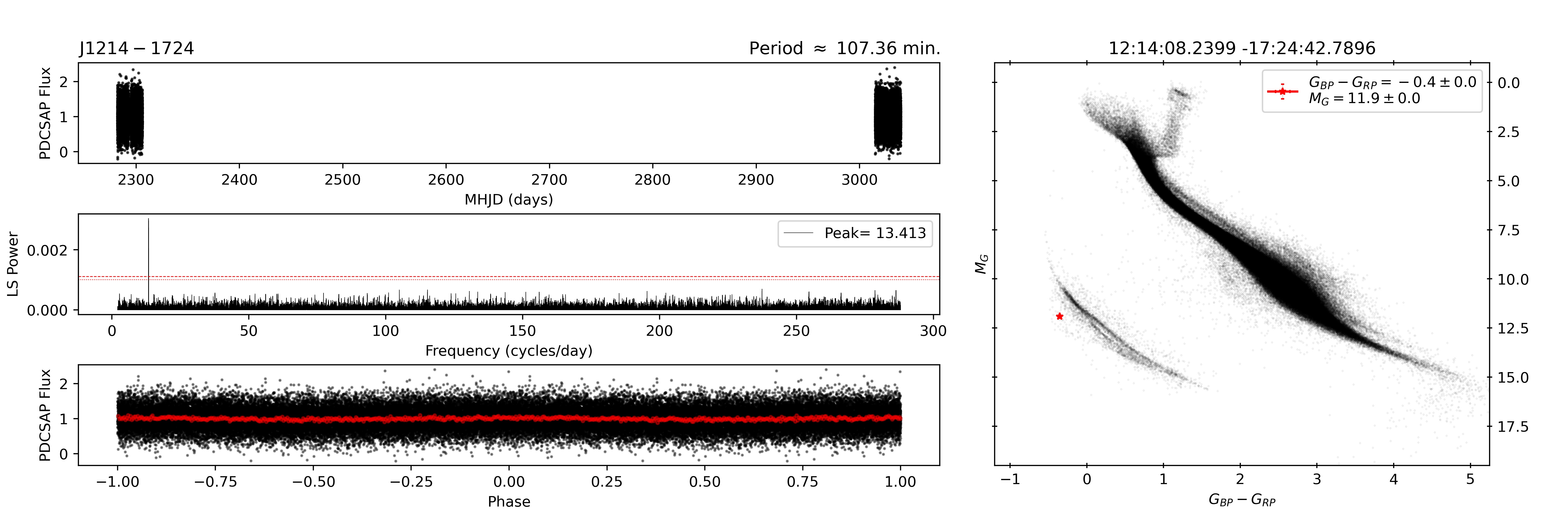}};
\node at (6.6,-7.65) {\textsf{\footnotesize J1214$-$1724}};
\node at (12.4, -7.65) {\textsf{\footnotesize Period=107.36 min}};
\node at (0,-15) {\includegraphics[width = 3.6in, clip, trim={0in 0in 5.8in .6in}]{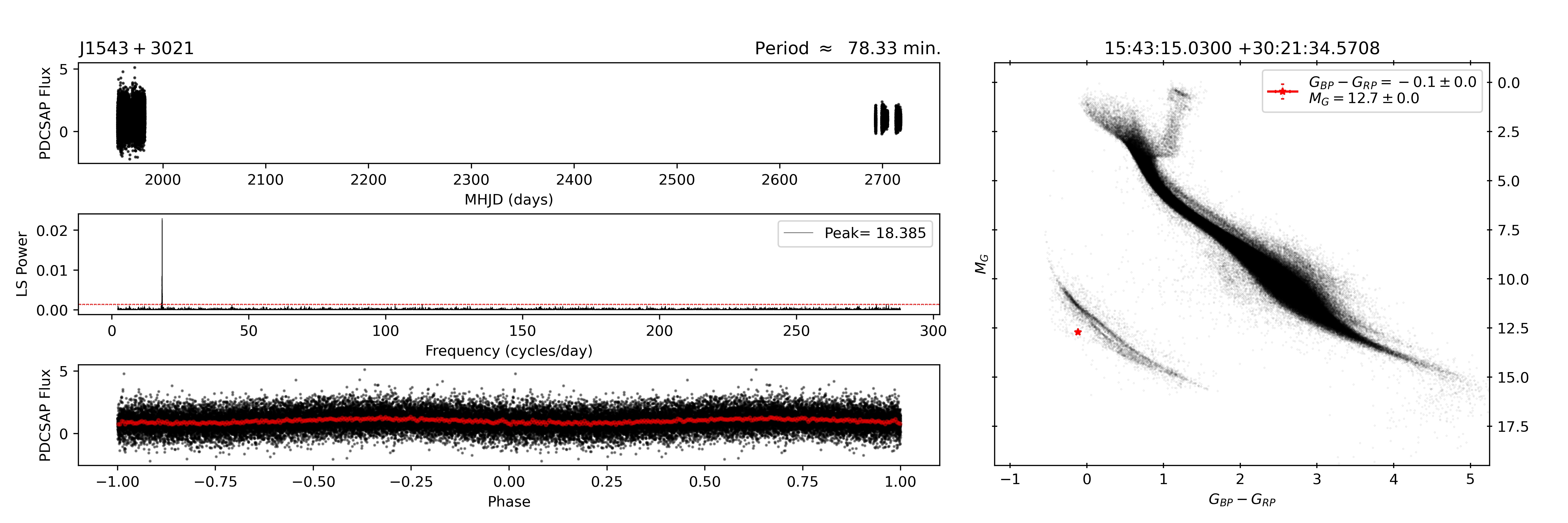}};
\node at (-2.8,-12.65) {\textsf{\footnotesize J1543+3021}};
\node at (3,-12.65) {\textsf{\footnotesize Period=78.32 min}};
\node at (9.5,-15) {\includegraphics[width = 3.6in, clip, trim={0in 0in 5.8in .6in}]{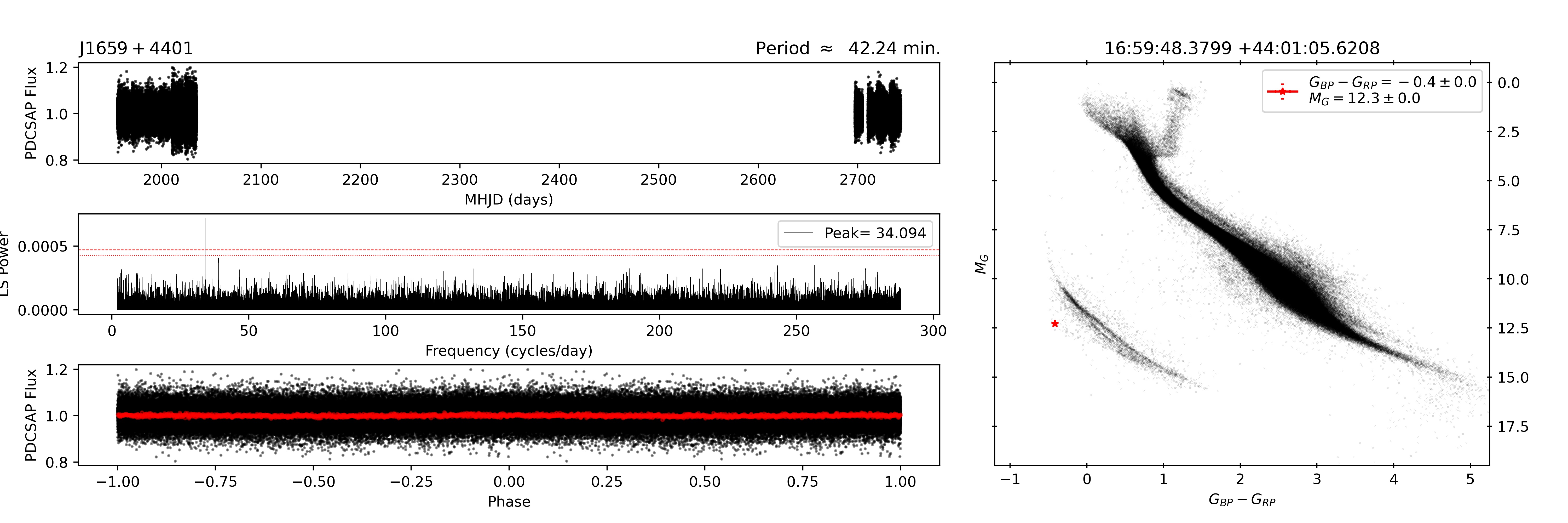}};
\node at (6.6,-12.65) {\textsf{\footnotesize J1659+4401}};
\node at (12.4,-12.65) {\textsf{\footnotesize Period=42.24 min}};
\end{tikzpicture}
\caption{Light curves (top panels), Lomb–Scargle periodograms (middle panels), and phase-folded light curves based on the highest peak in the periodograms (bottom panels) for 15 newly identified variable white dwarfs using the TESS 20 s or 2 min cadence data and ZTF photometry. TESS light curves for 10 stars are shown first, followed by 5 targets with ZTF data. Red data points in the TESS frames represent the original data binned by 100, whereas the green, red, and yellow symbols in the ZTF panels show the $g$-, $r$-, and $i$-band photometry. The dashed line is the 1\% false-alarm probability rate, and the dotted line is the 5\% false-alarm probability rate.}
\label{fig:tess}
\end{figure*} 

\addtocounter{figure}{-1}
\begin{figure*}
\begin{tikzpicture}    
\node at (-2,0) {\includegraphics[width = 3.7in, clip, trim={0in 0in 5.8in 0.6in}]{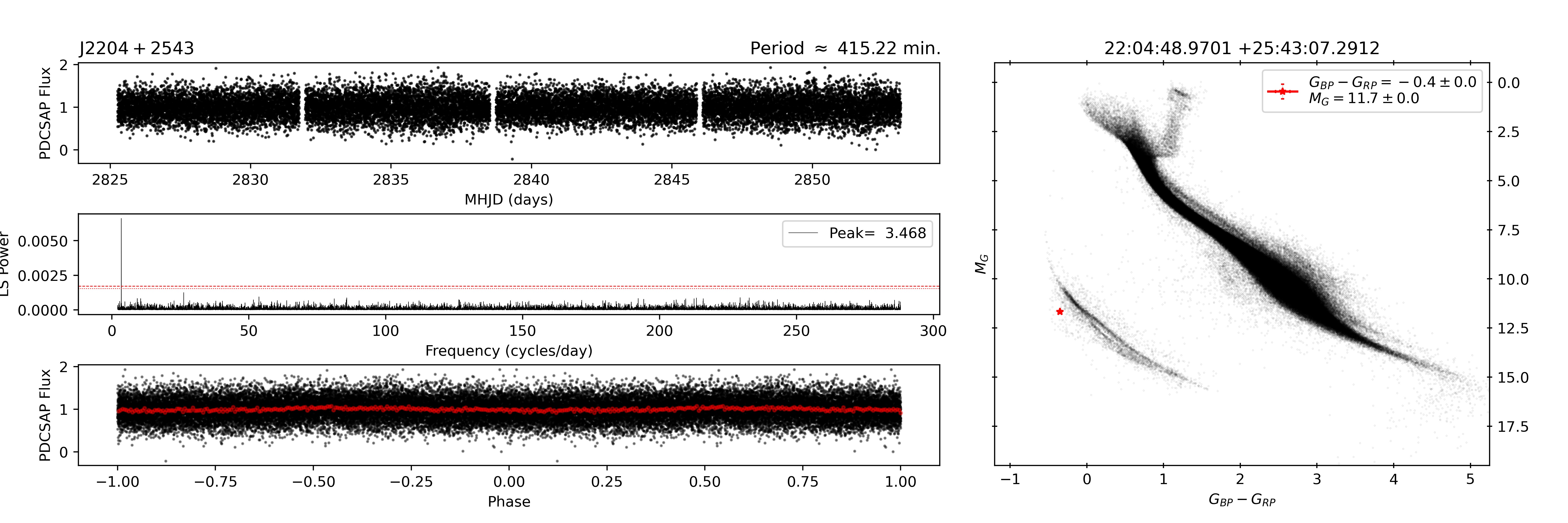}};
\node at (-5, 2.4) {\textsf{\footnotesize J2204+2543}};
\node at (1,2.4) {\textsf{\footnotesize Period=415.22 min}};
\node at (7.5,0) {\includegraphics[width = 3.7in, clip, trim={0in 0in 5.8in .6in}]{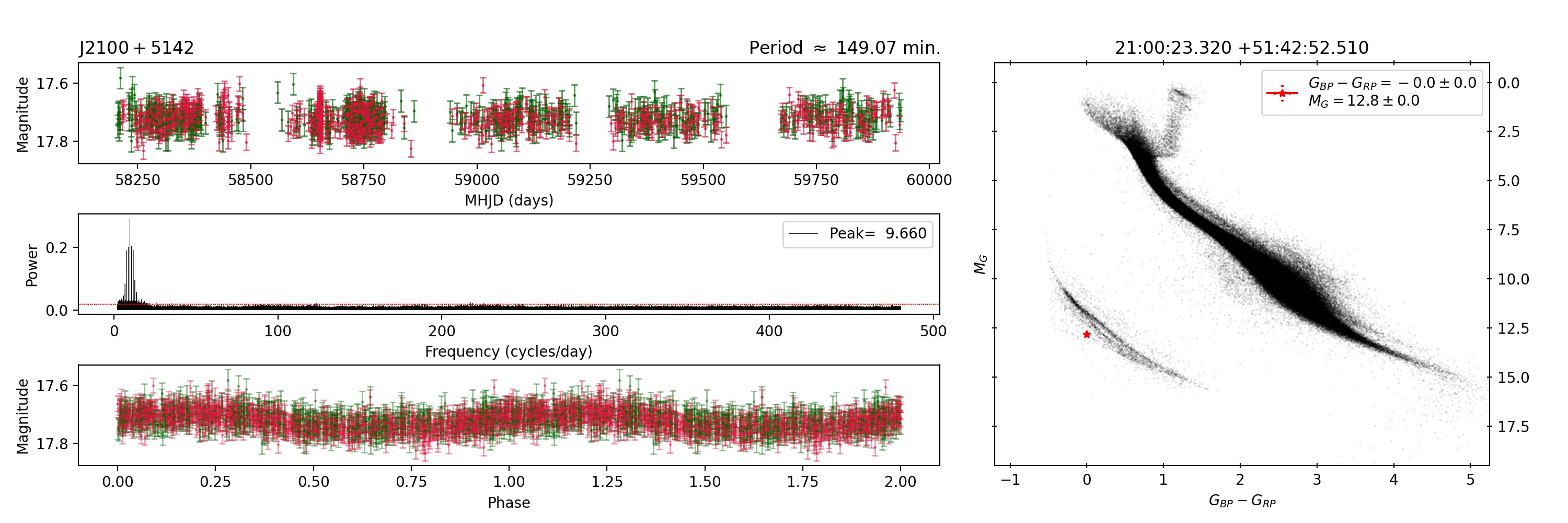}};
\node at (4.5,2.4) {\textsf{\footnotesize J2100+5142}};
\node at (10.4,2.4) {\textsf{\footnotesize Period=149.07 min}};
\node at (-2,-5) {\includegraphics[width = 3.7in, clip, trim={0in 0in 5.8in 0.6in}]{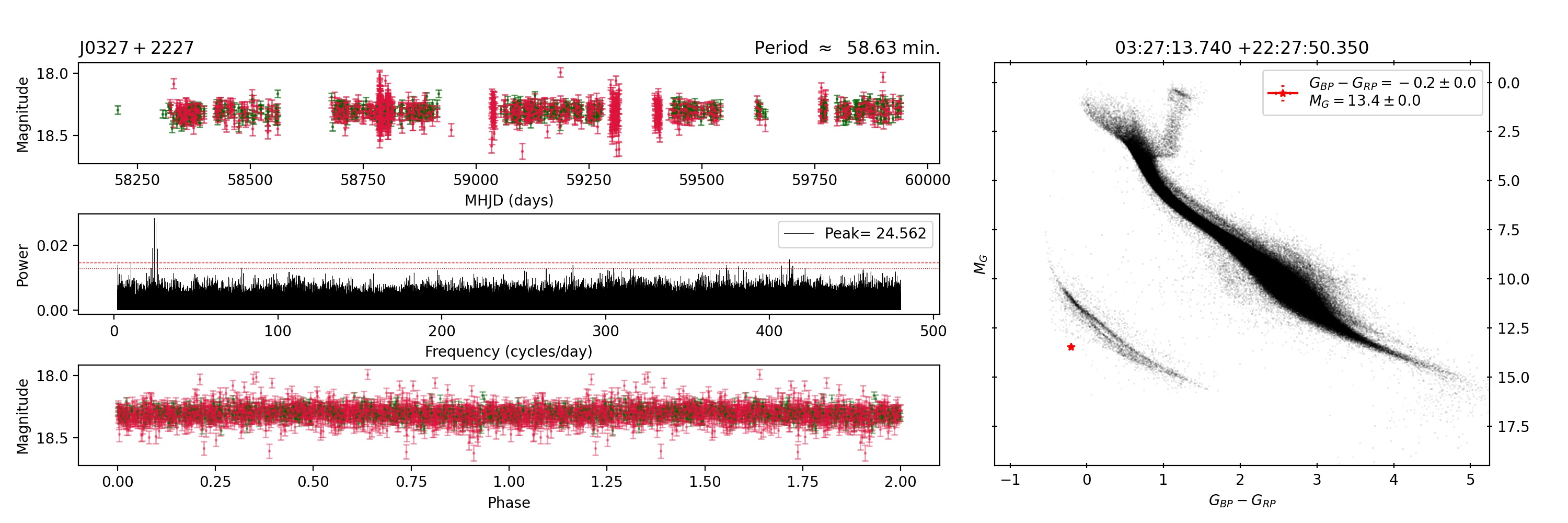}};
\node at (-5,-2.6) {\textsf{\footnotesize J0327+2227}};
\node at (1,-2.6) {\textsf{\footnotesize Period=58.63 min}};
\node at (7.5,-5) {\includegraphics[width = 3.7in, clip, trim={0in 0in 5.8in .6in}]{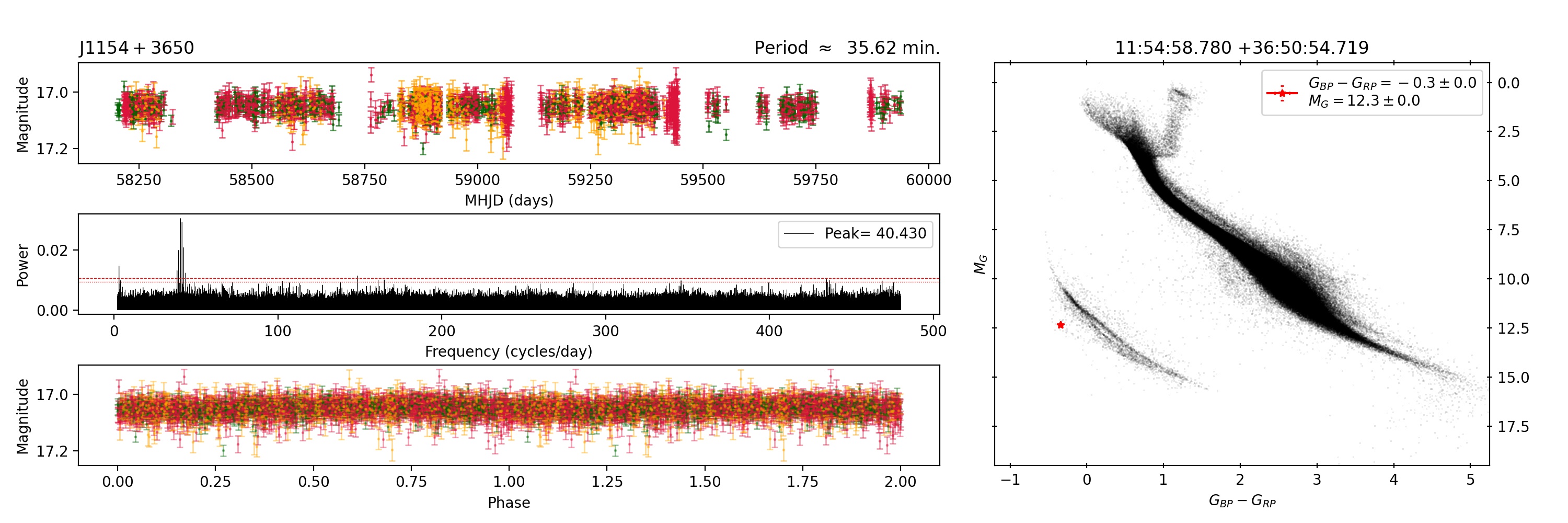}};
\node at (4.5,-2.6) {\textsf{\footnotesize J1154+3650}};
\node at (10.4,-2.6) {\textsf{\footnotesize Period=35.62 min}};
\node at (-2,-10) {\includegraphics[width = 3.7in, clip, trim={0in 0in 5.8in .6in}]{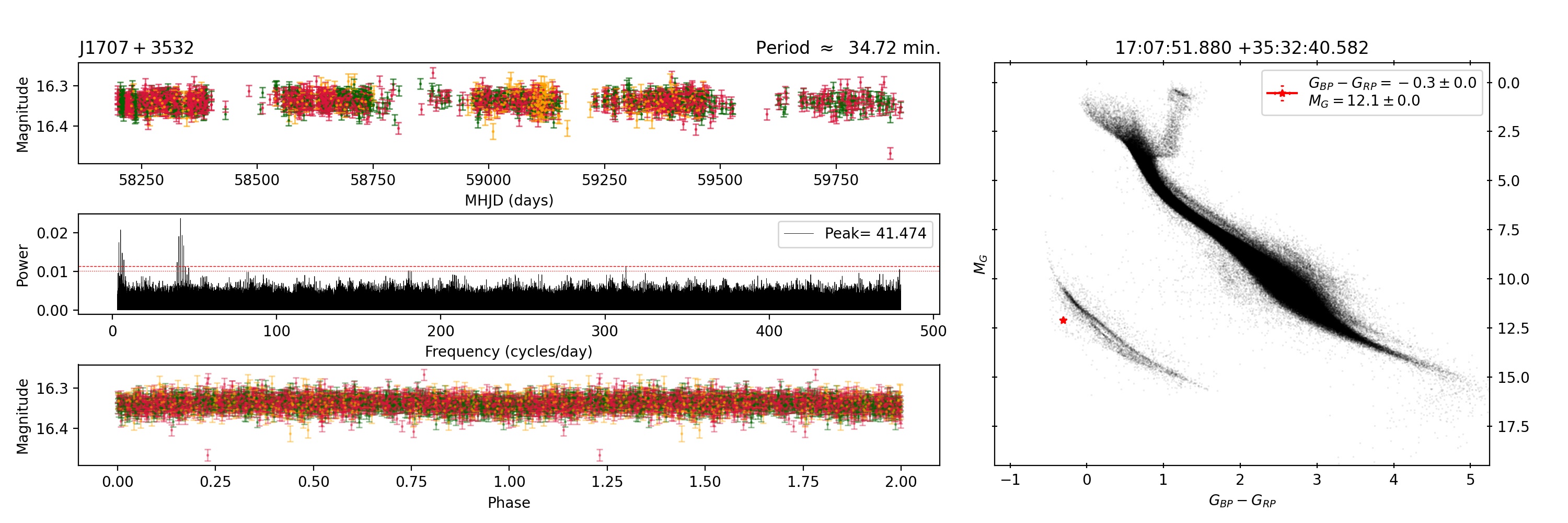}};
\node at (-5,-7.6) {\textsf{\footnotesize J1707+3532}};
\node at (1,-7.6) {\textsf{\footnotesize Period=34.72 min}};
\node at (7.5,-10) {\includegraphics[width = 3.7in, clip, trim={0in 0in 5.8in .6in}]{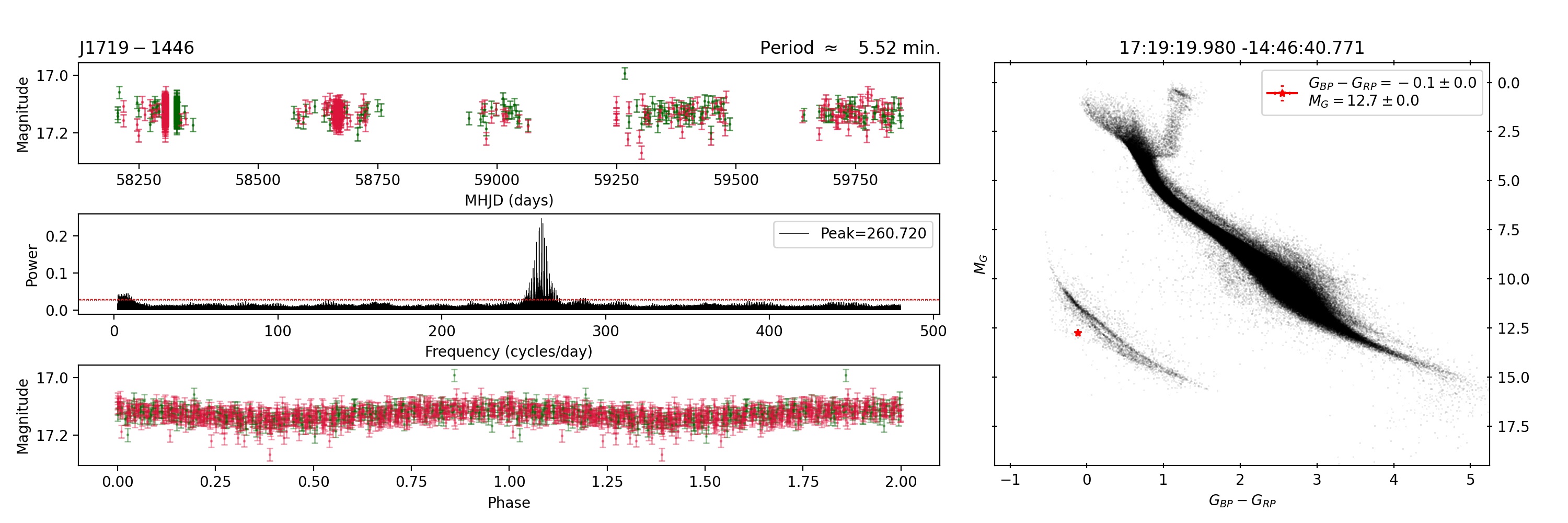}};
\node at (4.5,-7.6) {\textsf{\footnotesize J1719$-$1446}};
\node at (10.4,-7.6) {\textsf{\footnotesize Period=5.52 min}};
%
%
\end{tikzpicture}
\caption{Continued.}
\label{fig:ztf}
\end{figure*}

We identify 14 additional photometrically variable systems using TESS and ZTF. Figure \ref{fig:tess} shows the light curves, Lomb-Scargle periodograms, and phase folded light curves for these targets. In TESS, we searched for periods ranging from 1 to 684 min using the 20 s cadence data, and 5 to 684 min using the 2 min cadence data. In ZTF, we searched for periods ranging from 3 to 684 min. With TESS short cadence data, we found 8 objects with periods ranging from 11 to 107 min, and one more longer period system, J2204+2543, which has a rotation period of 415.22 minutes ($\approx$7 hours). With ZTF, we identified five additional variable targets with periods ranging from 5.5 to 149 min. Four objects show clear variations in both TESS and ZTF data. In all four cases, due to worse aliasing in ground-based observations, the ZTF data favor the first harmonic of the period measured from the TESS data. We adopt the TESS value as the true period for those objects.  

Out of the 22 objects shown in Table \ref{tab:rotation}, 20 are either magnetic or have unusual atmospheric composition, and the remaining 2 are DA white dwarfs. J1529+2928 is discussed above. The remaining DA, J1154+3650 has $T_{\rm eff} = 26,115 \pm 411$ K and $M = 1.251 \pm 0.006~M_{\odot}$, and it shows significant variability at a period of about 35.6 min. It is clearly outside of the ZZ Ceti instability strip. Hence, the variability in J1154+3650 and the rest of the objects in this table (and Figure \ref{fig:tess}) are clearly due to rotation. J1154+3650 appears to be a spotted DA white dwarf just like J1529+2928, where the variability is likely caused by the rotation of a star with a relatively weak magnetic field ($B<100$ kG) and/or an inhomogeneous atmosphere.   

\begin{deluxetable}{cccc}
\tablecolumns{4} 
\tablecaption{Previously known pulsating white dwarfs and NOVs (Not Observed to Vary) in our sample.}
\label{tab:pulsators}
\tablehead{\colhead{Pulsators} & \colhead{Reference} & \colhead{NOVs} & \colhead{Reference}}
\startdata
J0049$-$2525 & \citet{kilic23b} & J0135+5722 & \citet{vincent}\\
J0204+8713 & \citet{vincent} & J0234$-$0511 & \citet{gianninas}\\
J0448$-$1053 & \citet{romero} & J0347$-$1802 & \citet{guidry} \\
J0551+4135 & \citet{vincent} & J0408+2323 & \citet{vincent}\\
J0856+6206 & \citet{vincent} & J0538+3212 & \citet{vincent}\\
J1106+1802 & \citet{guidry} & J0634+3848 & \citet{vincent}\\
J1659+6610 & \citet{hermes} & J0657+7341 & \citet{vincent}\\
J1812+4321 & \citet{romero} &  J1140+2322 & \citet{kilic23a}\\
& & J1243+4805 & \citet{vincent}\\
& & J1626+2533 & \citet{vincent}\\
& & J1655+2533 & \citet{curd}\\
& & J1813+4427 & \citet{vincent}\\
& & J1910+7334 & \citet{vincent}\\
& & J1928+1526 & \citet{vincent}\\
\enddata
\tablecomments{\citet{guidry} detected long term variability in J0347$-$1802 from transiting debris, but no pulsations.}
\end{deluxetable}

Short rotation periods can indicate a merger origin. \citet{hermes17} found that white dwarfs with masses $0.51\leq M/M_\odot \leq 0.73$ have an average rotation period of 35 hours. These average-mass white dwarfs likely formed via single star evolution. On the other hand, \citet{Schwab} predicts merger products from two CO white dwarfs to have a rotation period as short as 10-20 minutes. Magnetic white dwarfs tend to have higher masses ($\sim$0.8 $M_\odot$, \citealt{Ferrario15}) than non-magnetic objects, as well as shorter rotation periods ($\sim$ less than 10 hours, \citealt{kawka20}). Magnetism in relatively hot (young) and massive white dwarfs is therefore a strong indicator of a merger origin \citep{bagnulo22}. All  of the objects shown in Table \ref{tab:rotation} have rotation periods less than 8 h. Hence, these relatively hot, young, and restless stars are consistent with being white dwarf merger remnants (see Sec.\ref{sec:mag}). 

\begin{figure}
\centering
\includegraphics[width=3.4in]{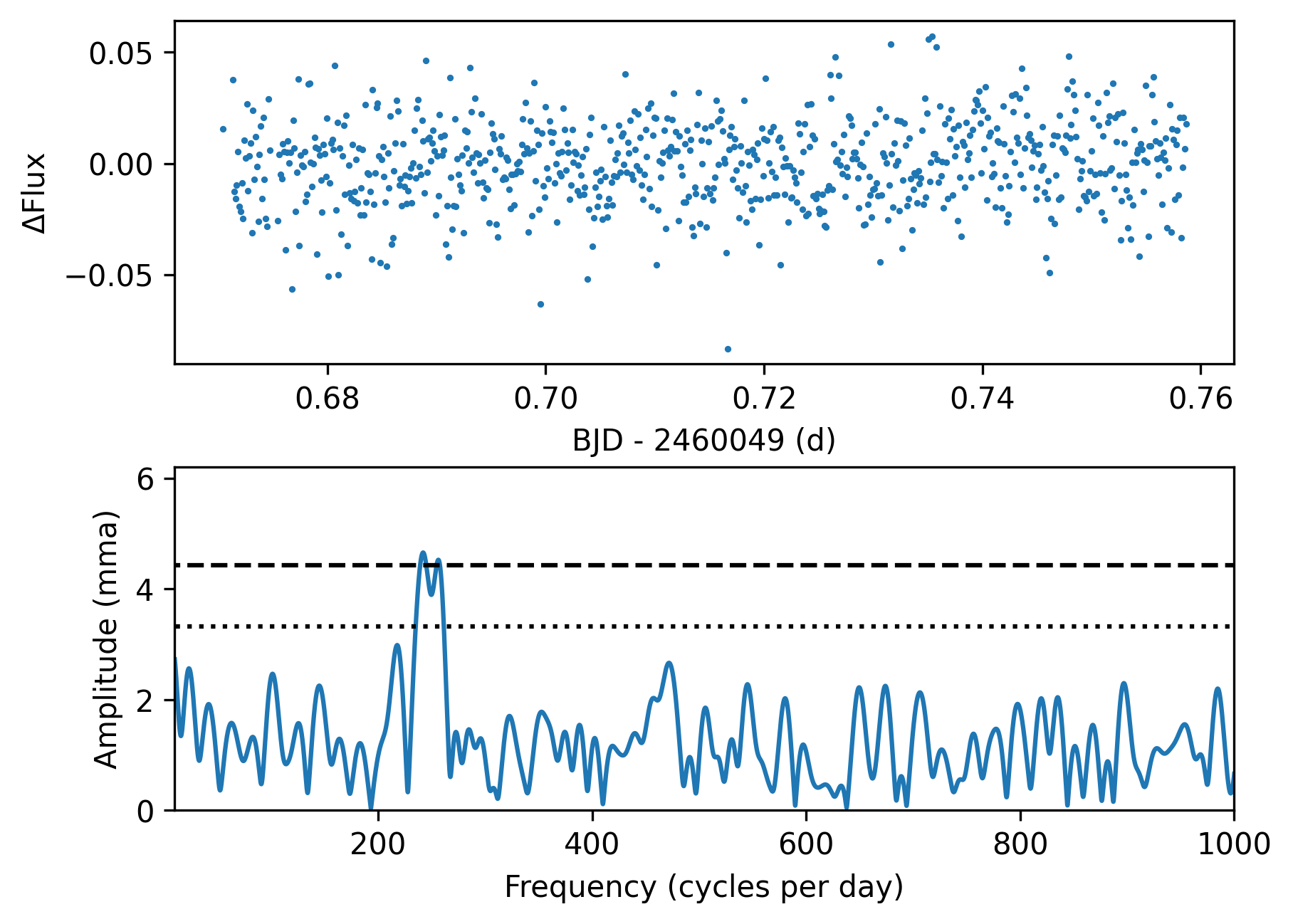}
\caption{APO time-series photometry of J0204+8713 (top panel) and its Fourier transform (bottom panel).
The dotted and dashed lines show the 3$\langle {\rm A}\rangle$ and  4$\langle {\rm A}\rangle$ levels.}
\label{fig:apoj0204}
\end{figure}

\subsection{Pulsating White Dwarfs}

\begin{figure*}
\centering 
\includegraphics[width=0.75\linewidth]{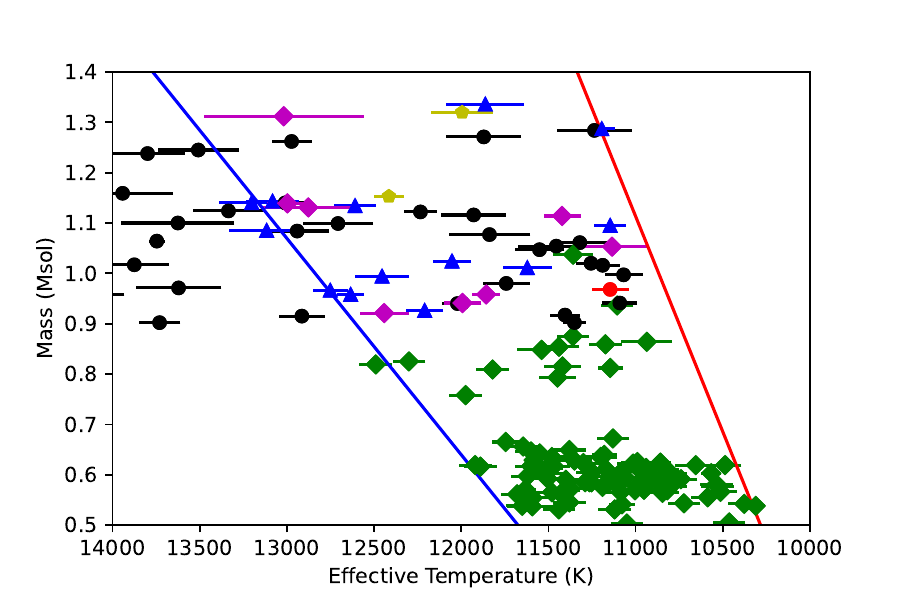}
\caption{Masses and effective temperatures for our massive DA white dwarf sample along with the previously known ZZ Ceti white dwarfs (green diamonds) from \citet{vincent}. The blue and red lines show the empirical boundaries of the ZZ Ceti instability strip from the same work. Magenta diamonds and blue triangles mark the massive DAVs and NOVs, respectively. J0959$-$1828 and J0135+5722 are potential variable white dwarfs marked by yellow pentagons, and objects in the sample with no follow up photometry are marked by black points. The red point marks the spotted white dwarf J1529+2928 that also falls within the instability strip \citep{kilic2015}.}
    \label{fig:zzceti}
\end{figure*}

Several objects in our sample have time-series photometry reported in the literature and have either been confirmed as pulsators or reported as an NOV (not observed to vary).
Table \ref{tab:pulsators} presents the list of eight previously known pulsating massive white dwarfs and 14 NOVs in this sample. All but one of these pulsators had spectral classification available
in the literature. J0204+8713 was previously classified as a pulsating ZZ Ceti based on the detection of a single mode at 330 s \citep{vincent}, but there was no spectroscopy available. Our follow-up spectroscopy confirms that J0204+8713 is indeed a massive DA white dwarf. 

In addition, to confirm that the variability is due to pulsations, we acquired high-speed photometry of J0204+8713 using the APO 3.5m telescope with the Agile frame transfer camera \citep{mukadam11} and the BG40 filter on UT 2023 April 15. We obtained back-to-back exposures of 10s over 2.1 hours. We binned the CCD by 2x2, which resulted in a plate scale of $0.258\arcsec$ per pixel. Figure \ref{fig:apoj0204} shows the APO light curve for J0204+8713 along with its Fourier transform. We detect two significant modes at frequencies of 245.0 and 253.5 cycles d$^{-1}$ with amplitudes of 5 mma, confirming that the variability is due to pulsations, and not rotation.

Figure \ref{fig:zzceti} shows the masses and effective temperatures for our massive DA white dwarf sample along with the ZZ Ceti white dwarfs from \citet{vincent}. There are 45 DAs within or near the boundaries of the ZZ Ceti instability strip, including the spotted white dwarf J1529+2928 discussed above, and the eight pulsators and 14 NOVs presented in Table \ref{tab:pulsators}. The most massive pulsating white dwarf known is J0049$-$2525. J0959$-$1828 is also close in mass, and potentially variable, but previous observations were inconclusive \citep{kilic23b,kilic23a}. 

\begin{deluxetable}{ccc}
\tablecolumns{3} 
\tablecaption{DA white dwarfs in our sample that are in/near the ZZ Ceti instability strip with no time-series follow up.}
\label{tab:zz_ceti}
\tablehead{\colhead{} & \colhead{Object Name} & \colhead{}}
\startdata
J0039$-$0357 & J0949$-$0730 & J1656+5719\\
J0050$-$2826 & J0950$-$2841 & J1722+3958\\
J0127$-$2436 & J1052+1610 & J1819+1225\\
J0154+4700 & J1107+0405 & J1929$-$2926\\
J0158$-$2503 & J1342$-$1413 & J2026$-$2254\\
J0712$-$1815 & J1451$-$2502 & J2107+7831\\
J0725+0411 & J1552+0039 & J2208+2059\\
J0912$-$2642 & & \\
\enddata
\end{deluxetable}

Interestingly, there are 22 ZZ Ceti candidates with no time-series follow-up as of yet. Table \ref{tab:zz_ceti} provides a list of these massive ZZ Ceti candidates. Four of these objects have $M>1.2~M_{\odot}$ under the assumption of CO cores. These objects are J0039$-$0357 (Gaia DR3 2527618112309283456) with $T_{\rm eff} = 11,871 \pm 214$ K and  $M = 1.271 \pm 0.009~M_{\odot}$, J0127$-$2436 (Gaia DR3 5040290528701395456) with $T_{\rm eff} = 11,236 \pm 214$ K and  $M = 1.284 \pm 0.009~M_{\odot}$, J0912$-$2642 (Gaia DR3 5649808720867457664) with $T_{\rm eff} = 12,973 \pm 115$ K and  $M = 1.262 \pm 0.002~M_{\odot}$, and J1552+0039 (Gaia DR3 4410623858974488832) with $T_{\rm eff} = 13,508 \pm 233$ K and  $M= 1.245 \pm 0.011~M_{\odot}$. 

\begin{figure*}
\centering
\includegraphics[width=8in, clip, trim={.5in 1in .5in 1in}]{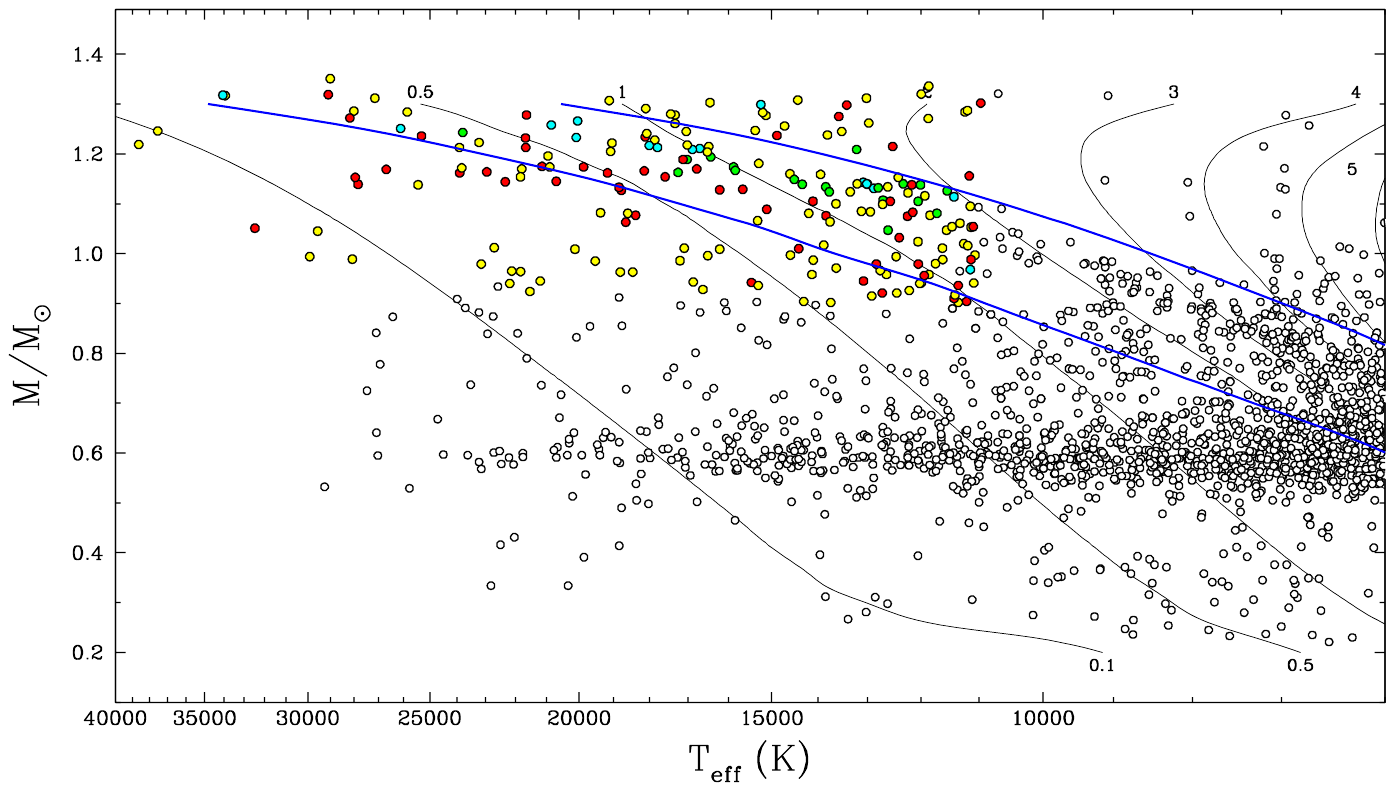}
\caption{Stellar masses as a function of effective temperature for the MWDD 100 pc sample (white dots) along with our massive white dwarf sample (yellow dots). Red, green, and cyan symbols mark the magnetic white dwarfs, hot/warm DQs, and objects with large tangential velocities or rapid rotation, respectively. Solid curves are theoretical isochrones, labeled in units of Gyr, obtained from standard cooling sequences with CO-core compositions, $q({\rm He}) \equiv M({\rm He})/M_{\star} = 10^{-2}$, and $q$(H) = $10^{-4}$. The lower blue solid curve indicates the onset of crystallization at the center of evolving models, while the upper one indicates the locations where 80\% of the total mass has solidified.}
\label{fig:correltm}
\end{figure*}

In addition, there are several NOVs from \citet{vincent} that fall right in the middle of the ZZ Ceti strip; one of the most interesting is J0135+5722 (Gaia DR3 412839403319209600) with $T_{\rm eff} = 12,415 \pm 87$ K and $M= 1.153 \pm 0.004~M_{\odot}$. \citet{vincent} did not detect any significant variations in this star at the 7.8\% level, hence low level variability could easily be missed in those initial observations. In fact, preliminary observations of this target at the APO 3.5m reveal significant variability at the 10 mma (1\%) level in a single mode at 2.6 min period. However, additional observations are needed to confirm and constrain multi-mode pulsations in this object. Follow-up time-series observations of these candidates would be invaluable in finding additional massive pulsating white dwarfs, and probe their interiors through asteroseismology. 

\section{Discussion}

Our results are summarized in Figure \ref{fig:correltm}, where we show the stellar masses as a function of effective temperature for our massive white dwarf sample (top left corner of this plot) along with the MWDD 100 pc sample \citep{kilic20}. Objects of particular astrophysical interest are also identified in this figure, which we discuss in turn.

\subsection{Magnetism}\label{sec:mag}

Our spectroscopic follow-up of the relatively hot and massive ($T_{\rm eff}>11,000$ K and $M>0.9~M_{\odot}$) white dwarfs in the MWDD 100 pc sample within the Pan-STARRS footprint reveals an unusual mix of spectral types. We find only 6 massive DB white dwarfs, but none are normal. One of these is a rapidly rotating DBA \citep{pshirkov20}, and the remaining five are magnetic. Massive DBs seem to be very rare. For example, \citet{obrien24} find only 2 massive DBs (both near $1.1~M_{\odot}$, see their Figure 5) in the 40 pc sample. On the other hand, massive DQs are more common: there are 20 massive DQ white dwarfs in our sample, including 5 DAQs. In total, we find 66 magnetic white dwarfs (32\% of the sample); there are 50 DA and 5 DB white dwarfs in our sample that are either confirmed or suspected to be magnetic, 8 DC white dwarfs that must be magnetic to have featureless spectra at $T_{\rm eff}>11,000$ K, and 3 magnetic DQs. 

Several theories exist in the literature regarding how magnetic fields are produced in white dwarfs. Magnetic fields could be fossil in origin in remnants from highly magnetic Ap/Bp main-sequence stars \citep{Tout04}. However, these progenitors typically have main-sequence masses of 2--3 $M_\odot$, which would produce lower mass white dwarfs than what are analyzed in this work. In addition, the fraction of magnetic stars among the more massive O and B stars is relatively low \citep[$6\pm3$\%,][]{scholler17}. A dynamo generated via crystallization has been invoked for many cooler white dwarfs \citep{isern17}. \citet{bagnulo22} detail a scenario where the magnetic field takes 2--3 Gyr to propagate to the surface after forming in the interior. They find that strong magnetic fields are very common in massive white dwarfs and appear immediately after the formation of the star, whereas magnetic fields appear in lower-mass white dwarfs only when they are older. They attribute the former to mergers and the latter to a crystallization induced dynamo in lower-mass white dwarfs. \citet{blatman24} estimate a delay between the onset of crystallization and field breakout of order a few Gyr. 

\begin{figure}
\centering
\includegraphics[width=3.4in, clip, trim={0in 0in 0.4in 0.4in}]{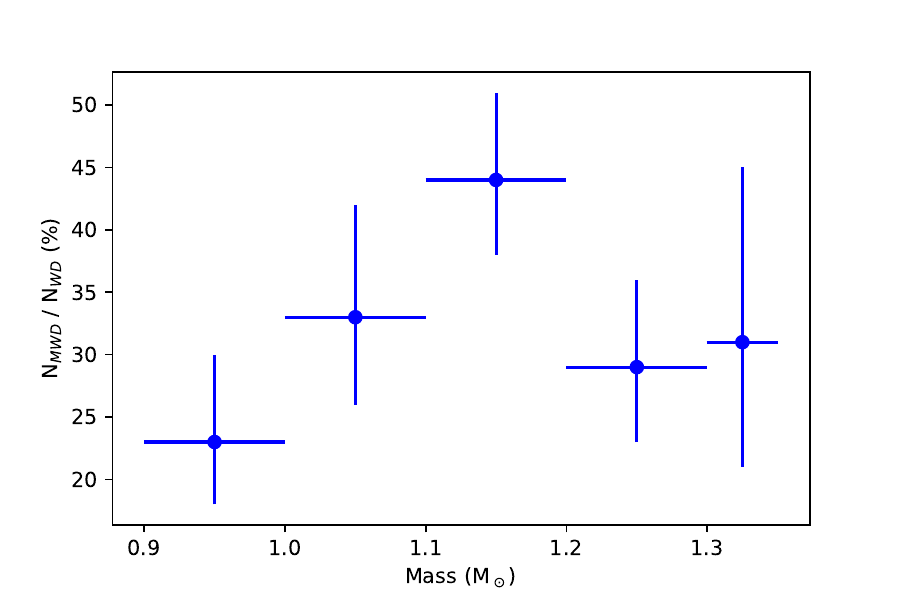}
\caption{The fraction of magnetic white dwarfs as a function of mass for our massive white dwarf sample. The horizontal error bars in this plot show the width of the mass bins and the vertical error bars were calculated using the binomial probability distribution detailed in \cite{burgasser}. The mass bins are equal in width except for $M>1.3\  M_\odot$.}
\label{fig:40pc}
\end{figure}

Our massive magnetic white dwarf sample has a median cooling age of 0.7 Gyr. In fact, all but one (J2255+0710) of the magnetic white dwarfs in our sample have a cooling age $\leq1.8$ Gyr based on the standard cooling tracks. In addition, roughly 1/3 of the magnetic white dwarfs in our sample also show rapid rotation and/or large tangential velocities. Therefore, mergers are more likely to explain the strongly magnetic white dwarfs in our sample \citep{Garcia,Briggs}. This channel gives rise to both higher mass objects and typically stronger magnetic fields ($\sim$MG scales) than the aforementioned methods. However, there is a caveat in this argument; the majority of the magnetic white dwarfs in our sample are found in the crystallization sequence (see Figure \ref{fig:correltm}). If these stars also suffer from extra cooling delays due to distillation \citep{bedard24}, then their cooling ages could be much longer than estimated, and we cannot rule out the crystallization induced dynamo as the source of magnetism. In addition, ultramassive white dwarfs with ONe cores crystallize much earlier, and the magnetic fields could be visible at the surface within $<1$ Gyr. Hence, it is plausible that crystallization induced dynamos may explain at least a fraction of the magnetic white dwarfs in our sample.

We investigate what fraction of our sample is magnetic and if there are trends between magnetism, mass, and fast rotation. To do this, we divide the sample into mass bins of equal width, except for the most massive bin. This leaves us with bin widths of 0.9--1.0 $M_\odot$, 1.0--1.1 $M_\odot$, 1.1--1.2 $M_\odot$, 1.3+ $M_\odot$. There are 48, 36, 58, 49, and 13 white dwarfs in each bin, respectively. When calculating the fraction of magnetic objects, we include all 66 objects either confirmed or suspected of magnetism. 

\begin{figure}
\includegraphics[width=3.4in, clip, trim={0.2in 0.05in 0.4in 0.4in}]{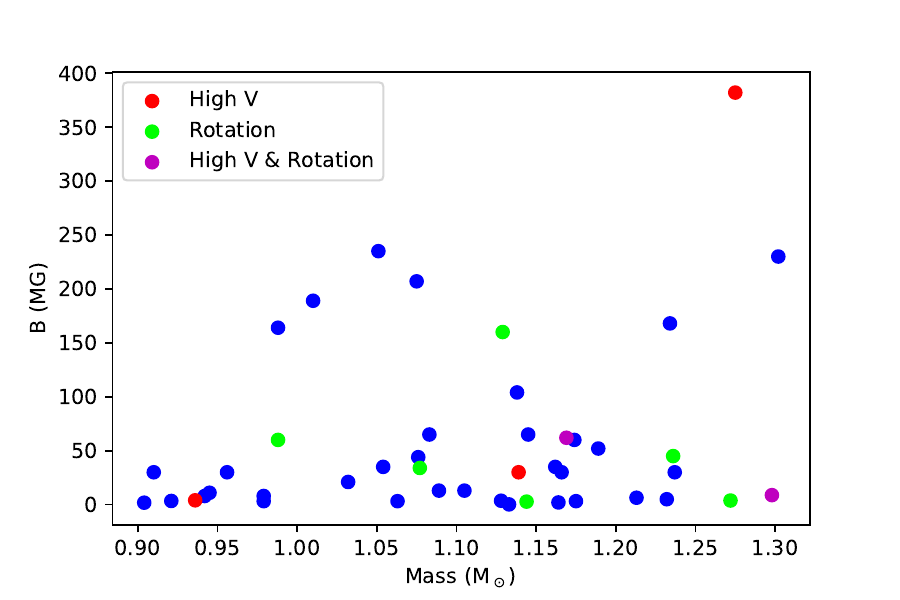}
\caption{The magnetic field strength as a function of mass for the objects we are able to successfully fit. Objects with rapid rotation and large tangential velocities are marked by green and red dots, respectively.}
\label{fig:mag}
\end{figure}

Figure \ref{fig:40pc} shows the fraction of magnetic white dwarfs as a function of mass. Given the errors, we do not see a significant increase in the magnetic fraction as a function of mass, except for the 1.1--1.2 $M_\odot$ bin, which has a magnetic fraction of {$43_{-6}^{+7}$}\%. 
This is likely caused by mergers of average mass white dwarfs \citep{Garcia}, as the mass distribution of DA white dwarfs strongly peaks at 0.59 $M_\odot$ \citep[e.g.,][]{kilic20,obrien24}. 
 
\citet{bagnulo22} find that $\sim$10\% of the 40 pc white dwarfs younger than 0.6 Gyr are magnetic. This is significantly lower than the fraction of magnetic white dwarfs in our massive white dwarf sample. Magnetic white dwarfs tend to be more massive in general, hence the higher fraction of magnetic objects in our sample is not surprising \citep{vennes}. Looking at the fraction of magnetic white dwarfs in the 40 pc sample as a function of mass, \citet{obrien24} find that the magnetic fraction goes up from about 6\% for $0.6~M_{\odot}$ white dwarfs to about 14, 18, and 40\% for 0.8, 1.0, and $1.2~M_{\odot}$ white dwarfs, respectively. For comparison, \cite{kilic23a} also found a magnetic fraction of 40\% among the most massive white dwarfs ($M\sim1.3~M_{\odot}$) in the solar neighborhood. Our results shown in Figure \ref{fig:40pc} are consistent with the previous estimates within the errors, but provide better constraints on the fraction of magnetic white dwarfs at these relatively large masses given the larger sample size for massive white dwarfs. 

Figure \ref{fig:mag} shows the magnetic field strength as a function of mass for the 45 magnetic objects that we were able to successfully fit. We do not see a trend in the magnetic field strength as a function of mass. \citet{hardy23} analyzed 185 magnetic DA white dwarfs and similarly found no correlation between mass and field strength, except for that the strongest fields occur in the higher mass white dwarfs.  However their sample included objects down to 0.4 $M_\odot$, whereas we restrict our sample specifically to those higher mass objects. 

If a significant portion of our magnetic objects come from mergers, it appears that the masses of the binary components does not affect the resulting magnetic field strength. Some other mechanism would need to be attributed to why some targets have field strengths on the order of a few MG, while others have strengths on the order of tens or hundreds of MG. \citet{hardy23} also looked for a correlation between effective temperature and magnetic field strength to potentially point to an origin of the magnetic field, but found no relation. We also do not see any correlations between the surface temperature and the field strength in our smaller sample.

\subsection{Kinematics}
\label{kin}

\begin{figure}
\includegraphics[width=3.4in, clip, trim={0.2in 0.05in 0.4in 0.4in}]{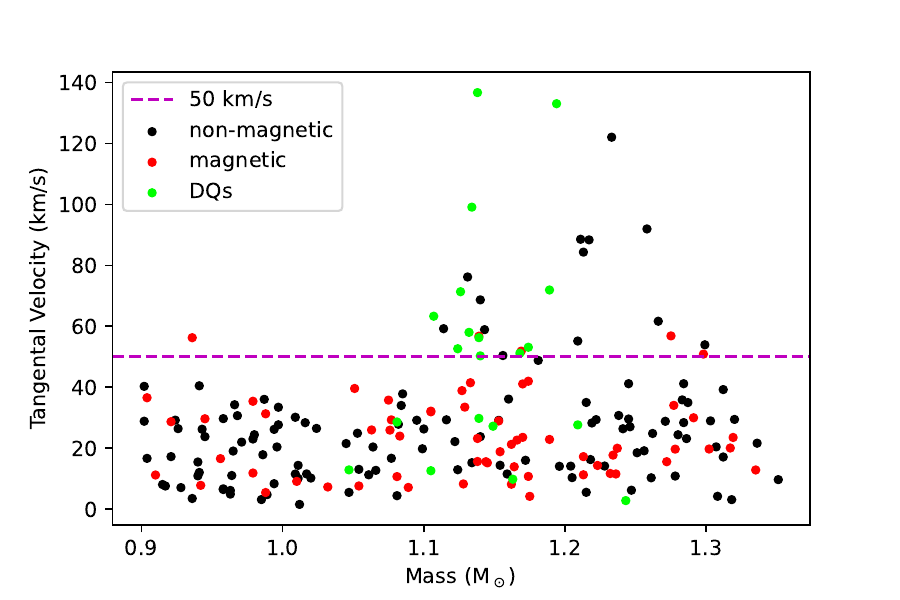}
\caption{Tangential velocity versus mass for our sample. Red and green points mark the magnetic and DQ (including three magnetic DQ) white dwarfs, respectively. The black points show the rest of the objects in the sample. The dashed magenta line shows the 50 km s$^{-1}$ limit. }
\label{fig:vtan}
 \end{figure}

\begin{figure}
\includegraphics[width=3.4in, clip, trim={0.2in 0.05in 0.4in 0.4in}]{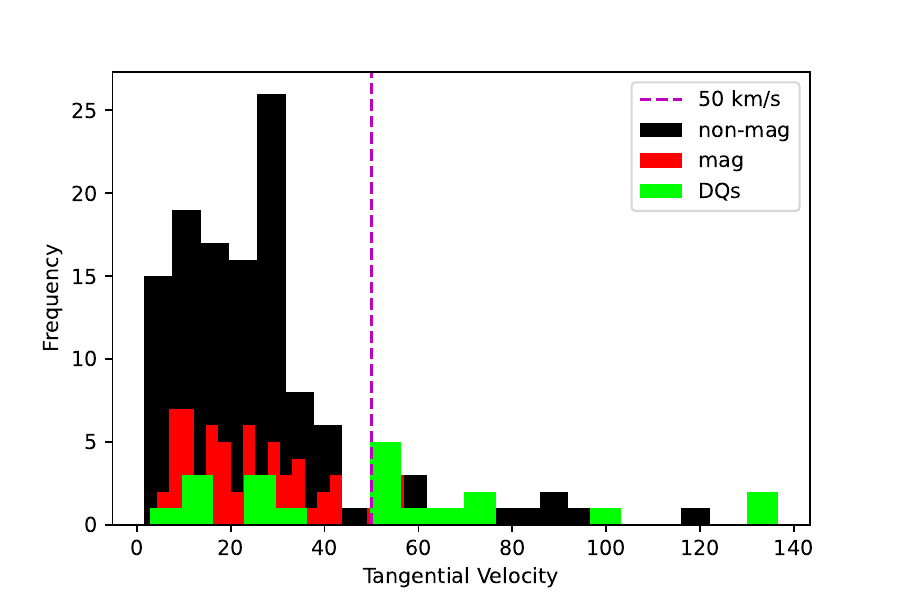}
\caption{Tangential velocity distribution for our sample. We use the same color scheme to highlight the magnetic, DQ, and the rest of the sample as in Figure \ref{fig:vtan}.}
\label{fig:histogram}
\end{figure} 

Kinematics can be helpful in identifying unusual objects in the solar neighborhood; thick disk and halo white dwarfs have a higher velocity dispersion, and therefore can be identified based on their transverse velocities. Given the relatively short main-sequence lifetimes of their progenitors, massive white dwarfs that formed through single star evolution in the thick disk or halo should have cooled below $T_{\rm eff} = 11,000$ K in the distant past. Hence, the only way for thick disk or halo white dwarfs to be included in our sample is if their evolution is reset by a merger event in their recent history. 

\citet{wegg} noted that the smoking gun signature of merger remnants would be high-mass white dwarfs travelling at $>50$ km s$^{-1}$. However, instead of an increase in velocity, they found that the velocity dispersion of the Palomar-Green and the SDSS white dwarf samples actually decreases with increasing white dwarf mass, which prompted them to conclude that the observed kinematics are consistent with the majority of high-mass white dwarfs forming through single star evolution. In addition, they did not find any high-mass white dwarfs moving with velocities above 50 km s$^{-1}$.

Figures \ref{fig:vtan} and \ref{fig:histogram} show the tangential velocity distribution of our sample of massive white dwarfs. We mark the magnetic white dwarfs with red symbols, and DQs (including 3 magnetic DQs) with green symbols. The average velocity of the sample excluding 3$\sigma$ outliers is 28 km s$^{-1}$, consistent with the disk population. The magenta line shows the velocity limit from \citet{wegg}, 50 km s$^{-1}$. 

Interestingly, we find 30 massive white dwarfs (15\% of the sample) with velocities larger than this limit. 
These include 12 warm DQs, 1 He-DA, 5 magnetic white dwarfs, and 12 normal DA white dwarfs. Hence, it appears that some of the merger products in the solar neighborhood hide among the normal DA population. There are eight objects with $V_{\rm tan}>80$ km s$^{-1}$, three of these are DAQ/DQA, and 5 are normal DAs. Surprisingly, these five DAs (J0401+2140, J0447+4224, J0455$-$0058, J0529+5239, and J1924$-$2717) are found in a relatively narrow mass and temperature range of $M=1.21-1.26~M_{\odot}$ and $T_{\rm eff}=16,700-20,800$ K with estimated cooling ages of $\leq1$ Gyr. Hence, with masses roughly twice the mass of the most common white dwarfs in the solar neighborhood, and with unusual kinematics for their ages, these massive DA white dwarfs must be merger remnants. Mergers can reset the white dwarf cooling clock, and they can also change the composition of the remnant star in a way that it then undergoes distillation and is therefore kinematically much older than other warm white dwarfs \citep[e.g.,][]{cheng19,bedard24}.

On the other hand, not all merger remnants show large tangential velocities. Out of the 20 warm DQs in our sample, 12 move faster than the 50 km s$^{-1}$ limit \citep[see also][]{kawka23,kilic24}. Those must be merger remnants in the thick disk or the halo. Even though eight of these warm DQs have relatively small tangential velocities, at least one has a large radial velocity \citep[J0551+4135,][]{hollands} that also indicates a kinematically old population. Hence, 
not all merger remnants can be identified based on their tangential velocities. 

\begin{deluxetable}{cccc}
\tabletypesize{\normalsize}
\tablecolumns{4} \tablewidth{0pt}
\tablecaption{Number of magnetic white dwarfs and all merger remnants as a function of mass. } \label{tab:mergers}
\tablehead{\colhead{Mass} & {\# of} & \colhead{Magnetic} & \colhead{Mergers} \\
($M_{\odot}$) & stars & stars & (All) }
\startdata 
    $0.9-1.0$ & 48 & 11 ($23_{-5}^{+7}$\%) & 12 ($25_{-5}^{+7}$\%) \\
    $1.0-1.1$ & 36 & 12 ($33_{-6}^{+9}$\%) & 13 ($36_{-7}^{+9}$\%)\\
    $1.1-1.2$ & 58 & 25 ($43_{-6}^{+7}$\%) & 45 ($78_{-7}^{+4}$\%) \\
    $1.2-1.3$ & 49 & 14 ($29_{-6}^{+7}$\%) & 24 ($49_{-7}^{+7}$\%) \\
    $1.3+$    & 13 &  4 ($31_{-10}^{+14}$\%) & 5 ($38_{-11}^{+15}$\%) \\
    Sample    & 204 & 66 ($32_{-3}^{+4}$\%) & 99 ($49_{-4}^{+3}$\%)\\
\enddata
\end{deluxetable}

Since the nearby white dwarf population is dominated by the thin disk, merger
remnants in the disk would have a range of formation times and cooling ages; some of these merger remnants would be hot and young enough to be included in our sample.
In fact, the majority of the magnetic white dwarfs are indistinguishable from the non-magnetic objects in terms of their kinematics, providing further evidence that a significant fraction of merger remnants (like warm DQs or magnetic white dwarfs) may not stand out in their kinematics.

\subsection{The merger fraction}

We seek to determine the merger fraction for the massive white dwarfs in the MWDD 100 pc sample in the Pan-STARRS footprint. We can identify possible merger products by signs of unusual atmospheric composition, magnetism, rapid rotation, and high tangential velocity. For a full list of the objects we identify as likely merger products, see Table \ref{tab:phys}, where we use the following code to identify merger evidence; ``A" for an unusual atmospheric composition, ``M" for magnetism, ``R" for rapid rotation, and ``V" for a high tangential velocity. 

Starting with the atmospheric composition, we classify the 20 massive hot and warm DQs in our sample as stars with an unusual atmospheric composition. As discussed in detail by several authors 
\citep{dufour08, coutu19, koester19, hollands, kawka23,kilic24}, hot DQs and warm DQs (including DAQ and DQA white dwarfs) stand out among the solar neighborhood white dwarfs in terms of their
atmospheric composition, rapid rotation, and large tangential velocities. 

While the source of magnetism in white dwarfs is unclear, it is likely that several channels contribute to the emergence of strong magnetic fields in white dwarfs. The 20 and 40 pc samples discussed in \citet{bagnulo22} are most relevant for our purposes, as they found that large magnetic fields in the most-massive white dwarfs emerge at the stellar surface shortly after the start of the cooling phase, whereas the frequency of the magnetic white dwarfs grows slowly with time for lower-mass white dwarfs. Hence, \citet{bagnulo22} favor a merger origin for relatively hot and massive white dwarfs with strong fields. In total, we identify 66 magnetic white dwarfs in our sample, where all but one of them have fields ranging in strength from 2 MG to hundreds of MG. Out of these 66 objects, 20 also show rapid rotation and/or large tangential velocities, and three are magnetic DQs. 

Our best candidates for merger products show all of the signatures discussed above. \cite{kilic21} presented J2211+1136, which has an unusual mixed hydrogen/helium atmosphere, a high magnetic field strength, a rotation period of only 70 s, and a large transverse velocity. There are four objects in our sample that show three different symptoms of merger remnants. Two of these are the DAQ white dwarfs J0831$-$2231 and J2340$-$1819 with carbon + hydrogen atmospheres, rapid rotation (11--12 min periods), and unusual kinematics \citep{kilic24}. The other two, J1214$-$1724 and J2257+0755, are strongly magnetic, have large tangential velocities, and show photometric variability at {$\approx$}1.8 h and 22.8 min \citep{williams}, respectively. 

Binary mergers are expected to contribute 10 to 30\% of all observable single white dwarfs in the solar neighborhood \citep{toonen17,Temmink}. These mergers are dominated by binaries involving post-main-sequence + main-sequence stars, with a contribution of $\leq$15\% from double white dwarf mergers. However, binary mergers are even more important for $M>0.9~M_{\odot}$ white dwarfs. \citet{Temmink} predict that 30--50\% of single massive white dwarfs form through mergers, with the dominant contribution from double white dwarfs. 
In most of their simulated populations, double white dwarf mergers contribute $\sim$45\% of the mergers that lead to a single massive white dwarf. 

\begin{figure}
\centering
\includegraphics[width=3.5in, clip, trim={0.2in 0.05in 0.4in 0.4in}]{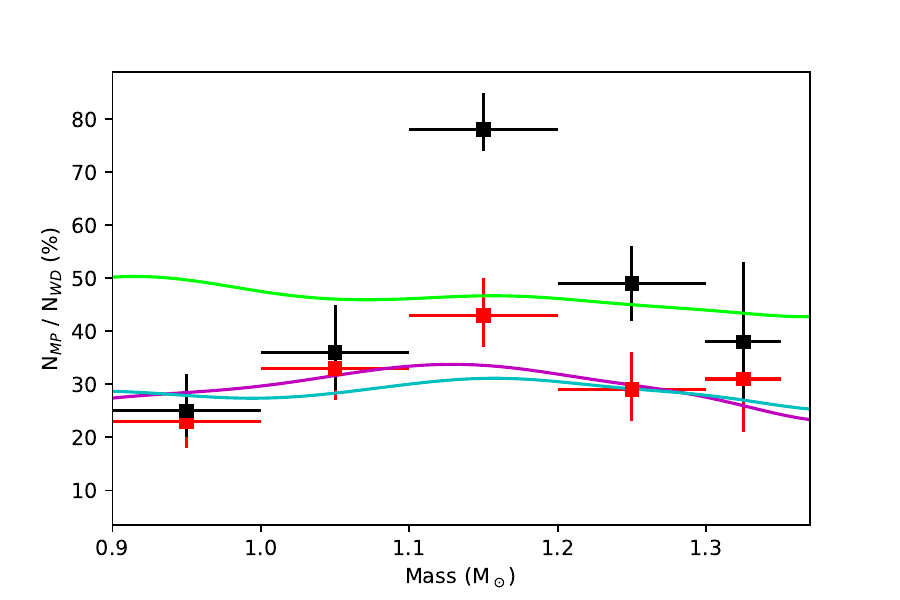}
\caption{The merger fraction (black points) as a function of mass. For comparison, the red points show the the magnetic fraction. The horizontal error bars indicate the size of the mass bin and the vertical error bars represent the upper and lower 1$\sigma$ limits. Also shown are the predicted merger fractions from \citet{Temmink} for 3 different models. The DM91 model is shown in purple, the $\alpha$-efficient in blue, and the $\alpha$-inefficient in green.}
\label{fig:summary}
\end{figure}

Table \ref{tab:mergers} and Figure \ref{fig:summary} present the fraction of mergers in our sample along with the predictions from binary population synthesis models. These models depend heavily on the input assumptions about the initial conditions and the common-envelope evolution. The $\alpha$ prescription is commonly used to model the common-envelope evolution, where $\alpha$ is the fraction of the orbital energy that is used to unbind the common envelope. To demonstrate the range of predictions from the population synthesis models, here we show three models from \citet{Temmink} with various assumptions, $\alpha$-efficient, $\alpha$-inefficient, and DM91. 

A more efficient $\alpha$ means that a larger fraction of the orbital energy can be used to unbind the common envelope, and therefore a larger number of binaries survive this phase. This model ($\alpha$-efficient) predicts a merger fraction of $\sim$30\% among the massive white dwarf population with a slightly higher contribution among 1.1--1.2 $M_{\odot}$ white dwarfs (blue line). On the other hand, a more inefficient common envelope leads a larger number of systems to merge to form single massive white dwarfs. In this case, the merger fraction is significantly higher, roughly 50\%, for 0.9--1.3 $M_{\odot}$ white dwarfs (green line). Finally, the DM91 model has similar input parameters to the default model from \citet{Temmink}, but here the initial periods are drawn from a log-normal distribution from \citet{duquennoy91}. 

\begin{figure*}
\centering
\includegraphics[width=3.4in]{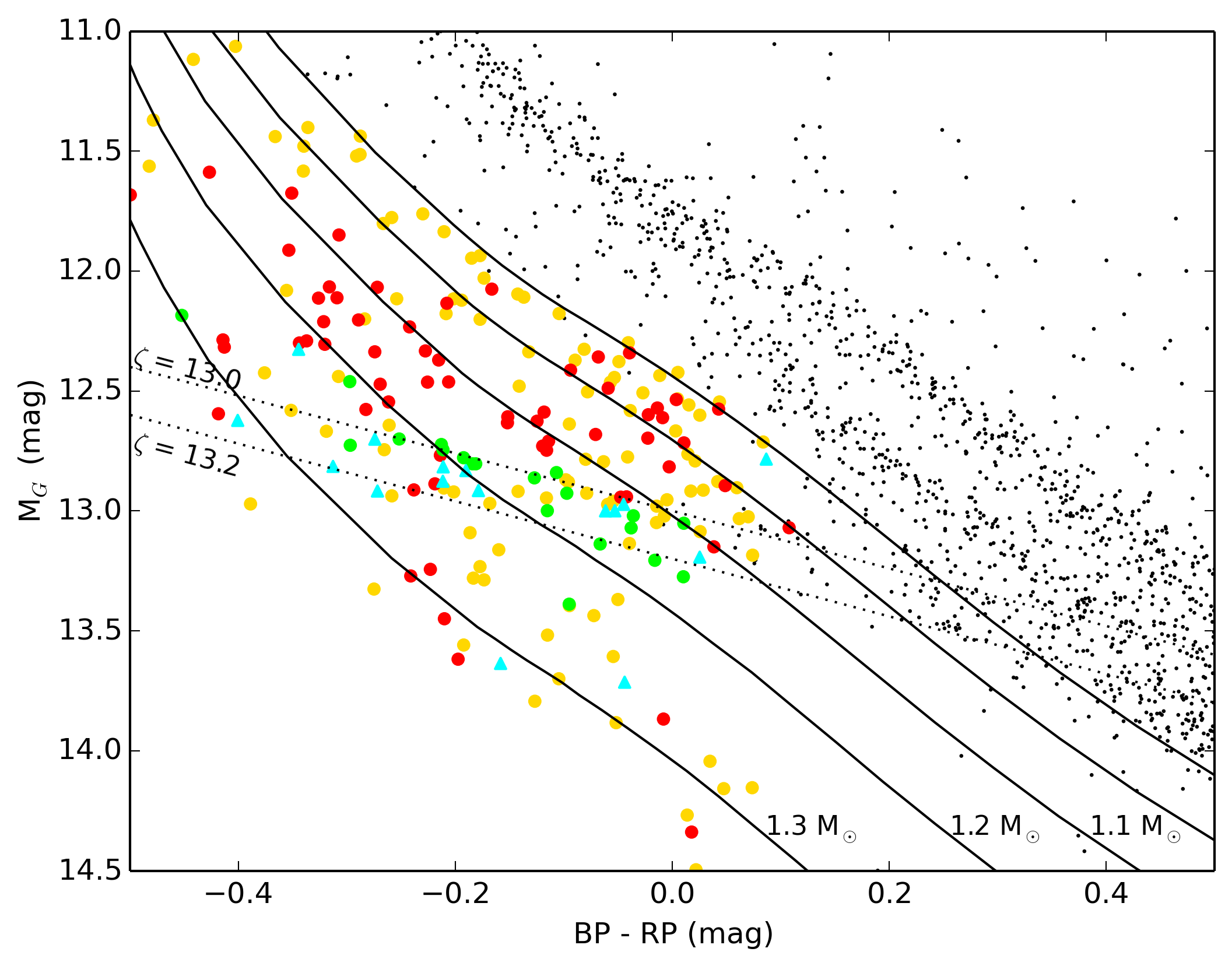}
\includegraphics[width=3.4in]{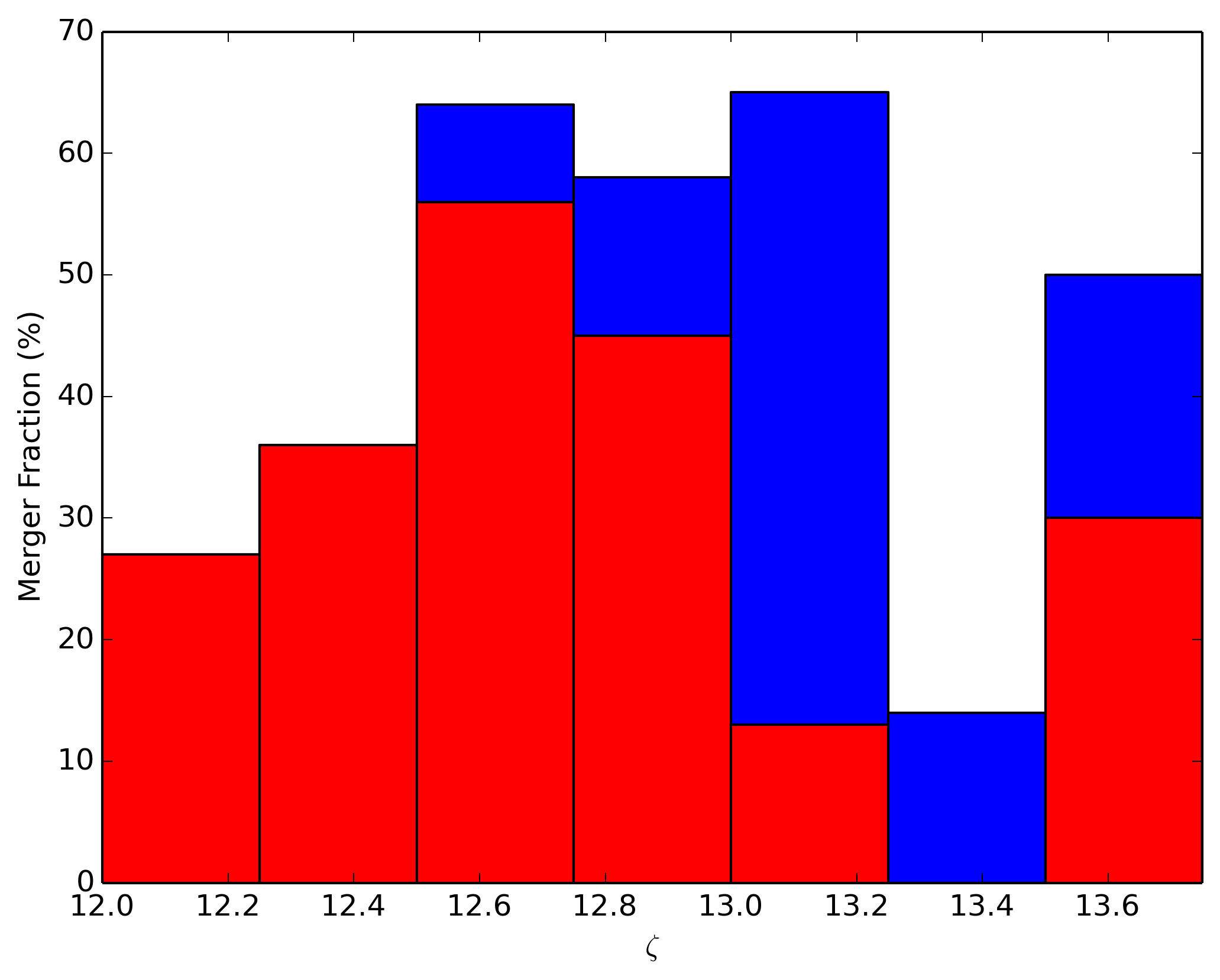}
\caption{{\it Left:} Color-magnitude diagram of the 100 pc white dwarfs along with our massive white dwarf sample (colored dots).
Red, green, and cyan symbols mark the magnetic white dwarfs, hot/warm DQs, and objects with large tangential velocities or rapid rotation,
respectively. The rest of our sample is marked by yellow dots. The lines of constant $\zeta$ are also shown to highlight the Q-branch overdensity at $M_G\approx13$ mag. The solid lines show the cooling sequences for 0.9 to $1.3~M_{\odot}$ (from top to bottom) pure H atmosphere white dwarfs for reference. {\it Right:} The fraction of magnetic white dwarfs (red histogram) and the total fraction of merger systems (blue histogram) as a function of $\zeta$.}
\label{figzeta}
\end{figure*}

A comparison between the merger fraction from the 100 pc sample and the binary population synthesis models shows that in broad lines there is an agreement on the fraction of mergers amongst white dwarfs per mass bin, but that there is no model that can explain all of the data simultaneously.
 At low masses (0.9--1.0 $M_{\odot}$) the observed merger fraction favors models with an efficient common envelope (e.g., $\alpha$-efficient), whereas at high masses the merger fraction is significantly higher and more consistent with the $\alpha$-inefficient models. More strikingly, we find a merger fraction of $78_{-7}^{+4}$\% among the 1.1--1.2 $M_{\odot}$ white dwarfs, which is significantly higher than predicted by the models presented in \citet{Temmink}. There are a few caveats in the population synthesis models; they are based on the evolutionary models for CO core white dwarfs for all objects, and they do not include any cooling delays from distillation. In addition, the models also assume a constant binary fraction across the entire mass range of the initial population, which likely underestimates the number of mergers at higher masses. However, even with these caveats, we show below that the discrepancy between the observed and predicted merger fractions between 1.1 and 1.2 $M_{\odot}$ is most likely explained by our selection bias including the Q-branch white dwarfs.

Massive DQs are marked by green dots in Figure \ref{fig:correltm}. Remarkably, 16 of the 20 hot/warm DQs in our sample are found in the 1.1--1.2 $M_{\odot}$ mass range, and all hot/warm DQs in the 100 pc sample are also found in the crystallization sequence, likely because they are stuck there due to the multigigayear cooling delays from $^{22}$Ne distillation \citep{blouin21,bedard24}. Hence, they are likely over-represented in our sample because of these cooling delays. 

The left panel in Figure \ref{figzeta} shows the color-magnitude diagram of the 100 pc sample centered on the Q-branch overdensity at $M_G \approx 13$ mag. The colored symbols mark our massive white dwarf sample. Specifically, magnetic white dwarfs are shown in red, hot/warm DQs in green, and objects with unusual kinematics or rapid rotation in cyan. The dotted lines show the tracks for constant $\zeta = M_G - 1.2 \times (BP-RP)$ \citep{camisassa21}, which delineate the Q-branch overdensity at $\zeta=13-13.2$. The solid lines show the 0.9 to $1.3~M_{\odot}$ pure H atmosphere white dwarf cooling sequences for reference. 

Clearly, there is an overdensity of warm DQs and objects with large tangential velocities on the Q-branch, which is located where CO white dwarfs with thin helium envelopes, $q({\rm He}) \sim 10^{-6}$, crystallize \citep{bedard24}. This is key evidence for the merger origin of the delayed population on the Q-branch, as most of the helium is assumed to be burned during the merger.
The majority of these objects also fall between the evolutionary sequences for 1.1--1.2 $M_{\odot}$ white dwarfs. 
The right panel in Figure \ref{figzeta} shows the merger fraction as a function of $\zeta$ (blue histogram). For comparison, the fraction of magnetic white dwarfs is also shown as a red histogram. Even though the merger fraction estimate is dominated by magnetic white dwarfs over the most of the parameter space, the same is not true for the Q-branch overdensity at $\zeta = 13-13.2$. The latter is dominated by contribution from warm DQs and objects with unusual kinematics (green and cyan symbols in the left panel). Hence, it is not only the warm DQs, but also massive DA white dwarfs on the Q-branch that inflate the merger fraction estimate for the 1.1--1.2 $M_{\odot}$ white dwarfs. 

Because our sample is limited to objects with $T_{\rm eff}\geq11,000$ K, this temperature cut-off implies that the 1.1--1.2 $M_{\odot}$ bin has lots of Q-branch stars, while the lower- and higher-mass bins have much fewer. For example, between 0.9 and $1.0~M_{\odot}$, our 11,000 K cut-off implies that we have no stars on the Q-branch included in our sample. Similarly, the bins at 1.0--1.1 and $M>1.3~M_{\odot}$ have only a few stars on the Q-branch. This is in contrast to the 1.1--1.2 and 1.2--1.3 $M_{\odot}$ bins that have many Q-branch objects. This is significant because
about 50\% of stars on the Q-branch are delayed due to the $^{22}$Ne distillation \citep{cheng19,blouin21,bedard24}, they inflate our merger rate estimate considerably. Based on the population synthesis models presented in \citet{bedard24}, distillation approximately doubles the number of merger products in the 1.08--1.23 $M_{\odot}$ range with $T_{\rm eff}>11,000$ K. This estimate depends on the fraction of stars undergoing distillation, which is somewhere between 5 and 9\%. Hence, the true merger fraction for 1.1--1.2 $M_{\odot}$ white dwarfs is likely half of what is observed, $\sim$40\%, more in line with the predictions from the population synthesis models.

Even though the majority of the magnetic white dwarfs in our sample are found in the crystallization sequence in Figure \ref{fig:correltm}, there are only a few magnetic white dwarfs on the Q-branch in Figure \ref{figzeta}. It is possible that the magnetic white dwarfs simply come from double white dwarf mergers, and the Q-branch objects (mostly warm DQs and DAs) come from white dwarf + subgiant star mergers. The latter may produce ultramassive CO white dwarfs with enough neutron-rich impurities that can power the $^{22}$Ne distillation mechanism and lead to multigigayear cooling delays \citep{shen23}. We suspect that double white dwarf mergers cannot explain the delayed population because they do not produce extra neutron-rich species as white dwarf + subgiant mergers do, so they do not cluster on the Q-branch. 

\section{Conclusions}

We have presented the results from our detailed spectroscopic analysis of the massive ($M>0.9\ M_{\odot}$) white dwarfs with $T_{\rm eff}\geq11,000$ K in the MWDD 100 pc sample and the Pan-STARRS footprint. Our sample contains 204 objects, 109 of which had no previous spectral classification in the literature. Our spectroscopic follow-up is complete for this sample; we find
118 normal DA white dwarfs, but no normal DBs. There are 3 DBs that are magnetic, 2 additional DBs that are both strongly magnetic and rapidly rotating (J0043$-$1000 and J1214$-$1724), and 1 DBA \citep[J1832+0856,][]{pshirkov20} that rotates rapidly. 
In total, there are 66 objects that are magnetic. We also find 20 warm/hot DQs, including 14 warm DQ/DQAs, 5 DAQs, and 1 hot DQ. Previously, there was only 1 known DAQ in the literature \citep{hollands}, so we have quintupled the sample of DAQs known within the 100 pc sample and the Pan-STARRS footprint. 

Our main goal with analyzing this sample is to constrain the merger fraction as a function of mass. The signatures of a merger origin are unusual atmospheric composition, magnetism, rapid rotation, and unusual kinematics. Massive DQs stand out in terms of their composition. 
Interestingly, we also find all of them in the crystallization sequence. \citet{cheng19} demonstrated that 5--9\% of high-mass white
dwarfs in the crystallization sequence show a multi-Gyr cooling anomaly, which implies 50\% of Q-branch white dwarfs exhibit longer cooling delays. \citet{blouin21} and \citet{bedard24} show that this cooling
delay is likely due to the $^{22}$Ne distillation process, which can lead to 7--10 Gyr delays in cooling. 
The relatively large tangential velocities and rapid rotation rates in massive DQs favor a merger origin, which could lead to massive CO core white dwarfs with enough neutron-rich impurities that can power the distillation mechanism.  

\citet{wegg} identify high tangential velocity as the “smoking gun” signature for merger products. We find 30 massive white dwarfs (15\% of our sample) with $V_{\rm tan}>50$ km s$^{-1}$. However, we also show that not all merger products have high velocities, as six of the DQs within 100 pc (which are merger products) have velocities below this limit. We also do not find a  trend in the kinematics of the magnetic versus non-magnetic objects in the sample. 

The fraction of magnetic white dwarfs in our sample is 32\%. This is significantly higher than the fraction ($\sim$10\%) found for white dwarfs younger than 0.6 Gyr and within 40 pc \citep{bagnulo22}. However, even in the 40 pc sample, the mass distribution of the magnetic white dwarfs is skewed toward the highest masses. Hence, it is not surprising that the magnetic fraction is relatively high in our massive white dwarf sample. 

More interestingly, both the magnetic fraction and the merger fraction show a peak in the distribution between 1.1 and $1.2~M_{\odot}$. A comparison with the binary population synthesis calculations show that there is no single model that can explain all of the observations, though the predictions from the different population synthesis models overlap with the observed merger fraction for the most of the mass range studied. The exception is the 1.1--1.2 $M_{\odot}$ range, where the observed fraction is significantly higher than predicted. Given that this mass range corresponds to roughly twice the mass of the most common white dwarfs in the solar neighborhood, and that the majority of the warm DQs are also found in this range indicate a white dwarf merger origin for these systems. We discuss the most likely explanation for the higher merger fraction in the observed population, and demonstrate that multi-Gyr cooling delays from $^{22}$Ne distillation could also explain a larger than expected contribution from merger remnants among the single white dwarfs in the local white dwarf population.

\begin{acknowledgments}

This work is based on observations obtained at the MDM Observatory,  operated by Dartmouth College, Columbia University, Ohio State University, Ohio University, and the University of Michigan. The authors are honored to be permitted to conduct astronomical research on Iolkam Du'ag (Kitt Peak), a mountain with particular significance to the Tohono O'odham. We thank Justin Rupert for performing the MDM queue observations. 

This work is supported in part by the NSF under grant  AST-2205736, NASA under grants 80NSSC22K0479, 80NSSC24K0380, and 80NSSC24K0436, NSERC Canada, Fund FRQ-NT (Qu\'ebec), Canadian Institute for Theoretical Astrophysics (CITA) National Fellowship Program, and the Smithsonian Institution. ST acknowledges support from the Netherlands Research Council NWO (VIDI 203.061 grants). \\

The Apache Point Observatory 3.5-meter telescope is owned and operated by the Astrophysical Research Consortium.

Based on observations obtained at the MMT Observatory, a joint facility of the Smithsonian  Institution and the University of Arizona.

Based on observations obtained at the international Gemini Observatory, a program of NSF’s NOIRLab, which is managed by the Association of Universities for Research in Astronomy (AURA) under a cooperative agreement with the National Science Foundation on behalf of the Gemini Observatory partnership: the National Science Foundation (United States), National Research Council (Canada), Agencia Nacional de Investigaci\'{o}n y Desarrollo (Chile), Ministerio de Ciencia, Tecnolog\'{i}a e Innovaci\'{o}n (Argentina), Minist\'{e}rio da Ci\^{e}ncia, Tecnologia, Inova\c{c}\~{o}es e Comunica\c{c}\~{o}es (Brazil), and Korea Astronomy and Space Science Institute (Republic of Korea).

\end{acknowledgments}

\facilities{MMT (Blue Channel spectrograph), ARC 3.5m (KOSMOS spectrograph, Agile imager), Gemini (GMOS), FLWO:1.5m (FAST spectrograph), Magellan:Baade (MagE spectrograph), 	Hiltner (OSMOS spectrograph)}

\bibliography{references}{}
\bibliographystyle{aasjournal}

\appendix

\begin{longtable}{cccccccccc}
\caption{Observational properties of our massive white dwarf sample based on Gaia Data Release 3.}\\
\hline 
\hline 
Object name & Gaia ID & RA & DEC & Parallax & $\mu$$_{RA}$ & $\mu$$_{DEC}$ & G & $G_{BP}$ & $G_{RP}$\\
 & &  ($^{\circ}$) & ($^{\circ}$) & (mas) & (mas yr$^{-1}$) & (mas yr$^{-1}$) & (mag) & (mag) & (mag)\\
\hline
J0006$+$3104 & 2861452348130844160 & 1.65808 & 31.07098 & 10.19 $\pm{0.06}$ & 21.7 & $-$25.8 & 16.80 & 16.69 & 17.00 \\
J0012$-$0606 & 2443419990050464128 & 3.08603 & $-$6.10606 & 14.62 $\pm{0.07}$ & 123.5 & 14.2 & 16.35 & 16.31 & 16.42 \\
J0029$+$3648 & 366784816895496064 & 7.49632 & 36.80948 & 17.01 $\pm{0.06}$ & 90.9 & $-$45.9 & 16.42 & 16.31 & 16.66 \\
J0039$-$0357 & 2527618112309283456 & 9.78326 & $-$3.95607 & 11.30 $\pm{0.22}$ & 54.5 & $-$41.8 & 18.61 & 18.62 & 18.67 \\
J0043$-$1000 & 2377863773908424448 & 10.94092 & $-$10.00754 & 32.12 $\pm{0.04}$ & $-$145.6 & $-$134.9 & 14.53 & 14.43 & 14.70 \\
J0045$-$2336 & 2348747743931814656 & 11.36596 & $-$23.60878 & 21.17 $\pm{0.07}$ & 283.6 & $-$145.4 & 16.64 & 16.65 & 16.64 \\
J0049$-$2525 & 2345323551189913600 & 12.32153 & $-$25.43257 & 10.03 $\pm{0.25}$ & 22.5 & $-$28.3 & 19.03 & 19.08 & 19.04 \\
J0050$+$3138 & 360858960322547968 & 12.57920 & 31.64609 & 13.27 $\pm{0.13}$ & $-$90.1 & $-$38.7 & 18.08 & 18.06 & 18.17 \\
J0050$-$0326 & 2529337507976700928 & 12.69082 & $-$3.44882 & 12.58 $\pm{0.08}$ & $-$23.6 & $-$18.3 & 16.79 & 16.67 & 17.01 \\
J0050$-$2826 & 2342438501397962112 & 12.71700 & $-$28.43495 & 11.07 $\pm{0.11}$ & 69.3 & 16.1 & 17.81 & 17.86 & 17.80 \\
J0104$+$4650 & 401215160231429120 & 16.05720 & 46.84480 & 12.01 $\pm{0.11}$ & 50.6 & $-$117.0 & 17.74 & 17.71 & 17.78 \\
J0107$+$2518 & 306350606950880128 & 16.85952 & 25.30977 & 11.16 $\pm{0.06}$ & 56.2 & $-$25.6 & 16.70 & 16.65 & 16.83 \\
J0107$+$2904 & 308383019835259008 & 16.93868 & 29.07230 & 17.00 $\pm{0.09}$ & 38.5 & $-$33.0 & 16.75 & 16.69 & 16.90 \\
J0118$-$0156 & 2533306985471073920 & 19.54309 & $-$1.93677 & 9.98 $\pm{0.07}$ & 25.4 & $-$9.3 & 16.68 & 16.50 & 17.00 \\
J0127$-$2436 & 5040290528701395456 & 21.89464 & $-$24.60557 & 12.36 $\pm{0.19}$ & 107.3 & 3.0 & 18.69 & 18.74 & 18.69 \\
J0135$+$2229 & 289247527487337344 & 23.77272 & 22.49331 & 9.90 $\pm{0.09}$ & 33.9 & $-$7.9 & 17.12 & 17.10 & 17.24 \\
J0135$+$5722 & 412839403319209600 & 23.82370 & 57.37989 & 19.66 $\pm{0.05}$ & 60.0 & $-$104.9 & 16.66 & 16.67 & 16.71 \\
J0138$+$5124 & 406267557895785728 & 24.70586 & 51.40391 & 11.72 $\pm{0.07}$ & $-$7.2 & 14.2 & 16.97 & 16.93 & 17.01 \\
J0138$+$2523 & 292454841560140032 & 24.72100 & 25.38939 & 12.32 $\pm{0.06}$ & 52.8 & $-$51.2 & 15.91 & 15.75 & 16.23 \\
J0150$+$2835 & 299265624604662656 & 27.55507 & 28.59888 & 13.92 $\pm{0.08}$ & 102.0 & $-$20.0 & 16.89 & 16.90 & 16.91 \\
J0151$+$2435 & 291231871097124736 & 27.93047 & 24.59737 & 10.87 $\pm{0.09}$ & $-$24.3 & $-$3.9 & 17.28 & 17.23 & 17.45 \\
J0154$+$4700 & 356186555597277440 & 28.64365 & 47.01313 & 10.21 $\pm{0.12}$ & $-$25.8 & $-$25.0 & 17.93 & 17.95 & 17.96 \\
J0158$-$2503 & 5121833510769131136 & 29.69913 & $-$25.05308 & 11.21 $\pm{0.10}$ & $-$7.1 & $-$52.0 & 17.79 & 17.79 & 17.80 \\
J0204$+$8713 & 575585919005741184 & 31.12914 & 87.22579 & 11.11 $\pm{0.08}$ & $-$43.8 & 38.56 & 17.79 & 17.85 & 17.78 \\
J0205$+$2057 & 94276941624384000 & 31.45605 & 20.95109 & 11.71 $\pm{0.11}$ & $-$213.1 & $-$250.4 & 17.38 & 17.28 & 17.58 \\
J0211$+$2115 & 99498964725981440 & 32.95131 & 21.26327 & 16.81 $\pm{0.08}$ & 120.3 & $-$40.3 & 16.76 & 16.79 & 16.74 \\
J0216$+$3541 & 328009783428208256 & 34.00489 & 35.68819 & 11.47 $\pm{0.05}$ & $-$30.6 & $-$90.8 & 15.63 & 15.48 & 15.93 \\
J0230$+$3842 & 334723160910027136 & 37.73357 & 38.70531 & 10.80 $\pm{0.08}$ & $-$34.6 & $-$1.0 & 17.04 & 16.93 & 17.25 \\
J0234$-$0511 & 2488960249844340352 & 38.53328 & $-$5.19384 & 41.64 $\pm{0.02}$ & 244.1 & 92.1 & 14.34 & 14.34 & 14.39 \\
J0248$+$1600 & 33656531962611072 & 42.094794 & 16.00365 & 12.16 $\pm{0.07}$ & 55.1 & $-$45.6 & 16.68 & 16.61 & 16.87 \\
J0249$-$1831 & 5129157117202633216 & 42.25439 & $-$18.52126 & 17.75 $\pm{0.06}$ & 83.6 & 14.8 & 16.29 & 16.21 & 16.47 \\
J0256$-$1515 & 5154220209880074752 & 44.11563 & $-$15.25928 & 11.89 $\pm{0.07}$ & 23.3 & $-$17.2 & 16.92 & 16.81 & 17.13 \\
J0257$+$0308 & 1973795170585088 & 44.37398 & 3.13848 & 10.90 $\pm{0.14}$ & $-$1.5 & $-$55.1 & 17.75 & 17.75 & 17.80 \\
J0307$+$0313 & 2555303678311296 & 46.84852 & 3.22814 & 18.14 $\pm{0.07}$ & 112.8 & $-$33.4 & 17.07 & 17.07 & 17.12 \\
J0311$-$2254 & 5075443981321647744 & 47.77964 & $-$22.90163 & 17.93 $\pm{0.03}$ & $-$16.7 & $-$37.8 & 15.13 & 15.01 & 15.35 \\
J0317$-$2916 & 5058587261181078400 & 49.36961 & $-$29.27257 & 9.81 $\pm{0.17}$ & 63.8 & 82.6 & 18.75 & 18.75 & 18.79 \\
J0319$+$4628 & 243001626047882368 & 49.89805 & 46.47374 & 9.91 $\pm{0.09}$ & $-$18.1 & $-$5.8 & 17.43 & 17.38 & 17.48 \\
J0323$+$3501 & 126017157963829120 & 50.88220 & 35.02105 & 10.46 $\pm{0.09}$ & 39.6 & $-$28.2 & 17.27 & 17.27 & 17.32 \\
J0323$+$3457 & 126015822230372352 & 50.95486 & 34.96002 & 11.15 $\pm{0.11}$ & 7.8 & $-$26.0 & 17.26 & 17.24 & 17.32 \\
J0325$-$0815 & 5168129306849447552 & 51.44795 & $-$8.26361 & 11.45 $\pm{0.10}$ & $-$17.9 & 21.4 & 17.66 & 17.66 & 17.71 \\
J0326$+$1331 & 17709047809907584 & 51.58057 & 13.51891 & 11.31 $\pm{0.09}$ & 124.9 & $-$49.1 & 17.33 & 17.33 & 17.36 \\
J0327$+$2227 & 61856496954301696 & 51.80726 & 22.46398 & 10.16 $\pm{0.20}$ & 70.3 & $-$19.8 & 18.41 & 18.35 & 18.56 \\
J0332$+$3005 & 120664082524436096 & 53.19928 & 30.09952 & 11.49 $\pm{0.15}$ & 16.7 & $-$85.3 & 18.21 & 18.23 & 18.34 \\
J0347$-$1802 & 5107322396824711680 & 56.76401 & $-$18.04825 & 13.21 $\pm{0.10}$ & 163.5 & $-$14.8 & 17.39 & 17.38 & 17.44 \\
J0348$-$0058 & 3251244858154433536 & 57.20948 & $-$0.97636 & 32.21 $\pm{0.04}$ & 84.6 & $-$163.0 & 14.02 & 13.85 & 14.33 \\
J0401$+$2140 & 53042472446551424 & 60.26763 & 21.67249 & 14.03 $\pm{0.08}$ & $-$52.4 & $-$256.3 & 17.09 & 17.04 & 17.23 \\
J0408$+$2323 & 53716846734195328 & 62.01259 & 23.39512 & 12.40 $\pm{0.09}$ & 57.4 & $-$38.9 & 17.29 & 17.32 & 17.30 \\
J0422$-$2407 & 4896474240286811648 & 65.60905 & $-$24.12346 & 13.95 $\pm{0.06}$ & 83.4 & $-$22.8 & 17.02 & 16.94 & 17.21 \\
J0439$+$4543 & 253936196167057664 & 69.96957 & 45.71726 & 10.43 $\pm{0.14}$ & $-$15.1 & $-$42.3 & 18.23 & 18.17 & 18.44 \\
J0447$+$4224 & 203678825329613312 & 71.94959 & 42.41020 & 10.26 $\pm{0.11}$ & 147.1 & $-$123.0 & 17.85 & 17.82 & 18.00 \\
J0448$-$1053 & 3181589319065856384 & 72.13359 & $-$10.8972 & 18.19 $\pm{0.05}$ & $-$43.5 & $-$15.4 & 16.23 & 16.24 & 16.24 \\
J0455$-$0058 & 3226519762223501696 & 73.89203 & $-$0.97170 & 10.72 $\pm{0.11}$ & 112.1 & $-$175.3 & 17.66 & 17.57 & 17.88 \\
J0507$+$2645 & 3421894079307215744 & 76.79028 & 26.75400 & 18.62 $\pm{0.05}$ & 6.8 & $-$45.4 & 16.22 & 16.14 & 16.42 \\
J0521$-$1029 & 3014049448078210304 & 80.32910 & $-$10.48843 & 11.86 $\pm{0.04}$ & 13.5 & $-$52.2 & 15.74 & 15.59 & 16.03 \\
J0529$+$5239 & 263082591016645504 & 82.46206 & 52.66239 & 25.56 $\pm{0.04}$ & 364.2 & $-$548.3 & 15.66 & 15.57 & 15.85 \\
J0533$-$2713 & 2908425653829891712 & 83.430768 & $-$27.23082 & 23.17 $\pm{0.03}$ & $-$40.8 & $-$55.3 & 15.61 & 15.51 & 15.82 \\
J0534$-$0214 & 3216947242193857024 & 83.58469 & $-$2.24198 & 10.67 $\pm{0.04}$ & 9.7 & 4.5 & 15.92 & 15.78 & 16.18 \\
J0538$+$3212 & 3447991090873280000 & 84.74184 & 32.20788 & 10.30 $\pm{0.10}$ & $-$16.6 & $-$7.3 & 17.51 & 17.53 & 17.57 \\
J0547$+$1501 & 3347953532952671360 & 86.75002 & 15.03041 & 14.50 $\pm{0.05}$ & $-$11.2 & $-$21.1 & 16.26 & 16.22 & 16.39 \\
J0547$-$1250 & 2997076768116088576 & 86.78517 & $-$12.84371 & 18.12 $\pm{0.04}$ & 23.0 & $-$15.6 & 16.42 & 16.43 & 16.42 \\
J0550$-$1630 & 2994934026112726528 & 87.51337 & $-$16.50561 & 11.93 $\pm{0.05}$ & 4.3 & $-$10.2 & 16.79 & 16.73 & 16.94 \\
J0551$+$4135 & 192275966334956672 & 87.89424 & 41.59197 & 21.61 $\pm{0.04}$ & 114.3 & 73.2 & 16.34 & 16.35 & 16.38 \\
J0601$+$3726 & 3456777730670779776 & 90.49467 & 37.43344 & 13.79 $\pm{0.07}$ & 11.7 & $-$82.6 & 16.64 & 16.64 & 16.68 \\
J0602$+$4652 & 963601717022097152 & 90.56218 & 46.87762 & 12.03 $\pm{0.11}$ & $-$16.3 & $-$-68.3 & 17.98 & 17.96 & 18.06 \\
J0607$+$3415 & 3452373568124842752 & 91.98229 & 34.25727 & 13.50 $\pm{0.06}$ & $-$10.7 & 5.3 & 16.68 & 16.60 & 16.87 \\
J0608$-$0059 & 3121385658671190784 & 92.21438 & $-$0.99745 & 16.16 $\pm{0.08}$ & 2.8 & $-$36.9 & 17.23 & 17.20 & 17.38 \\
J0625$+$1902 & 3372355922219793536 & 96.39696 & 19.04540 & 10.51 $\pm{0.15}$ & $-$21.2 & $-$33.1 & 17.77 & 17.71 & 17.92 \\
J0634$+$3848 & 945007674022721280 & 98.56907 & 38.81528 & 21.88 $\pm{0.04}$ & $-$58.3 & $-$107.0 & 15.73 & 15.74 & 15.75 \\
J0655$+$2939 & 887758130788405504 & 103.89756 & 29.65195 & 12.70 $\pm{0.07}$ & 101.0 & $-$164.1 & 17.22 & 17.15 & 17.37 \\
J0657$+$7341 & 1114813977776610944 & 104.29631 & 73.69572 & 11.86 $\pm{0.09}$ & 30.9 & $-$22.3 & 17.65 & 17.67 & 17.68 \\
J0705$-$2046 & 2929364817693200640 & 106.33599 & $-$20.77872 & 10.39 $\pm{0.06}$ & 99.1 & $-$75.3 & 16.50 & 16.36 & 16.78 \\
J0707$+$5611 & 988421680189764224 & 106.970556 & 56.19978 & 11.51 $\pm{0.12}$ & $-$36.3 & $-$63.1 & 17.96 & 17.89 & 18.13 \\
J0712$-$1815 & 2934281803636268416 & 108.08535 & $-$18.26537 & 17.87 $\pm{0.05}$ & $-$59.0 & 70.6 & 16.40 & 16.41 & 16.41 \\
J0718$+$3731 & 898348313253395968 & 109.56823 & 37.52740 & 11.81 $\pm{0.09}$ & $-$34.9 & $-$35.8 & 16.95 & 16.82 & 17.23 \\
J0725$+$0411 & 3139633462883694976 & 111.45264 & 4.19282 & 11.37 $\pm{0.08}$ & $-$35.9 & $-$9.5 & 17.22 & 17.23 & 17.26 \\
J0733$+$5750 & 989376331161515008 & 113.30400 & 57.84982 & 12.44 $\pm{0.06}$ & 24.8 & 17.2 & 16.72 & 16.67 & 16.85 \\
J0734$+$4841 & 976040702520790400 & 113.61355 & 48.68672 & 30.76 $\pm{0.03}$ & $-$132.1 & $-$192.9 & 14.93 & 14.90 & 14.99 \\
J0737$-$0610 & 3055999768051212416 & 114.43533 & $-$6.17176 & 12.88 $\pm{0.11}$ & $-$5.3 & 33.8 & 17.51 & 17.50 & 17.54 \\
J0747$+$4527 & 927205339521671552 & 116.80249 & 45.45371 & 11.26 $\pm{0.10}$ & $-$35.3 & $-$51.5 & 17.52 & 17.51 & 17.59 \\
J0801$+$7749 & 1138353770109251200 & 120.34454 & 77.83191 & 12.52 $\pm{0.06}$ & $-$23.4 & $-$143.9 & 17.38 & 17.32 & 17.54 \\
J0803$+$1229 & 653051877600013312 & 120.99977 & 12.49551 & 10.31 $\pm{0.08}$ & 9.6 & $-$14.9 & 17.39 & 17.34 & 17.55 \\
J0811$+$4323 & 928079146323608320 & 122.88316 & 43.38686 & 14.52 $\pm{0.05}$ & $-$51.7 & $-$51.3 & 15.70 & 15.61 & 15.90 \\
J0811$+$4212 & 921804126089222784 & 122.95537 & 42.20216 & 10.89 $\pm{0.11}$ & $-$39.6 & $-$72.9 & 17.74 & 17.70 & 17.78 \\
J0831$-$2231 & 5702793425999272576 & 127.89758 & $-$22.52503 & 12.22 $\pm{0.08}$ & $-$133.2 & 218.2 & 17.42 & 17.39 & 17.52 \\
J0835$+$1042 & 601566038739612160 & 128.826931 & 10.71620 & 10.12 $\pm{0.10}$ & $-$29.8 & $-$39.4 & 17.39 & 17.27 & 17.65 \\
J0842$-$0222 & 3072348715677121280 & 130.56226 & $-$2.37407 & 16.32 $\pm{0.06}$ & $-$72.7 & 8.5 & 16.00 & 15.89 & 16.21 \\
J0851$+$1201 & 604972428842238080 & 132.77523 & 12.03266 & 12.82 $\pm{0.07}$ & $-$95.3 & $-$26.3 & 17.03 & 17.07 & 17.03 \\
J0856$+$6206 & 1042926292644833024 & 134.08056 & 62.10905 & 13.46 $\pm{0.06}$ & $-$8.2 & 17.0 & 16.95 & 16.97 & 16.95 \\
J0912$-$2642 & 5649808720867457664 & 138.11673 & $-$26.70103 & 27.48 $\pm{0.04}$ & $-$33.9 & $-$139.9 & 16.41 & 16.40 & 16.46 \\
J0940$-$1034 & 5740410334419524864 & 145.04940 & $-$10.57385 & 16.97 $\pm{0.04}$ & $-$32.4 & $-$65.5 & 15.97 & 15.92 & 16.11 \\
J0949$-$0730 & 3819743428284589696 & 147.41109 & $-$7.501786 & 12.42 $\pm{0.08}$ & $-$88.1 & 14.0 & 17.32 & 17.33 & 17.39 \\
J0950$-$2841 & 5464929134894103808 & 147.74005 & $-$28.68750 & 12.76 $\pm{0.09}$ & 25.9 & 23.3 & 17.34 & 17.31 & 17.41 \\
J0951$-$1517 & 5686960183679450624 & 147.96133 & $-$15.28683 & 9.67 $\pm{0.18}$ & $-$113.7 & $-$22.8 & 18.69 & 18.64 & 18.84 \\
J0959$-$1828 & 5671878015177884032 & 149.88858 & $-$18.47373 & 16.83 $\pm{0.14}$ & $-$65.6 & $-$81.5 & 18.13 & 18.15 & 18.14 \\
J1010$-$2427 & 5473389537569200896 & 152.69909 & $-$24.45327 & 25.18 $\pm{0.03}$ & $-$113.0 & $-$53.9 & 15.19 & 15.09 & 15.38 \\
J1014$-$0417 & 3780111080689647488 & 153.58589 & $-$4.28924 & 14.20 $\pm{0.06}$ & $-$39.7 & $-$12.5 & 16.35 & 16.25 & 16.57 \\
J1034$+$0327 & 3856950175919062144 & 158.62563 & 3.46002 & 12.80 $\pm{0.07}$ & $-$6.5 & $-$18.1 & 17.05 & 17.03 & 17.14 \\
J1046$-$0518 & 3776918275016618112 & 161.54037 & $-$5.30430 & 12.10 $\pm{0.08}$ & $-$97.7 & 17.5 & 16.81 & 16.74 & 16.98 \\
J1052$+$1610 & 3981634249048141440 & 163.15010 & 16.17275 & 13.39 $\pm{0.12}$ & $-$-10.5 & $-$26.8 & 17.27 & 17.29 & 17.26 \\
J1054$+$5523 & 850037352876685568 & 163.70665 & 55.38547 & 10.49 $\pm{0.06}$ & $-$90.8 & $-$13.8 & 17.26 & 17.20 & 17.42 \\
J1101$-$1314 & 3564376149017654272 & 165.30100 & $-$13.24509 & 21.41 $\pm{0.04}$ & 0.8 & $-$6.9 & 14.92 & 14.82 & 15.16 \\
J1101$+$1741 & 3983530391209588480 & 165.40356 & 17.69886 & 10.04 $\pm{0.08}$ & 9.8 & $-$3.6 & 16.82 & 16.78 & 16.99 \\
J1104$-$2826 & 5453539813634370432 & 166.18348 & $-$28.43749 & 13.48 $\pm{0.05}$ & $-$39.2 & $-$11.7 & 16.46 & 16.41 & 16.61 \\
J1105$+$5225 & 842469143761181824 & 166.27427 & 52.42282 & 11.06 $\pm{0.08}$ & 43.8 & $-$4.0 & 17.38 & 17.33 & 17.48 \\
J1106$+$1802 & 3983606596814071680 & 166.51878 & 18.04182 & 12.33 $\pm{0.09}$ & $-$27.8 & $-$196.2 & 17.51 & 17.51 & 17.55 \\
J1107$+$0405 & 3815200997858084480 & 166.76553 & 4.09445 & 15.92 $\pm{0.09}$ & $-$59.2 & $-$223.0 & 16.98 & 16.98 & 17.03 \\
J1107$-$1607 & 3559496413333223936 & 166.96996 & $-$16.11815 & 11.65 $\pm{0.10}$ & $-$120.0 & 2.0 & 17.61 & 17.58 & 17.69 \\
J1107$+$8122 & 1133601302897245568 & 166.99677 & 81.37026 & 9.94 $\pm{0.11}$ & $-$2.5 & $-$41.8 & 18.25 & 18.20 & 18.42 \\
J1135$-$2430 & 3533885454629567872 & 173.75880 & $-$24.50540 & 13.14 $\pm{0.11}$ & $-$107.3 & 39.0 & 17.49 & 17.45 & 17.64 \\
J1137$+$2947 & 4019458647338779648 & 174.27051 & 29.79946 & 63.73 $\pm{0.03}$ & $-$147.7 & $-$12.5 & 12.49 & 12.40& 12.69 \\
J1140$+$2322 & 3980865789203927680 & 175.05330 & 23.36794 & 13.76 $\pm{0.28}$ & $-$30.8 & $-$54.7 & 18.80 & 18.84 & 18.82 \\
J1154$+$3650 & 4031828157446980608 & 178.74492 & 36.84853 & 10.55 $\pm{0.07}$ & $-$40.0 & 9.9 & 17.20 & 17.09 & 17.43 \\
J1203$+$6450 & 1585063422960992256 & 180.88227 & 64.84976 & 11.47 $\pm{0.08}$ & $-$61.9 & $-$140.1 & 17.62 & 17.59 & 17.69 \\
J1213$+$1140 & 3908299636678454784 & 183.31018 & 11.68057 & 11.54 $\pm{0.07}$ & $-$30.6 & $-$5.1 & 17.16 & 17.14 & 17.28 \\
J1214$-$1724 & 3520974164461518592 & 183.53432 & $-$17.41188 & 11.01 $\pm{0.08}$ & $-$28.2 & $-$117.2 & 16.70 & 16.58 & 16.93 \\
J1217$+$0828 & 3902183809407583872 & 184.39664 & 8.46944 & 12.34 $\pm{0.06}$ & $-$67.6 & $-$1.5 & 16.67 & 16.61 & 16.81 \\
J1243$+$4805 & 1543370904111505408 & 190.92339 & 48.09304 & 12.51 $\pm{0.05}$ & $-$84.8 & 31.4 & 16.97 & 16.96 & 17.02 \\
J1254$-$0452 & 3678497445865946624 & 193.62031 & $-$4.87429 & 11.10 $\pm{0.26}$ & 9.6 & 2.0 & 18.56 & 18.54 & 18.66 \\
J1256$-$0619 & 3677303170080009856 & 194.04533 & $-$6.32515 & 15.59 $\pm{0.13}$ & $-$5.4 & $-$62.7 & 17.43 & 17.42 & 17.51 \\
J1329$+$2549 & 1448232907440917760 & 202.35849 & 25.82683 & 11.58 $\pm{0.09}$ & $-$16.9 & 16.5 & 17.65 & 17.53 & 17.92 \\
J1333$+$6406 & 1665858350572796672 & 203.41871 & 64.10746 & 9.92 $\pm{0.10}$ & 60.1 & $-$29.3 & 17.95 & 17.98 & 18.02 \\
J1339$-$0713 & 3630648387747801088 & 204.91582 & $-$7.22127 & 17.75 $\pm{0.05}$ & $-$54.4 & $-$29.8 & 16.32 & 16.32 & 16.34 \\
J1342$-$1413 & 3606080968656361344 & 205.67123 & $-$14.22819 & 20.54 $\pm{0.04}$ & $-$103.0 & $-$70.8 & 15.98 & 16.00 & 15.96 \\
J1413$+$0755 & 3674476639217656576 & 213.27725 & 7.92306 & 19.28 $\pm{0.04}$ & 14.1 & 14.2 & $-$0.8 & 15.63 & 15.77 \\
J1440$-$1951 & 6281684373876096512 & 220.16041 & $-$19.86583 & 14.28 $\pm{0.07}$ & $-$84.8 & $-$46.9 & 16.95 & 16.91 & 17.03 \\
J1451$+$5110 & 1593473346884952448 & 222.82799 & 51.17986 & 10.96 $\pm{0.04}$ & $-$5.8 & 15.3 & 16.73 & 16.68 & 16.86 \\
J1451$-$2502 & 6229330032504603136 & 222.96896 & $-$25.04411 & 12.63 $\pm{0.15}$ & $-$67.1 & $-$35.0 & 17.39 & 17.43 & 17.37 \\
J1459$-$0411 & 6338900661178928896 & 224.76100 & $-$4.19951 & 16.50 $\pm{0.07}$ & $-$73.4 & $-$30.7 & 16.68 & 16.60 & 16.82 \\
J1501$+$3431 & 1289986903300047744 & 225.42112 & 34.53314 & 13.10 $\pm{0.03}$ & $-$72.3 & 36.0 & 15.84 & 15.76 & 16.04 \\
J1514$+$2313 & 1263434109805753984 & 228.58014 & 23.22941 & 16.23 $\pm{0.08}$ & $-$81.5 & $-$18.5 & 17.23 & 17.19 & 17.36 \\
J1519$+$6329 & 1643814211883928320 & 229.95518 & 63.49845 & 18.02 $\pm{0.04}$ & $-$170.9 & 160.5 & 16.63 & 16.55 & 16.82 \\
J1529$+$2928 & 1273456463234876288 & 232.39546 & 29.46718 & 11.47 $\pm{0.07}$ & $-$74.0 & $-$5.4 & 17.48 & 17.54 & 17.45 \\
J1537$+$8419 & 1724098901677145984 & 234.48658 & 84.32789 & 11.56 $\pm{0.05}$ & $-$66.1 & 78.2 & 16.98 & 16.85 & 17.20 \\
J1543$+$3021 & 1273088783971336576 & 235.81261 & 30.35960 & 14.65 $\pm{0.05}$ & $-$70.4 & 75.8 & 16.87 & 16.83 & 16.95 \\
J1548$+$2451 & 1219699145026398848 & 237.22955 & 24.85361 & 13.60 $\pm{0.06}$ & 49.3 & 3.6 & 16.80 & 16.71 & 16.98 \\
J1552$+$0039 & 4410623858974488832 & 238.15940 & 0.65268 & 9.72 $\pm{0.25}$ & 41.4 & $-$44.3 & 18.49 & 18.48 & 18.55 \\
J1621$+$0432 & 4436905352274528896 & 245.49049 & 4.53860 & 16.23 $\pm{0.07}$ & $-$66.4 & 11.7 & 16.85 & 16.77 & 17.01 \\
J1622$+$3004 & 1318204460477280512 & 245.65034 & 30.08172 & 13.38 $\pm{0.06}$ & $-$134.45 & $-$66.5 & 17.06 & 16.97 & 17.23 \\
J1626$+$2533 & 1304081783374935680 & 246.74813 & 25.55767 & 11.88 $\pm{0.07}$ & $-$28.0 & 52.6 & 17.59 & 17.60 & 17.66 \\
J1630$+$2724 & 1304733106575117056 & 247.65339 & 27.41430 & 10.52 $\pm{0.09}$ & 16.6 & $-$2.9 & 17.95 & 18.02 & 17.92 \\
J1655$+$2533 & 1306197930941817984 & 253.91237 & 25.56207 & 27.40 $\pm{0.05}$ & 49.0 & $-$196.1 & 16.96 & 17.00 & 16.93 \\
J1656$+$5719 & 1433629812475800320 & 254.16291 & 57.31760 & 12.69 $\pm{0.07}$ & $-$10.5 & 10.2 & 17.40 & 17.43 & 17.41 \\
J1659$+$6610 & 1635687790163070976 & 254.81293 & 66.17646 & 15.48 $\pm{0.05}$ & $-$104.1 & 162.9 & 17.24 & 17.27 & 17.24 \\
J1659$+$4401 & 1358301480583401728 & 254.95157 & 44.01822 & 31.59 $\pm{0.02}$ & $-$31.7 & 98.7 & 14.78 & 14.65 & 15.06 \\
J1706$-$0837 & 4336571785203401472 & 256.57684 & $-$8.63127 & 14.67 $\pm{0.10}$ & $-$260.0 & $-$333.8 & 17.37 & 17.37 & 17.38 \\
J1707$+$3532 & 1338455643596995072 & 256.96615 & 35.54460 & 13.29 $\pm{0.04}$ & $-$46.44 & 116.1 & 16.49 & 16.39 & 16.70 \\
J1709$-$2632 & 4108890281768798336 & 257.42013 & $-$26.54623 & 12.04 $\pm{0.11}$ & 12.4 & $-$9.7 & 17.82 & 17.74 & 17.92 \\
J1710$-$2005 & 4128167950420485632 & 257.64415 & $-$20.09529 & 11.57 $\pm{0.13}$ & $-$108.4 & $-$69.0 & 17.48 & 17.42 & 17.61 \\
J1719$-$1446 & 4137811178473465216 & 259.83326 & $-$14.77799 & 13.05 $\pm{0.08}$ & $-$42.4 & $-$105.0 & 17.16 & 17.13 & 17.24 \\
J1722$+$3958 & 1346883876962000000 & 260.66923 & 39.96965 & 10.30 $\pm{0.07}$ & $-$11.7 & $-$71.7 & 17.81 & 17.82 & 17.78 \\
J1723$+$0836 & 4490483248223416576 & 260.92344 & 8.61467 & 11.41 $\pm{0.08}$ & 24.5 & $-$57.4 & 17.39 & 17.38 & 17.45 \\
J1728$+$5558 & 1422012892308493568 & 262.23341 & 55.97397 & 21.24 $\pm{0.03}$ & $-$105.4 & 229.0 & 16.14 & 16.07 & 16.26 \\
J1744$-$2035 & 4118923497232723072 & 266.17306 & $-$20.59731 & 10.05 $\pm{0.12}$ & $-$17.9 & $-$81.2 & 17.65 & 17.57 & 17.89 \\
J1758$+$5906 & 1422782516088307840 & 269.58789 & 59.11256 & 10.46 $\pm{0.06}$ & $-$10.9 & 18.7 & 17.36 & 17.27 & 17.57 \\
J1800$+$4517 & 2115952197141317888 & 270.00492 & 45.29049 & 11.85 $\pm{0.09}$ & $-$23.3 & 68.5 & 18.18 & 18.13 & 18.32 \\
J1812$+$4321 & 2114811453822316160 & 273.09480 & 43.35229 & 17.21 $\pm{0.03}$ & 5.1 & 62.4 & 16.24 & 16.25 & 16.24 \\
J1813$+$4427 & 2114985726416563072 & 273.25474 & 44.45528 & 12.36 $\pm{0.06}$ & 48.5 & $-$58.6 & 17.72 & 17.75 & 17.68 \\
J1819$-$1208 & 4153618204302689920 & 274.80568 & $-$12.14896 & 19.41 $\pm{0.06}$ & 4.9 & 10.4 & 15.74 & 15.57 & 16.02 \\
J1819$+$1225 & 4484736543328721792 & 274.85822 & 12.43255 & 9.77 $\pm{0.13}$ & 17.9 & $-$57.7 & 18.13 & 18.13 & 18.11 \\
J1822$+$5323 & 2148495031195001728 & 275.61525 & 53.39238 & 13.28 $\pm{0.04}$ & 25.5 & 36.9 & 16.46 & 16.35 & 16.71 \\
J1832$+$0856 & 4479342339285057408 & 278.01181 & 8.94338 & 13.23 $\pm{0.09}$ & 7.8 & $-$3.8 & 17.01 & 16.87 & 17.27 \\
J1839$-$0448 & 4256698798129293568 & 279.88082 & $-$4.80528 & 12.45 $\pm{0.07}$ & 12.7 & $-$9.9 & 16.28 & 16.20 & 16.43 \\
J1849$+$6458 & 2253826832091026560 & 282.35943 & 64.96981 & 11.59 $\pm{0.06}$ & 22.4 & 22.1 & 17.51 & 17.47 & 17.58 \\
J1857$+$3147 & 2042089793425728640 & 284.48562 & 31.79556 & 11.20 $\pm{0.04}$ & $-$0.6 & 7.4 & 16.53 & 16.45 & 16.71 \\
J1900$+$7039 & 2262849634963004416 & 285.04387 & 70.66652 & 77.66 $\pm{0.01}$ & 85.8 & 505.1 & 13.24 & 13.23 & 13.25 \\
J1901$+$1458 & 4506869128279648512 & 285.38685 & 14.96898 & 24.15 $\pm{0.04}$ & 95.4 & 72.6 & 15.68 & 15.52 & 15.94 \\
J1902$+$7728 & 2292799369631237760 & 285.54989 & 77.46983 & 10.67 $\pm{0.09}$ & $-$22.9 & 3.0 & 18.02 & 17.96 & 18.12 \\
J1903$+$4657 & 2130610306341954688 & 285.98337 & 46.96014 & 9.99 $\pm{0.04}$ & 9.2 & 62.9 & 16.80 & 16.72 & 16.99 \\
J1910$+$7334 & 2265100885021724032 & 287.68078 & 73.57752 & 10.62 $\pm{0.06}$ & 29.9 & 79.3 & 17.64 & 17.64 & 17.68 \\
J1923$-$2328 & 6770033150551933824 & 290.94843 & $-$23.47892 & 14.42 $\pm{0.08}$ & 36.5 & $-$24.2 & 16.40 & 16.25 & 16.53 \\
J1924$-$2717 & 6765861019327924736 & 291.03084 & $-$27.29837 & 14.20 $\pm{0.11}$ & 101.6 & $-$231.5 & 17.05 & 17.02 & 17.23 \\
J1924$-$2913 & 6758742819699643392 & 291.12252 & $-$29.22637 & 11.60 $\pm{0.08}$ & 27.2 & 3.4 & 17.21 & 17.20 & 17.20 \\
J1925$-$0346 & 4213471120498390784 & 291.47959 & $-$3.77422 & 17.66 $\pm{0.06}$ & $-$93.2 & $-$39.8 & 16.56 & 16.48 & 16.67 \\
J1928$+$1526 & 4321498378443922816 & 292.06065 & 15.44412 & 10.26 $\pm{0.10}$ & $-$8.2 & 20.2 & 17.73 & 17.75 & 17.73 \\
J1928$+$5429 & 2141466403116486912 & 292.24538 & 54.49718 & 14.69 $\pm{0.03}$ & 21.6 & $-$12.6 & 16.46 & 16.46 & 16.50 \\
J1929$-$2926 & 6764486080026955136 & 292.30264 & $-$29.44520 & 11.28 $\pm{0.16}$ & 25.2 & $-$18.2 & 17.69 & 17.71 & 17.71 \\
J2011$+$4910 & 2087569060381096960 & 302.81466 & 49.17752 & 13.52 $\pm{0.06}$ & 144.2 & 81.42 & 17.34 & 17.32 & 17.43 \\
J2012$+$3113 & 2053953008490747392 & 303.09295 & 31.23065 & 32.50 $\pm{0.02}$ & 34.0 & 92.2 & 14.77 & 14.69 & 14.92 \\
J2026$+$1848 & 1815431209434186880 & 306.59529 & 18.81293 & 11.20 $\pm{0.07}$ & 25.1 & $-$2.5 & 17.37 & 17.35 & 17.48 \\
J2026$-$2254 & 6849850998873128704 & 306.69792 & $-$22.91429 & 14.20 $\pm{0.08}$ & $-$14.8 & $-$17.2 & 16.79 & 16.80 & 16.79 \\
J2035$-$1835 & 6861525956933302912 & 308.85149 & $-$18.58783 & 14.78 $\pm{0.10}$ & 65.5 & $-$30.6 & 17.30 & 17.33 & 17.29 \\
J2037$-$2857 & 6798304038337345280 & 309.46254 & $-$28.96513 & 15.60 $\pm{0.05}$ & 82.8 & $-$38.0 & 16.37 & 16.34 & 16.47 \\
J2054$-$2039 & 6857295585945072128 & 313.67865 & $-$20.65715 & 31.97 $\pm{0.03}$ & 68.9 & 10.2 & 15.11 & 15.03 & 15.29 \\
J2100$+$5142 & 2170187399180847872 & 315.09715 & 51.71458 & 10.31 $\pm{0.07}$ & $-$3.4 & 11.3 & 17.74 & 17.76 & 17.76 \\
J2105$+$5903 & 2190645256129430144 & 316.44914 & 59.05332 & 10.57 $\pm{0.07}$ & 44.4 & 38.6 & 17.81 & 17.74 & 18.00 \\
J2107$+$7831 & 2284856630775525376 & 316.96357 & 78.53158 & 10.15 $\pm{0.08}$ & $-$22.9 & $-$35.7 & 17.83 & 17.82 & 17.92 \\
J2111$+$1102 & 1745011677261492608 & 317.85762 & 11.03875 & 10.48 $\pm{0.08}$ & $-$12.8 & $-$13.0 & 17.52 & 17.50 & 17.65 \\
J2133$+$3529 & 1950617847697475584 & 323.38906 & 35.49040 & 10.61 $\pm{0.04}$ & 40.8 & $-$12.5 & 16.35 & 16.24 & 16.58 \\
J2148$-$1629 & 6837525469227801088 & 327.17892 & $-$16.48758 & 10.30 $\pm{0.10}$ & $-$12.4 & 22.5 & 17.42 & 17.40 & 17.46 \\
J2153$-$2628 & 6811977801160882944 & 328.38141 & $-$26.48211 & 12.93 $\pm{0.08}$ & $-$44.0 & $-$6.0 & 17.36 & 17.31 & 17.51 \\
J2204$+$2543 & 1891820737544168576 & 331.20404 & 25.71869 & 9.98 $\pm{0.06}$ & 60.0 & 10.9 & 16.67 & 16.54 & 16.89 \\
J2208$+$2059 & 1781605382738862592 & 332.13117 & 20.98618 & 11.09 $\pm{0.09}$ & 80.4 & 50.2 & 17.48 & 17.55 & 17.46 \\
J2221$+$4406 & 1959002792086235648 & 335.33318 & 44.10008 & 11.60 $\pm{0.06}$ & 67.3 & 27.4 & 17.03 & 17.03 & 17.10 \\
J2231$-$1546 & 2595983649380303360 & 337.99254 & $-$15.78257 & 11.39 $\pm{0.18}$ & $-$11.9 & $-$129.1 & 18.35 & 18.32 & 18.48 \\
J2250$+$3231 & 1890542971890672896 & 342.60121 & 32.52864 & 11.68 $\pm{0.04}$ & $-$44.9 & $-$46.3 & 15.60 & 15.47 & 15.88 \\
J2255$+$0710 & 2712093451662656256 & 343.80630 & 7.16687 & 10.81 $\pm{0.36}$ & 40.7 & $-$19.3 & 19.16 & 19.14 & 19.12 \\
J2257$+$0755 & 2712240064671438720 & 344.35807 & 7.92775 & 21.96 $\pm{0.10}$ & $-$165.6 & $-$167.8 & 17.15 & 17.16 & 17.17 \\
J2304$-$2658 & 2382415648967442432 & 346.20948 & $-$26.96853 & 12.47 $\pm{0.07}$ & 39.6 & 36.2 & 17.15 & 17.14 & 17.24 \\
J2306$-$2905 & 6606362529696816896 & 346.74467 & $-$29.08750 & 10.46 $\pm{0.12}$ & $-$7.2 & $-$9.8 & 17.86 & 17.82 & 17.99 \\
J2308$+$0347 & 2662208372887759744 & 347.18455 & 3.78839 & 15.34 $\pm{0.08}$ & $-$20.7 & $-$90.2 & 17.12 & 17.13 & 17.12 \\
J2309$+$0940 & 2714218433977373440 & 347.46133 & 9.66878 & 10.03 $\pm{0.09}$ & $-$21.9 & $-$30.8 & 17.02 & 16.98 & 17.15 \\
J2314$-$2208 & 2385217170235210496 & 348.74691 & $-$22.13932 & 10.66 $\pm{0.06}$ & 51.2 & 7.5 & 16.30 & 16.18 & 16.55 \\
J2340$-$1819 & 2393834386459511680 & 355.18326 & $-$18.32984 & 10.58 $\pm{0.15}$ & $-$0.6 & $-$114.2 & 17.60 & 17.54 & 17.75 \\
J2349$+$0907 & 2758938385082121216 & 357.34116 & 9.12072 & 11.52 $\pm{0.11}$ & 33.7 & $-$6.4 & 17.61 & 17.58 & 17.73 \\
\hline 
\label{tab:gaia}
\end{longtable}

\begin{longtable}{cccccccc}
\caption{The physical parameters for our massive white dwarf sample.}\\
\hline
\hline 
Object name & Composition & Spectral Type & $T_{eff}$ & Mass & Cooling Age &  Merger\\
& & & (K) & ($M_\odot$) & (Gyr) & Evidence\\
\hline 
J0006$+$3104 & H                 & DC     & 25442 $\pm$522 & 1.138 $\pm$0.010 & 0.20 $\pm$0.01 &  MR\\
J0012$-$0606 & H                 & DA     & 13730 $\pm$119 & 0.902 $\pm$0.006 & 0.55 $\pm$0.01  & \\
J0029$+$3648 & H                 & DA     & 25858 $\pm$313 & 1.284 $\pm$0.004 & 0.42 $\pm$0.02  & \\
J0039$-$0357 & H                 & DA     & 11871 $\pm$214 & 1.271 $\pm$0.009 & 2.09 $\pm$0.06  & \\
J0043$-$1000 & log(H/He)=$-$2.00 &  DBAH    & 18381 $\pm$371 & 1.077 $\pm$0.014 & 0.43 $\pm$0.03 & MR\\
J0045$-$2336 & log(H/He)=$-$4.00 & DQ     & 11540 $\pm$43 & 1.126 $\pm$0.004 & 1.80 $\pm$0.02 &  AV\\
J0049$-$2525 & H                 & DA     & 13018 $\pm$460 & 1.312 $\pm$0.010 & 1.72 $\pm$0.10  & \\
J0050$+$3138 & log(H/He)=$-$3.00 & He-DA & 12519 $\pm$221 & 1.215 $\pm$0.009 & 1.67 $\pm$0.05 &  \\
J0050$-$0326 & H                 & DC     & 23916 $\pm$355 & 1.213 $\pm$0.006 & 0.33 $\pm$0.02 &  MR\\
J0050$-$2826 & H                 & DA     & 11320 $\pm$155 & 1.061 $\pm$0.011 & 1.72 $\pm$0.08   & \\
J0104$+$4650 & log(H/He)=$-$1.64 & DQA    & 12321 $\pm$111 & 1.140 $\pm$0.008 & 1.61 $\pm$0.03 &  AV\\
J0107$+$2518 & H                 & DA     & 16866 $\pm$273 & 0.943 $\pm$0.011 & 0.35 $\pm$0.02  & \\
J0107$+$2904 & H                 & DA     & 17860 $\pm$186 & 1.228 $\pm$0.004 & 0.90 $\pm$0.03  & \\
J0118$-$0156 & H                 & DAH    & 79556 $\pm$9435 & 1.335 $\pm$0.010 & \nodata &  MR\\
J0127$-$2436 & H                 & DA     & 11236 $\pm$214 & 1.284 $\pm$0.009 & 2.22 $\pm$0.06  & \\
J0135$+$2229 & H                 & DA     & 14299 $\pm$208 & 0.904 $\pm$0.011 & 0.50 $\pm$0.02  & \\
J0135$+$5722 & H                 & DA     & 12415 $\pm$87 & 1.153 $\pm$0.004 & 1.75 $\pm$0.03  & \\
J0138$+$2523 & H                 & DA     & 38615 $\pm$2617 & 1.219 $\pm$0.015 & 0.08 $\pm$0.02  & \\
J0138$+$5124 & H                 & DA     & 14123 $\pm$191 & 0.958 $\pm$0.008 & 0.60 $\pm$0.02 & \\
J0150$+$2835 & H                 & DAH    & 12046 $\pm$101 & 0.979 $\pm$0.008 & 1.00 $\pm$0.03 &  M\\
J0151$+$2435 & H                 & DAH    & 19868 $\pm$364 & 1.174 $\pm$0.008 & 0.48 $\pm$0.03 &  M\\
J0154$+$4700 & H                 & DA     & 11838 $\pm$235 & 1.077 $\pm$0.014 & 1.59 $\pm$0.10  & \\
J0158$-$2503 & H                 & DA     & 12234 $\pm$94 & 1.122 $\pm$0.007 & 1.68 $\pm$0.04  & \\
J0204$+$8713 & H                 & DA     & 11135 $\pm$207 & 1.053 $\pm$0.015 & 1.75 $\pm$0.11  & \\
J0205$+$2057 & log(C/H)=+0.97    & DAQ    & 16427 $\pm$228 & 1.194 $\pm$0.008 & 0.96 $\pm$0.04 &  AV\\
J0211$+$2115 & H                 & DAH    & 12244 $\pm$135 & 1.075 $\pm$0.007 & 1.44 $\pm$0.05 &  M\\
J0216$+$3541 & H & DBAH & 32464 $\pm$1481 & 1.051 $\pm$0.024 & 0.06 $\pm$0.01 &  M\\
J0230$+$3842 & H                 & DAH    & 20695 $\pm$417 & 1.145 $\pm$0.009 & 0.37 $\pm$0.02  & M\\
J0234$-$0511 & H                 & DA     & 12634 $\pm$77 & 0.958 $\pm$0.004 & 0.81 $\pm$0.01 &  \\
J0248$+$1600 & H                 & DA     & 19369 $\pm$331 & 1.082 $\pm$0.009 & 0.36 $\pm$0.02 &  \\
J0249$-$1831 & H                 & DAH    & 18144 $\pm$243 & 1.166 $\pm$0.006 & 0.60 $\pm$0.03 &  M\\
J0256$-$1515 & H                 & DAH    & 25312 $\pm$294 & 1.236 $\pm$0.005 & 0.31 $\pm$0.01 &  MR\\
J0257$+$0308 & H                 & DAH    & 12149 $\pm$139 & 1.083 $\pm$0.011 & 1.51 $\pm$0.07  & M\\
J0307$+$0313 & H                 & DA     & 13799 $\pm$212 & 1.238 $\pm$0.004 & 1.63 $\pm$0.05  & \\
J0311$-$2254 & H                 & DA     & 22184 $\pm$392 & 0.940 $\pm$0.010 & 0.15 $\pm$0.01  & \\
J0317$-$2916 & log(H/He)=$-$3.00 & He-DA & 11165 $\pm$346 & 1.156 $\pm$0.020 & 1.95 $\pm$0.11 &  V\\
J0319$+$4628 & H                 & DAH    & 14403 $\pm$210 & 1.010 $\pm$0.010 & 0.65 $\pm$0.03 & M\\
J0323$+$3457 & H                 & DA     & 13875 $\pm$198 & 1.017 $\pm$0.009 & 0.74 $\pm$0.04 & \\
J0323$+$3501 & H                 & DA     & 13621 $\pm$246 & 0.971 $\pm$0.011 & 0.68 $\pm$0.04  & \\
J0325$-$0815 & H                 & DA     & 13942 $\pm$289 & 1.159 $\pm$0.008 & 1.35 $\pm$0.08  & \\
J0326$+$1331 & H                 & DAH    & 11351 $\pm$101 & 0.936 $\pm$0.01 & 1.02 $\pm$0.04  & MV\\
J0327$+$2227 & H                 & DC     & 15128 $\pm$317 & 1.277 $\pm$0.007 & 1.44 $\pm$0.05  & MR\\
J0332$+$3005 & H                 & DA     & 15200 $\pm$382 & 1.283 $\pm$0.006 & 1.43 $\pm$0.06  & \\
J0347$-$1802 & H                 & DA     & 13082 $\pm$150 & 1.143 $\pm$0.007 & 1.51 $\pm$0.05 &  V\\
J0348$-$0058 & H                 & DA     & 37535 $\pm$1428 & 1.246 $\pm$0.008 & 0.10 $\pm$0.01  & \\
J0401$+$2140 & H                 & DA     & 18009 $\pm$416 & 1.217 $\pm$0.007 & 0.84 $\pm$0.06  & V\\
J0408$+$2323 & H                 & DA     & 12053 $\pm$110 & 1.024 $\pm$0.008 & 1.22 $\pm$0.06  & \\
J0422$-$2407 & H                 & DA     & 19042 $\pm$90 & 1.222 $\pm$0.005 & 0.72 $\pm$0.04  & \\
J0439$+$4543 & H                 & DA     & 19119 $\pm$625 & 1.307 $\pm$0.007 & 0.96 $\pm$0.06  & \\
J0447$+$4224 & H                 & DA     & 16702 $\pm$341 & 1.211 $\pm$0.008 & 1.00 $\pm$0.06 &  V\\
J0448$-$1053 & H                 & DA     & 11993 $\pm$108 & 0.941 $\pm$0.007 & 0.89 $\pm$0.02  & \\
J0455$-$0058 & H                 & DA     & 20846 $\pm$405 & 1.258 $\pm$0.006 & 0.68 $\pm$0.04 & V\\
J0507$+$2645 & H                 & DAH    & 21668 $\pm$350 & 1.232 $\pm$0.005 & 0.50 $\pm$0.03 & M\\
J0521$-$1029 & H                 & DA     & 29561 $\pm$862 & 1.045 $\pm$0.017 & 0.08 $\pm$0.01  & \\
J0529$+$5239 & H                 & DA     & 20087 $\pm$217 & 1.233 $\pm$0.003 & 0.65 $\pm$0.03  & V\\
J0533$-$2713 & H                 & DA     & 20951 $\pm$359 & 1.196 $\pm$0.005 & 0.45 $\pm$0.03  & \\
J0534$-$0214 & H                 & DA     & 28064 $\pm$560 & 0.989 $\pm$0.013 & 0.08 $\pm$0.01  & \\
J0538$+$3212 & H                 & DA     & 12454 $\pm$154 & 0.994 $\pm$0.010 & 0.95 $\pm$0.05  & \\
J0547$+$1501 & H                 & DAH    & 15460 $\pm$199 & 0.942 $\pm$0.008 & 0.44 $\pm$0.02  & M\\
J0547$-$1250 & H                 & DAH    & 12394 $\pm$139 & 1.032 $\pm$0.007 & 1.15 $\pm$0.06  & M\\
J0550$-$1630 & H                 & DA     & 18599 $\pm$450 & 1.081 $\pm$0.012 & 0.40 $\pm$0.03  & \\
J0551$+$4135 & log(C/H)=$-$0.48  & DAQ    & 12997 $\pm$115 & 1.139 $\pm$0.005 & 1.44 $\pm$0.03 &  A\\
J0601$+$3726 & H                 & DAH    & 12713 $\pm$82 & 0.921 $\pm$0.006 & 0.72 $\pm$0.02  & M\\
J0602$+$4652 & log(H/He)=$-$5.00 & DQH?    & 13211 $\pm$192 & 1.209 $\pm$0.008 & 1.52 $\pm$0.04  & M\\
J0607$+$3415 & H                 & DAH    & 21139 $\pm$340 & 1.175 $\pm$0.006 & 0.39 $\pm$0.02  & M\\
J0608$-$0059 & H                 & DA     & 17329 $\pm$180 & 1.278 $\pm$0.003 & 1.13 $\pm$0.02  & \\
J0625$+$1902 & H                 & DAH   & 18120 $\pm$397 & 1.234 $\pm$0.009 & 0.89 $\pm$0.06   & M\\
J0634$+$3848 & H                 & DA     & 12210 $\pm$106 & 0.926 $\pm$0.006 & 0.81 $\pm$0.02 &  \\
J0655$+$2939 & log(C/H)=$-$0.40     & DAQ     & 17020 $\pm$412 & 1.189 $\pm$0.009 & 0.87 $\pm$0.06  & AV\\
J0657$+$7341 & H                 & DA     & 12609 $\pm$119 & 1.134 $\pm$0.007 & 1.61 $\pm$0.05  & \\
J0705$-$2046 & H                 & DAH    & 27829 $\pm$515 & 1.139 $\pm$0.010 & 0.15 $\pm$0.01  & MV\\
J0707$+$5611 & H                 & DC     & 18101 $\pm$355 & 1.291 $\pm$0.005 & 1.06 $\pm$0.04  & MR\\
J0712$-$1815 & H                 & DA     & 11742 $\pm$136 & 0.980 $\pm$0.008 & 1.09 $\pm$0.05  & \\
J0718$+$3731 & H                 & DC     & 33942 $\pm$1411 & 1.317 $\pm$0.007 & 0.20 $\pm$0.03 &  MR\\
J0725$+$0411 & H                 & DA     & 12022 $\pm$88 & 0.940 $\pm$0.008 & 0.88 $\pm$0.02 &  \\
J0733$+$5750 & H                 & DA     & 16212 $\pm$160 & 1.009 $\pm$0.006 & 0.47 $\pm$0.01 &  \\
J0734$+$4841 & H                 & DA     & 14094 $\pm$88 & 0.987 $\pm$0.003 & 0.65 $\pm$0.01 &  \\
J0737$-$0610 & log(H/He)=$-$2.07 & DQA    & 12051 $\pm$66 & 1.105 $\pm$0.006 & 1.59 $\pm$0.02  &  A\\
J0747$+$4527 & H                 & DA     & 13625 $\pm$324 & 1.100 $\pm$0.010 & 1.15 $\pm$0.10  & \\
J0801$+$7749 & H                 & DA     & 16886 $\pm$306 & 1.209 $\pm$0.006 & 0.96 $\pm$0.05  & V\\
J0803$+$1229 & H                 & DAH    & 19168 $\pm$332 & 1.162 $\pm$0.008 & 0.50 $\pm$0.03  & M\\
J0811$+$4212 & H                 & DA     & 14595 $\pm$280 & 1.160 $\pm$0.008 & 1.20 $\pm$0.07 &  \\
J0811$+$4323 & H                 & DA     & 21195 $\pm$356 & 0.945 $\pm$0.010 & 0.18 $\pm$0.01 & \\
J0831$-$2231 & log(C/H)=+0.04    & DAQ    & 13836 $\pm$180 & 1.134 $\pm$0.007 & 1.23 $\pm$0.04 &  AVR\\
J0835$+$1042 & H                 & DA     & 28001 $\pm$512 & 1.286 $\pm$0.006 & 0.33 $\pm$0.02  & \\
J0842$-$0222 & H                 & DAH?   & 23910 $\pm$212 & 1.162 $\pm$0.004 & 0.26 $\pm$0.01 &  M\\
J0851$+$1201 & H                 & DAH    & 11209 $\pm$83 & 0.904 $\pm$0.008 & 0.97 $\pm$0.03  & M\\
J0856$+$6206 & H                 & DA     & 11857 $\pm$79 & 0.958 $\pm$ 0.007 & 0.97 $\pm$ 0.03  & \\
J0912$-$2642 & H                 & DA     & 12973 $\pm$115 & 1.262 $\pm$0.002 & 1.84 $\pm$0.03  & \\
J0940$-$1034 & H                 & DA     & 16503 $\pm$189 & 0.996 $\pm$0.007 & 0.43 $\pm$0.02  & \\
J0949$-$0730 & H                 & DA     & 12941 $\pm$184 & 1.084 $\pm$0.008 & 1.26 $\pm$0.07  & \\
J0950$-$2841 & H                 & DA     & 13335 $\pm$203 & 1.124 $\pm$0.006 & 1.35 $\pm$0.06  & \\
J0951$-$1517 & H                 & DAH    & 13566 $\pm$406 & 1.275 $\pm$0.010 & 1.72 $\pm$0.08  & MV\\
J0959$-$1828 & H                 & DA     & 11995 $\pm$176 & 1.320 $\pm$0.004 & 1.83 $\pm$0.05  & \\
J1010$-$2427 & H                 & DC     & 21788 $\pm$176 & 1.170 $\pm$0.004 & 0.35 $\pm$0.01  & M\\
J1014$-$0417 & H                 & DAH    & 22956 $\pm$605 & 1.164 $\pm$0.010 & 0.30 $\pm$0.02  & M\\
J1034$+$0327 & H                 & DAH    & 15110 $\pm$167 & 1.089 $\pm$0.006 & 0.76 $\pm$0.03 & M\\
J1046$-$0518 & log(H/He)=$-$5.00 & DBH    & 18776 $\pm$924 & 1.127 $\pm$0.024 & 0.48 $\pm$0.09  & M\\
J1052$+$1610 & H                 & DA     & 11256 $\pm$84 & 1.020 $\pm$0.009 & 1.52 $\pm$0.06  & \\
J1054$+$5523 & H                 & DAH:   & 18842 $\pm$206 & 1.133 $\pm$0.006 & 0.46 $\pm$0.02 &  M\\
J1101$+$1741 & H                 & DA     & 18458 $\pm$240 & 0.963 $\pm$0.010 & 0.28 $\pm$0.01 &  \\
J1101$-$1314 & H                 & DA     & 22705 $\pm$356 & 1.012 $\pm$0.008& 0.18 $\pm$0.01 &  \\
J1104$-$2826 & H                 & DA     & 17094 $\pm$287 & 1.011 $\pm$0.009 & 0.41 $\pm$0.02 &  \\
J1105$+$5225 & H                 & DAH?   & 17588 $\pm$140 & 1.154 $\pm$0.005 & 0.63 $\pm$0.02 &  M\\
J1106$+$1802 & H                 & DA     & 12877 $\pm$269 & 1.131 $\pm$0.011 & 1.51 $\pm$0.09  & V\\
J1107$+$0405 & H                 & DA     & 13010 $\pm$108 & 1.140 $\pm$0.005 & 1.51 $\pm$0.04  & V\\
J1107$+$8122 & H                 & DAH    & 14881 $\pm$594 & 1.237 $\pm$0.010 & 1.40 $\pm$0.12  & M\\
J1107$-$1607 & H                 & DA     & 15292 $\pm$234 & 1.181 $\pm$0.007 & 1.13 $\pm$0.05  & \\
J1135$-$2430 & H                 & DA     & 17039 $\pm$290 & 1.245 $\pm$0.006 & 1.07 $\pm$0.04 &  \\
J1137$+$2947 & H                 & DA     & 21827 $\pm$45 & 0.964 $\pm$0.001 & 0.17 $\pm$0.001 &  \\
J1140$+$2322 & H                 & DA     & 11862 $\pm$223 & 1.336 $\pm$0.006 & 1.71 $\pm$0.06 &  \\
J1154$+$3650 & H                 & DA     & 26115 $\pm$411 & 1.251 $\pm$0.006 & 0.30 $\pm$0.02  & R\\
J1203$+$6450 & log(H/He)=$-$2.00 & DQA    & 12700 $\pm$71 & 1.107 $\pm$0.006 & 1.42 $\pm$0.02   & AV\\
J1213$+$1140 & H                 & DA     & 15317 $\pm$166 & 1.066 $\pm$0.006 & 0.66 $\pm$0.03  & \\
J1214$-$1724 & log(H/He)=$-$5.00 & DBH    & 26677 $\pm$1877 & 1.169 $\pm$0.010 & 0.21 $\pm$0.05  & MRV\\
J1217$+$0828 & H                 & DAH    & 18654 $\pm$268 & 1.063 $\pm$0.008 & 0.38 $\pm$0.02  & M\\
J1243$+$4805 & H                 & DA     & 12751 $\pm$101 & 0.966 $\pm$0.006 & 0.81 $\pm$0.02  & \\
J1254$-$0452 & H                 & DA     & 14423 $\pm$393 & 1.308 $\pm$0.008 & 1.52 $\pm$0.07  & \\
J1256$-$0619 & H                 & DA     & 14712 $\pm$218 & 1.256 $\pm$0.004 & 1.47 $\pm$0.04  & \\
J1329$+$2549 & H                 & DA     & 29007 $\pm$748 & 1.351 $\pm$0.006 & 0.37 $\pm$0.03  & \\
J1333$+$6406 & H                 & DAH    & 12564 $\pm$181 & 1.105 $\pm$0.010 & 1.49 $\pm$0.07  & M\\
J1339$-$0713 & H                 & DAH    & 11942 $\pm$94 & 0.956 $\pm$0.006 & 0.94 $\pm$0.03  & M\\
J1342$-$1413 & H                 & DA     & 11351 $\pm$66 & 0.902 $\pm$0.006 & 0.93 $\pm$0.02  & \\
J1413$+$0755 & H                 & DA     & 15300 $\pm$131 & 0.936 $\pm$0.005 & 0.45 $\pm$0.01  & \\
J1440$-$1951 & H                 & DAH?   & 14102 $\pm$146 & 1.105 $\pm$0.005 & 1.05 $\pm$0.04  & M\\
J1451$+$5110 & H                 & DA     & 16621 $\pm$172 & 0.928 $\pm$0.007 & 0.35 $\pm$0.01  & \\
J1451$-$2502 & H                 & DA     & 11189 $\pm$97 & 1.016 $\pm$0.011 & 1.53 $\pm$0.07  & \\
J1459$-$0411 & H                 & DAH   & 17122 $\pm$214 & 1.189 $\pm$0.005 & 0.84 $\pm$0.04  & M \\
J1501$+$3431 & H                 & DA     & 21525 $\pm$214 & 0.924 $\pm$0.007 & 0.16 $\pm$0.01  & \\
J1514$+$2313 & H                 & DA     & 17440 $\pm$210 & 1.280 $\pm$0.003 & 1.11 $\pm$0.03  & \\
J1519$+$6329 & H                 & DA     & 20046 $\pm$411 & 1.266 $\pm$0.005 & 0.79 $\pm$0.04  & V\\
J1529$+$2928 & H                 & DA     & 11146 $\pm$106 & 0.968 $\pm$0.010 & 1.24 $\pm$0.06  & R\\
J1537$+$8419 & H                 & DC     & 20897 $\pm$482 & 1.174 $\pm$0.009 & 0.41 $\pm$0.03  & M\\
J1543$+$3021 & H                 & DAH    & 15661 $\pm$198 & 1.129 $\pm$0.006 & 0.82 $\pm$0.04  & MR\\
J1548$+$2451 & H                 & DAH    & 21662 $\pm$386 & 1.213 $\pm$0.006 & 0.44 $\pm$0.03  & M\\
J1552$+$0039 & H                 & DA     & 13508 $\pm$233 & 1.245 $\pm$0.011 & 1.70 $\pm$0.05 & \\
J1621$+$0432 & H                 & DAH    & 21636 $\pm$197 & 1.278 $\pm$0.003 & 0.69 $\pm$0.02  & M\\
J1622$+$3004 & log(H/He)=$-$1.39 & DQA    & 15878 $\pm$216 & 1.174 $\pm$0.007 & 0.99 $\pm$0.04 &   AV\\
J1626$+$2533 & H                 & DA     & 13202 $\pm$189 & 1.140 $\pm$0.007 & 1.46 $\pm$0.06  & \\
J1630$+$2724 & H                 & DAH    & 11099 $\pm$151 & 1.054 $\pm$0.013 & 1.77 $\pm$0.08  & M\\
J1655$+$2533 & H                 & DA     & 11194 $\pm$75 & 1.287 $\pm$0.002 & 2.22 $\pm$0.02  & \\
J1656$+$5719 & H                 & DA     & 11551 $\pm$143 & 1.047 $\pm$0.010 & 1.54 $\pm$0.07  & \\
J1659$+$4401 & H                 & DAH    & 28170 $\pm$327 & 1.272 $\pm$0.003 & 0.29 $\pm$0.01  & MR\\
J1659$+$6610 & H                 & DA     & 11422 $\pm$106 & 1.114 $\pm$0.006 & 1.94 $\pm$0.05  & V\\
J1706$-$0837 & log(H/He)=$-$1.88 & DQA    & 12008 $\pm$132 & 1.138 $\pm$0.008 & 1.69 $\pm$0.04 &  AV\\
J1707$+$3532 & H                 & DAH    & 22332 $\pm$413 & 1.144 $\pm$0.008 & 0.30 $\pm$0.02  & MR\\
J1709$-$2632 & H                 & DA     & 15374 $\pm$451 & 1.247 $\pm$0.007 & 1.34 $\pm$0.08  & \\
J1710$-$2005 & log(H/He)=$-$1.61 & DQA    & 13762 $\pm$190 & 1.124 $\pm$0.010 & 1.22 $\pm$0.05 &  AV\\
J1719$-$1446 & H                 & DAH?   & 16776 $\pm$150 & 1.170 $\pm$0.005 & 0.82 $\pm$0.03 &  MR\\
J1722$+$3958 & H                 & DA     & 11069 $\pm$109 & 0.997 $\pm$0.010 & 1.45 $\pm$0.07  & \\
J1723$+$0836 & H                 & DAH    & 13830 $\pm$195 & 1.076 $\pm$0.007 & 0.97 $\pm$0.06  & M\\
J1728$+$5558 & log(H/He)=$-$1.66 & DQA    & 14331 $\pm$170 & 1.139 $\pm$0.006 & 1.15 $\pm$0.04 &  AV\\
J1744$-$2035 & H                 & DA     & 27136 $\pm$888 & 1.312 $\pm$0.008 & 0.43 $\pm$0.05  & \\
J1758$+$5906 & log(C/H)=$-$2.00  & DQH    & 17251 $\pm$267 & 1.163 $\pm$0.008 & 0.76 $\pm$0.04 &  AM\\
J1800$+$4517 & H                 & DA     & 16447 $\pm$285 & 1.303 $\pm$0.004 & 1.26 $\pm$0.04  & \\
J1812$+$4321 & H                 & DA     & 12442 $\pm$141 & 0.921 $\pm$0.007 & 0.77 $\pm$0.03  & \\
J1813$+$4427 & H                 & DA     & 11146 $\pm$90 & 1.095 $\pm$0.007 & 1.96 $\pm$0.05  & \\
J1819$+$1225 & H                 & DA     & 11929 $\pm$185 & 1.116 $\pm$0.012 & 1.76 $\pm$0.08  & \\
J1819$-$1208 & CO                & hotDQ  & 23800 & 1.243 & 0.42  & A\\
J1822$+$5323 & H                 & DA     & 23836 $\pm$338 & 1.172 $\pm$0.006 & 0.28 $\pm$0.01  & \\
J1832$+$0856 & log(H/He)=$-$5.00 & DBA    & 34056 $\pm$1013 & 1.318 $\pm$0.004 & 0.21 $\pm$0.02  & R\\
J1839$-$0448 & H                 & DA     & 18804 $\pm$222 & 0.963 $\pm$0.008 & 0.27 $\pm$0.01  & \\
J1849$+$6458 & log(H/He)=$-$5.00 &  DQH?   & 12605 $\pm$156 & 1.047 $\pm$0.010 & 1.23 $\pm$0.05  & M\\
J1857$+$3147 & H                 & DA     & 19522 $\pm$337 & 0.985 $\pm$0.010 & 0.26 $\pm$0.01  & \\
J1900$+$7039 & H                 & DAP    & 11615 $\pm$100 & 0.988 $\pm$0.006 & 1.18 $\pm$0.04  & M\\
J1901$+$1458 & H                 & DAH    & 29100 $\pm$474 & 1.319 $\pm$0.004 & 0.35 $\pm$0.02  & MR\\
J1902$+$7728 & H                 & DA     & 17329 $\pm$300 & 1.261 $\pm$0.005 & 1.08 $\pm$0.04  & \\
J1903$+$4657 & H                 & DA     & 20124 $\pm$278 & 1.009 $\pm$0.008 & 0.25 $\pm$0.01  & \\
J1910$+$7334 & H                 & DA     & 13116 $\pm$214 & 1.085 $\pm$0.008 & 1.21 $\pm$0.07  & \\
J1923$-$2328 & H                 & DA     & 21830 $\pm$568 & 1.154 $\pm$0.010 & 0.33 $\pm$0.03  & \\
J1924$-$2717 & H                 & DA     & 17789 $\pm$219 & 1.213 $\pm$0.005 & 0.85 $\pm$0.04  & V\\
J1924$-$2913 & H                 & DAH    & 11421 $\pm$88 & 0.910 $\pm$0.009 & 0.94 $\pm$0.02  & M\\
J1925$-$0346 & log(H/He)=$-$1.84 & DQA    & 14499 $\pm$132 & 1.149 $\pm$0.006 & 1.15 $\pm$0.03 &  A\\
J1928$+$1526 & H                 & DA     & 11621 $\pm$141 & 1.011 $\pm$0.012 & 1.31 $\pm$0.08  & \\
J1928$+$5429 & H                 & DA     & 12914 $\pm$130 & 0.915 $\pm$0.006 & 0.68 $\pm$0.02 & \\
J1929$-$2926 & H                 & DA     & 11455 $\pm$220 & 1.054 $\pm$0.016 & 1.62 $\pm$0.12  & \\
J2011$+$4910 & log(H/He)=$-$2.14 & DQ     & 12789 $\pm$111 & 1.132 $\pm$0.005 & 1.47 $\pm$0.03 &  AV\\
J2012$+$3113 & log(H/He)=$-$5.00 & DBP    & 23225 $\pm$507 & 1.223 $\pm$0.005 & 0.40 $\pm$0.03  & M\\
J2026$+$1848 & H                 & DC     & 14190 $\pm$300 & 1.081 $\pm$0.009 & 0.91 $\pm$0.08  & M\\
J2026$-$2254 & H                 & DA     & 11405 $\pm$88 & 0.917 $\pm$0.008 & 0.95 $\pm$0.03 & \\
J2035$-$1835 & H                 & DAH    & 12164 $\pm$119 & 1.138 $\pm$0.007 & 1.77 $\pm$0.05  & M\\
J2037$-$2857 & H                 & DA     & 14584 $\pm$128 & 0.997 $\pm$0.004 & 0.61 $\pm$0.02  & \\
J2054$-$2039 & H                 & DA     & 19080 $\pm$264 & 1.205 $\pm$0.005 & 0.66 $\pm$0.03 & \\
J2100$+$5142 & H                 & DAH    & 11135 $\pm$126 & 0.988 $\pm$0.012 & 1.37 $\pm$0.08  & MR\\
J2105$+$5903 & H                 & DA     & 18078 $\pm$418 & 1.241 $\pm$0.007 & 0.92 $\pm$0.06  & \\
J2107$+$7831 & H                 & DA     & 12707 $\pm$202 & 1.099 $\pm$0.009 & 1.41 $\pm$0.08 & \\
J2111$+$1102 & H                 & DAH    & 16212 $\pm$203 & 1.128 $\pm$0.007 & 0.73 $\pm$0.04  & M\\
J2133$+$3529 & H                 & DA     & 22115 $\pm$476 & 0.965 $\pm$0.012 & 0.17 $\pm$0.01  & \\
J2148$-$1629 & H                 & DAH    & 12827 $\pm$156 & 0.979 $\pm$0.011 & 0.82 $\pm$0.03  & M\\
J2153$-$2628 & H                 & DA     & 17015 $\pm$254 & 1.218 $\pm$0.006 & 0.98 $\pm$0.04  & \\
J2204$+$2543 & H                 & DAH    & 27944 $\pm$579 & 1.153 $\pm$0.011 & 0.16 $\pm$0.01  & MR\\
J2208$+$2059 & H                 & DA     & 11091 $\pm$99 & 0.941 $\pm$0.011 & 1.12 $\pm$0.05  & \\
J2221$+$4406 & H                 & DAH    & 13076 $\pm$186 & 0.945 $\pm$0.008 & 0.71 $\pm$0.03  & M\\
J2231$-$1546 & H                 & DA     & 15242 $\pm$319 & 1.299 $\pm$0.005 & 1.42 $\pm$0.05 & V\\
J2250$+$3231 & H                 & DA     & 29913 $\pm$610 & 0.994 $\pm$0.013 & 0.06 $\pm$0.01  & \\
J2255$+$0710 & H                 & DAH    & 10974 $\pm$208 & 1.302 $\pm$0.011 & 2.19 $\pm$0.09  & M\\
J2257$+$0755 & H                 & DAH    & 13407 $\pm$128 & 1.298 $\pm$0.003 & 1.72 $\pm$0.02  & MRV\\
J2304$-$2658 & H                 & DA     & 13746 $\pm$46 & 1.064 $\pm$0.008 & 0.93 $\pm$0.07 & \\
J2306$-$2905 & H                 & DA     & 16475 $\pm$213 & 1.215 $\pm$0.006 & 1.05 $\pm$0.04  & \\
J2308$+$0347 & log(H/He)=$-$1.84 & DQA    & 11717 $\pm$40 & 1.081 $\pm$0.005 & 1.62 $\pm$0.02  &  A\\
J2309$+$0940 & H                 & DA     & 17202 $\pm$231 & 0.986 $\pm$0.010 & 0.37 $\pm$0.02  & \\
J2314$-$2208 & H                 & DA     & 23152 $\pm$466 & 0.979 $\pm$0.012 & 0.15 $\pm$0.01  & \\
J2340$-$1819 & log(C/H)=+0.36    & DAQ    & 15836 $\pm$291 & 1.167 $\pm$0.011 & 0.97 $\pm$0.05 &  AVR\\
J2349$+$0907 & H                 & DA     & 16511 $\pm$250 & 1.204 $\pm$0.006 & 1.00 $\pm$0.05  & \\
\hline 
\label{tab:phys}
\end{longtable}

\end{document}